\documentclass{aa}
\usepackage{graphicx,natbib,hyperref}
\usepackage[varg]{txfonts}
\usepackage{pdflscape}
\usepackage{afterpage}
\usepackage{color}

\newcommand{\vsi}{\ensuremath{v\,\sin i}}
\bibpunct[ ]{(}{)}{;}{a}{}{,}

\begin{document}

\title{Long period Ap stars discovered with TESS data}

\author{G.~Mathys\inst{1}
  \and D.~W.~Kurtz\inst{2,3}
  \and D.~L.~Holdsworth\inst{3}}

\institute{European Southern Observatory,
  Alonso de Cordova 3107, Vitacura, Santiago, Chile\\\email{gmathys@eso.org}
\and
Centre for Space Research, Physics Department, North West University, Mahikeng 2735, South Africa
\and
Jeremiah Horrocks Institute, University of Central Lancashire, Preston PR1 2HE, UK}

\date{Received $\ldots$ / Accepted $\ldots$}

\titlerunning{Long period Ap stars}

\abstract
{The TESS space mission has a primary goal to search for exoplanets around bright, nearby stars. Because of the high-precision photometry required for the main mission, it also is producing superb data for asteroseismology, eclipsing binary stars, gyrochronology -- any field of stellar astronomy where the data are variable light curves.}
{In this work we show that the TESS data are excellent for astrophysical inference from peculiar stars that show \textit {no} variability. The Ap stars have the strongest magnetic fields of any main-sequence stars. Some Ap stars have also been shown to have rotation periods of months, years, decades and even centuries. The astrophysical cause of their slow rotation -- the braking mechanism -- is not known with certainty. These stars are rare: there are currently about 3 dozen with known periods.}
{The magnetic Ap stars have long-lived spots that allow precise determination of their rotation periods. We argue, and show, that most Ap stars with TESS data that show no low-frequency variability must have rotation periods longer than, at least, a TESS sector of 27\,d.}
{From this we find 60 Ap stars in the southern ecliptic hemisphere TESS data with no rotational variability, of which at most a few can be pole-on, and six likely have nearly aligned magnetic and rotation axes.  Of the other 54, 31 were previously known to have long rotation periods or very low projected equatorial velocities, which proves our technique; 23 are new discoveries. These are now prime targets for long-term magnetic studies. We also find that 12 of the 54 (22~per~cent) long-period Ap stars are roAp stars, versus only 3~per~cent (29 out of 960) of the other Ap stars studied with TESS in sectors $1-13$, showing that the roAp phenomenon is correlated with rotation, although this correlation is not necessarily causal. In addition to probing rotation in Ap stars, these constant stars are also excellent targets to characterise the instrumental behaviour of the TESS cameras, as well as for the CHEOPS and PLATO missions.}
{This work demonstrates astrophysical inference from nonvariable stars -- we can get ``something for nothing''.} 


\maketitle

\section{Introduction}
\label{sec:intro}
It has long been known that, as a group, Ap stars rotate more slowly than superficially  normal stars of similar temperatures \citep[e.g.,][and references therein]{1974ARA&A..12..257P}. The existence of Ap stars with very long periods, from more than 100\,d to several years, was also recognised long ago \citep{1970stro.coll..254P}. For many years, the number of Ap stars known to have periods longer than 100\,d remained small. However, over the past two decades, many more such long-period Ap stars have been identified, mainly as a by-product of systematic searches for Ap stars with spectral lines resolved into their magnetically split components \citep{1997A&AS..123..353M,2008MNRAS.389..441F,2012MNRAS.420.2727E,2017A&A...601A..14M}. From consideration of the distribution of the periods of these stars, \citet{2017A&A...601A..14M} argued that several per\,cent of the Ap stars must have rotation periods longer than one year, that some of them  definitely  have periods of the order of 300\,yr, and that Ap stars with much longer periods (perhaps $\sim$1000\,yr) may exist. Since the fastest rotating Ap stars have periods of the order of half a day \citep{2004IAUS..215..270M,2017MNRAS.468.2745N}, this implies that the rotation periods of Ap stars span five to six orders of magnitude.

In a broader context, recent surveys made possible by the advent of a new generation of spectropolarimeters have established that large-scale organised magnetic fields similar to those of the Ap stars are present in 5 to 10 per~cent of all stars of spectral types O to early F that have radiative atmospheres \citep{2017MNRAS.465.2432G,2017A&A...599A..66S}. Similar to the Ap stars, the hotter magnetic stars of spectral types O and early B rotate on average more slowly than their non-magnetic counterparts \citep{2017MNRAS.465.2432G,2018MNRAS.475.5144S}, and their rotation periods also span more than five orders of magnitude. Actually, to first order, the cumulative distributions of the rotation periods of the Ap, magnetic early B, and magnetic O stars are very similar \citep[see Figure~21 of][]{2018MNRAS.475.5144S}.

There are many more magnetic Ap stars than magnetic O and early B stars known, both for physical reasons (initial mass function, dependence of stellar lifetime on stellar mass) and for historical reasons (the magnetic fields of Ap stars have been studied for much longer). Furthermore, the rotational variability of most Ap stars, which have very stable atmospheres, can be  diagnosed well not only from magnetic observations, but also from photometric and spectroscopic observations. By contrast, in hotter stars, long-lived inhomogeneities such as those present at the surface of the Ap stars cannot develop, and any rotational variability that may be present is in general vastly exceeded by irregular variations, from which it cannot be untangled. Accordingly, magnetic variations often are the only unambiguous diagnostic of the rotation period of the magnetic stars of spectral types earlier than $\sim$B3. 

Thus, while slow rotation and a spread of rotation rates over five to six orders of magnitude are characteristic of all early-type magnetic stars, the Ap stars represent the targets of choice for the study of these properties. Given the similarities between the magnetic and rotational properties of the Ap stars, on the one hand, and of the magnetic stars of the earlier types, on the other hand, one may reasonably expect the physical understanding gained by studying the former to be also applicable to the latter.

Ap stars experience at most a marginal loss of their angular momentum while on the main sequence \citep{2006A&A...450..763K,2007AN....328..475H}. The evolutionary changes of their rotation periods during their main-sequence lifetimes are small, of the order of a factor 2 at most. Thus, for the most part, period differentiation must have been achieved before Ap stars reached the main sequence, or even before their progenitors become observable as magnetic Herbig Ae/Be stars \citep{2013MNRAS.429.1027A}. The physical mechanisms responsible for this differentiation, and in particular for the existence of a significant population of extremely slowly rotating magnetic stars, remain elusive \citep[see e.g.][]{2014psce.conf...51D,2017RSOS....460271B}. Improving the characterisation of this population is essential to guide and constrain the theoretical developments.

Our current knowledge of the most slowly rotating Ap stars is limited by several factors. First and foremost, an accurate value of the rotation period of a star can be determined only if observations of this star have been obtained at suitable intervals over a time span significantly longer than this period. Otherwise, one may at most derive a lower limit of the period value, which may represent a more or less meaningful constraint according to the case. This limitation has been discussed in detail by \citet{2019A&A...624A..32M}. 

It is important to note here that the first determination of the rotation period of an Ap star was achieved just over 100 years ago \citep{1913AN....195..159B}; that hardly more than 70 years have elapsed since the first detection of a magnetic field in such a star \citep{1947ApJ...105..105B}; that for the star that has been observed in a sustained manner over the longest time base without covering yet a full rotation period, $\gamma$~Equ (= HD\,201601), the available magnetic measurements span $\sim$70 years \citep{2016MNRAS.455.2567B}; and that the longest rotation period that has been accurately determined up to now for an Ap star (HD\,50169 = BD\,$-1$\,1414) is 29~yr \citep{2019A&A...624A..32M}. These time scales are all considerably shorter than the longest periods of 300--1000\,yr that are expected for some Ap stars. The time spans over which the majority of the slowly rotating Ap stars have been followed up are actually even shorter. As a result, until now, periods longer than 1000\,d have been accurately determined for only eight stars \citep{2020A&A...636A...6M}. 

Obviously, the only way to overcome the above-described limitation is to continue to accumulate relevant observations of the stars of interest over the coming years and decades. In other words, this limitation is fundamental. By contrast, steps can be taken now to address other current deficiencies in our knowledge of the most slowly rotating Ap stars. The latter arise in great part from the way in which the very long period Ap stars that are presently known were identified and the reasons why they were studied. Indeed, one of the main interests of these stars is that, thanks to their low projected equatorial velocity (\vsi), some of their spectral lines are resolved into magnetically split components. 

From this magnetic splitting, a magnetic field moment, the mean magnetic field modulus (the line-intensity weighted average over the stellar disk of the modulus of the magnetic vector), can be determined in a very straightforward, mostly approximation-free and model-independent manner \citep{1997A&AS..123..353M}. The knowledge of this field moment is complementary to that of the mean longitudinal magnetic field (the line-intensity weighted average over the stellar disk of the component of the magnetic vector along the line of sight), which can be determined from the analysis of spectra recorded in circular polarisation. This complementarity was already recognised by \citet{1969ApJ...156..967P}, who showed how it could be exploited to derive a simple model of the geometrical structure of an Ap star magnetic field \citep{1970ApJ...160.1059P}. Fifty years on, a similar approach, albeit refined, is still used to build simple models of the magnetic fields of the Ap stars with the longest accurately determined periods, which simultaneously match the variation curves of their mean longitudinal fields and their mean magnetic field moduli (\citealt{2016A&A...586A..85M}; \citealt{2019A&A...624A..32M}; \citealt{2019A&A...629A..39M}; \citealt{2020A&A...636A...6M}).

The magnetic field modelling possibilities provided by observations of Ap stars with resolved magnetically split lines motivated a systematic search for such stars, as reported by \citet{1997A&AS..123..353M}. By definition, \vsi\ is low in these stars, and assuming a random distribution of the inclination, $i$, of their rotation axis to the line of sight, most of them must be expected to be slow rotators. This expectation is fully confirmed by the observed distribution of their rotation periods \citep{2017A&A...601A..14M}. As a result, most long period Ap stars that are known today are strongly magnetic. This bias is further strengthened by the fact that long rotation periods are best determined through the study of magnetic variations, as the ratio of the variation amplitude to the measurement uncertainties is generally much greater for the magnetic moments than for other observables. 

\citet{2017A&A...601A..14M} identified 35 stars that have rotation periods longer than 30\,d. Among them, only four do not show resolved magnetically split lines, that is, have magnetic fields weaker than $\sim$1.7\,kG \citep{1997A&AS..123..353M}. Thus, the occurrence of extremely slow rotation in weakly magnetic Ap stars is poorly characterised. This is especially unfortunate since it would be important to investigate the potential existence of correlations between rotation rate and magnetic field strength (and structure) in Ap stars. Such dependences, if they exist, may provide clues about the origin of the magnetic field and of the slow rotation in these stars.

\begin{landscape}
  \setlength\textwidth{705pt}
  \setlength\textheight{184mm}
  \begin{table*}
\small
\caption{List of long-period Ap stars found by our technique in the TESS Sectors $1-13$ data, obtained in the first year of mission operations and covering the southern ecliptic hemisphere. }
\begin{tabular*}{\textwidth}[]{@{}@{\extracolsep{\fill}}rrlrrrcclrlclll}
\hline\hline\\[-4pt]
  &  &Specrtral&   &  &    &   &   &   &  &    &  &  &TESS  \\
\multicolumn{1}{c}{HD}  &  \multicolumn{1}{c}{TIC}  &type&  \multicolumn{1}{c}{$V$}  & \multicolumn{1}{c}{$T_{\rm  eff}$}   & \multicolumn{1}{c}{$\log g$}   &  \multicolumn{1}{c}{roAp}  &  \multicolumn{1}{c}{$\langle B_z\rangle_{\mathrm{rms}}$/$B_0$/$Q_0$} &  Refs  &  \multicolumn{1}{c}{$P_{\rm rot}$}  &  Refs  & \multicolumn{1}{c}{Lines\tablefootmark{a}}  & Refs & Notes &sectors  \\
  &  &  &  \multicolumn{1}{c}{(mag)}  & \multicolumn{1}{c}{(K)}   & \multicolumn{1}{c}{(cm\,s$^{-2}$)}   &   &  \multicolumn{1}{c}{(kG)}  &   &  \multicolumn{1}{c}{(d)}  &    &  &  &  \\[4pt]
  \hline\\[-4pt]
12932  &  268751602  &  Ap~SrCrEu  &  10.17  &  7319  &    &  roAp  &  1.0/--/--  &  3  &    &    &  s  & 8/C &  BN\,Cet; Nd\,{\sc{iii}} 6145 strong  &  S3  \\
18610  &  280051011  &  A2p~CrEuSr  &  8.17  &  7994  &  3.76  &    &  --/5.7/--  &  1  &    &    &  r  & 1 &    &  S2,13  \\
19918  &  348717688  &  Ap~SrEuCr  &  9.35  &  7793  &  4.06  &  roAp  &  0.8/--/--  &  3 4  &    &    &  vs  & 8/C &  BT\,Hyi; Nd\,{\sc{iii}} 6145 strong &  S1,12-13  \\
27472  &  38586127  &  Ap~Sr  &  9.96  &  7309  &  3.93  &    &  --  &    &    &    & b   & 8/F &    &  S1-5,8-12  \\
29578  &  220399820  &  A4p~SrEuCr  &  8.52  &  7520  &  3.80  &    &  0.8/3.1/3.8  &  1  &  $>$5\,yr  &  1  &  r  & 1 &    &  S5,9  \\
33629  &  77743907  &  Ap~SrCr(Eu)  &  9.03  &  7266  &  3.90  &    &  --/4.8/--  &  1  &    &    &  r  & 1 &    &  S5  \\
42075  &  37354067  &  Ap~EuCrSr   &  8.98  &  7225  &  3.82  &    &  --/8.5/--  &  1  &    &    &  r  & 1 &    &  S6  \\
44226  &  38000192  &  A3p~SrCrEu  &  9.46  &  7707  &  3.86  &    &  --/5.0/--  &  1  &    &    &  r  & 1 &    &  S6    \\
44979  &  124988213  &  Ap~Si  &  6.53  &  12593  &    &    &  --  &    &    &    &    &   &  possible P$_{\rm rot}=3.3$\,d; A$= 17 $\,$\mu$mag &  S6,7  \\
  50169  &  282101454  &  A3p~SrCrEu  &  8.99  &  9556  &  3.75  &    &  1.3/5.1/6.4  &  1 2  &  10600  &  7  &  r  & 1 &    &  S6  \\
50345  &  237564008  &  Ap~EuCr  &  9.70  &  10026  &  4.22  &    &  --  &    &    &    &    &   &  one $\delta$\,Sct peak; $\nu = 13.5469$\,d$^{-1}$   &  S6  \\
  51684  &  156913148  &  F0p~SrEuCr  &  7.96  &  8747  &    &    &  1.4/6.0/6.6  &  1  &  371  &  7  &  r  & 1 &    &  S6-7  \\
52280  &  167751145  &  Ap~SrCrEu  &  9.82  &  7621  &  3.80  &    &  --  &    &    &    & vb & 8/F &    &  S1-8,10-13  \\
52847  &  80283159  &  Ap~CrEu(Sr)  &  8.13  &  8126  &  3.71  &    &  --/4.4/--  &  1  &    &    &  r  & 1 &    &  S7  \\
53722  &  177813458  &  Ap~EuCr(Sr)  &  9.02  &  7658  &  4.04  &    &  --  &    &    &    &    &   &    &  S7  \\
59071  &  49927077  &  Ap~Si  &  9.55  &  10127  &  4.13  &    &  --  &    &    &    &    &    &   &  S7  \\
59435  &  6176523  &  A6IIIp~Sr  &  7.97  &  6599  &    &    &  --/3.0/--  &  1  &  1360  &  7  &  r  & 1 &  SB2  &  S7  \\
61468  &  110709971  &  A3p~EuCr  &  9.85  &  9413  &  4.10  &    &  1.9/6.8/6.4  &  1  &  322  &  7  &  r  & 1 &    &  S7  \\
62261  &  280577153  &  Ap~Si  &  9.74  &  10488  &  4.05  &    &  --  &    &    &    &    &   &    &  S7  \\
62746  &  333819119  &  A0V~(+ApSi)  &  9.11  &  9427  &    &    &  --  &    &    &    &   &    &    &  S7  \\
66821  &  341550034  &  Ap~Si  &  9.99  &  10092  &  4.27  &    &  --  &    &  154.9  &  7  &    &   &    &  S7-9  \\
67337  &  133771486  &  Ap~Si  &  10.10  &    &    &    &  --  &    &    &    &    &  &    &  S7-8  \\
67658  &  80486647  &  Ap~Si  &  9.76  &  12018  &    &    &  --  &    &    &    &    &    & &  S7,9  \\
69013  &  125297016  &  A2p~SrEu  &  9.56  &  7013  &  3.98  &   roAp&  --/4.8/--  &  1  &    &    &  r  & 1 &    &  S7  \\
69578  &  307031171  &  F6III~Sr  &  9.56  &    &    &    &  --  &    &    &    & \phantom{mr}   &   &    &  S1-2,5-6,8-12  \\
  \phantom{1}69638  &  218710779  &  Ap~EuCr  &  9.43  &  8154  &  4.24  &    &  --  &    &    &    & vb & 8/F &    &  S7-8\\
  71376  &  50753571  &  A~Si  &  9.71  &  11191  &  4.11  &    &  --  &    &    &    &    &   &    &  S7  \\
76021  &  355850109  &  A~Si  &  10.34  &  11474  &    &    &  --  &    &    &    &    &    &   &  S9-11  \\
76460  &  356088697  &  A3p~Sr  &  9.80  &  7111  &  3.52  &  roAp  &  --/3.6/--  &  1  &    &    &  r  & 1 &  $\delta$~Sct? $\nu = 55.87$\,d$^{-1}$; ${\rm A} = 57$\,$\mu$mag  &  S9-11  \\
77438  &  401242170  &  Ap~EuCrSr   &  10.07  &  9944  &  4.14  &    &  --  &    &    &    &  s  & 8/F &    &  S9  \\[4pt]
 \hline\\[-4pt]
\end{tabular*}
\tablefoottext{a}{r = resolved; mr = marginally resolved; s = sharp; vs = very sharp; b = broad; vb = very broad}
\tablebib{(1)~\citet{2017A&A...601A..14M}; (2)~\citet{2019A&A...624A..32M}; (3)~\citet{2015A&A...583A.115B}; (4)~\citet{1997A&AS..124..475M}; (5)~\citet{2015MNRAS.446.1347M}; (6)~\citet{2002ASPC..279..365H}; (7)~\citet{2019arXiv191206107M}; (8)~This paper (8/C: based on ESO-CES spectra; 8/F: based on ESO-FEROS spectra; 8/H: based on SALT-HRS spectrum).} 
 \addtocounter{table}{-1}
 \label{table:1}
\end{table*}
\end{landscape}	

\begin{landscape}
  \setlength\textwidth{705pt}
  \setlength\textheight{184mm}
\begin{table*}
\small
\caption{continued.}
\begin{tabular*}{\textwidth}[]{@{}@{\extracolsep{\fill}}rrlrrrcclrlclll}
\hline\hline\\[-4pt]
  &  &Specrtral&   &  &    &   &   &   &  &    &  &  &TESS  \\
\multicolumn{1}{c}{HD}  &  \multicolumn{1}{c}{TIC}  &type&  \multicolumn{1}{c}{$V$}  & \multicolumn{1}{c}{$T_{\rm  eff}$}   & \multicolumn{1}{c}{$\log g$}   &  \multicolumn{1}{c}{roAp}  &  \multicolumn{1}{c}{$\langle B_z\rangle_{\mathrm{rms}}$/$B_0$/$Q_0$} &  Refs  &  \multicolumn{1}{c}{$P_{\rm rot}$}  &  Refs  & \multicolumn{1}{c}{Lines\tablefootmark{a}}   & Refs & Notes  &sectors  \\
  &  &  &  \multicolumn{1}{c}{(mag)}  & \multicolumn{1}{c}{(K)}   & \multicolumn{1}{c}{(cm\,s$^{-2}$)}   &   &  \multicolumn{1}{c}{(kG)}  &   &  \multicolumn{1}{c}{(d)}  &    &  &  &  \\[4pt]
  \hline\\[-4pt]
 85284  &  444094235  &  Ap~EuCr(Sr)  &  9.82  &  13640  &    &    &  --  &    &    &    & s & 8/F &    &  S9-10  \\
85564  &  131725540  &  Ap~(SiCrSr)  &  9.54  &  7667  &  4.10  &    &  --  &    &    &    & vb  & 8/F &    &  S9-10  \\
85766  &  48330947  &  Ap~(SiCr)  &  9.98  &  7428  &  4.09  &    &  --  &    &    &    &    &   &  $\delta$~Sct, $31-35$\,d$^{-1}$; three peaks  &  S9   \\
88241  &  72802368  &  F0p~SrEu  &  8.60  &  7038  &  3.69  &    &  --/3.6/--  &  1  &    &    &  r  & 1 &  possible roAp; 5$\sigma$  104.8\,d$^{-1}$ &  S9  \\
89075  &  168274937  &  A~(pEu)  &  8.53  &  10584  &  3.85  &    &  --  &    &    &    & vb & 8/F &  maybe Am  &  S9   \\
90131  &  73765625  &  Ap~EuCr(Sr)  &  9.49  &  8387  &  3.65  &    &  --  &    &    &    &    &   &    &  S9   \\
92499  &  146715928  &  A2p~SrEuCr  &  8.93  &  7732  &  3.98  &  roAp  &  1.1/8.3/--  &  1  &  $>$5 yr?  &  1  &  r  & 1 &    &  S9-10  \\
93507  &  398501076  &  Ap~SiCr\,pec  &  8.46  &  9760  &  4.01  &    &  2.2/7.2/7.7  &  1  &  556  &  7  &  r  & 1 &    &  S10-11  \\
95811  &  461161123  &  Ap~SrCrEu  &  9.56  &  6925  &  3.77  &    &  --  &    &    &    &    &   &    &  S9  \\
97132  &  81554659  &  Ap~SrCrEu  &  9.84  &  9257  &  3.83  &    &  --  &    &    &    &    &   &    &  S10  \\
106322  &  334505323  &  Ap~EuCr  &  9.39  &  7683  &  3.65  &    &  --  &    &    &    &    &   &    &  S10  \\
119027  &  49332521  &  Ap~SrEu(Cr)  &  9.92  &  6936  &  3.98  &  roAp  &  0.5/3.2/3.7  &  1  &    &    &  r  & 1 &  LZ Hya  &  S11  \\
133792  &  454802988  &  Ap~SrCrEu  &  6.25  &  9700  &  3.61  &    &  --/--/1.1  &  1  &    &    &  vs  & 8/C &  some g\,modes?  &  S12  \\
135396  &  455599747  &  Ap~(SrCr)  &  7.99  &  6820  &  3.44  &    &  --  &    &    &    &  s  & 8/C &    &  S12  \\
139850  &  65313788  &  Ap~Si  &  10.27  &  8187  &    &    &  --  &    &    &    &   &    &    &  S12  \\
141668  &  281965562  &  Ap~Si  &  9.61  &    &    &    &  --  &    &    &    &    &   &    &  S12  \\
147516  &  223456276  &  Ap~Si  &  10.09  &  9446  &  3.71  &    &  --  &    &    &    &    &   &    &  S12  \\
150562  &  44827786  &  A/F~(pEu)  &  9.91  &  7348  &  4.08  &  roAp  &  1.2/4.9/5.7  &  1  &    &    &  r  & 1 &  $\nu_{\rm rot}$ = 0.19 d$^{-1}$?  &  S12  \\
151860  &  170419024  &  Ap~SrEu(Cr)  &  9.01  &  6625  &  3.75  &  roAp  &  --  &    &    &    &  s  & 8/F &    &  S12  \\
156808  &  196639668  &  Ap~EuCr  &  8.65  &  5816  &    &    &  --  &    &    &    &  b  & 8/F &  peak at 3.465 d$^{-1}$; g\,mode?  &  S12  \\
163231  &  260084368  &  Ap~Si  &  9.63  &  12369  &    &    &  --  &    &    &    &    &   &  $\nu_{\rm rot}$ = 0.306 d$^{-1}$?  &  S13  \\
171420  &  291561579  &  Ap~Si  &  10.67  &  6793  &  4.33  &    &  --  &    &    &    &    &   &    &  S13  \\
176196  &  387132889  &  Ap~EuCr(Sr)  &  7.51  &  11711  &  4.27  &    &  0.2/--/--  &  3  &    &    &  vs  & 8/C &    &  S13  \\
185204  &  369871758  &  Ap~SrEuCr  &  9.53  &  7666  &  3.85  &    &  --/5.6/--  &  1  &    &    &  r  & 1 &  $\nu_{\rm rot}$ = 0.080 d$^{-1}$?  &  S13  \\
209364  &  206461701  &  Ap~SrCrEu  &  10.03  &  7188  &  3.57  &    &  --  &    &    &    &    &   &    &  S1  \\
213637  &  69855370  &  F1p~EuSr  &  9.65  &  6607  &  4.08  &  roAp  &  0.2/5.2/5.4  &  1  &    &    &  r  & 1 &  MM Aqr  &  S2  \\
217522  &  139191168  &  Ap~SrSi:  &  7.54  &  6918  &  4.02  &  roAp  &  0.8/--/2.0  &  3 4 5 6   &    &    &  vs  & 8/C &  BP Gru; Nd\,{\sc{iii}} 6145 strong  &  S1  \\
217704  &  12968953  &  Ap~Sr   &  10.17  &  7883  &  3.94  &  roAp  &  --  &    &    &    &  mr  & 8/C &    &  S2  \\
225234  &  266905315  &  A3V  &  8.87  &  8084  &  4.27  &    &  --  &    &    &    &    &   &    &  S1,13  \\
J0651\tablefootmark{b}  &  167695608  & F0p~SrEu(Cr)   &  11.51  &  7185  &  4.05  &  roAp  &  --  &    &    &    & s  & 8/H &    &  S1-4,6-13  \\[4pt]
 \hline\\[-4pt]
\end{tabular*}
\tablefoottext{b}{J0651 = J06514218-6325495.} 
\end{table*}
\end{landscape}

As another consequence of the above-described bias, it is currently impossible to obtain an accurate estimate of the overall rate of occurrence of extremely slow rotation among Ap stars. This significantly hampers theoretical developments, as a complete and exact knowledge of the period distribution represents an essential constraint for the models. In this paper, we show how TESS data can be exploited to improve on our current knowledge of the long-period tail of the distribution of the Ap star rotation periods, and minimise the biases affecting it. Indeed, an exhaustive list of Ap stars was proposed for observation by TESS during the nominal mission \citep{2019MNRAS.487.3523C}. While not all were observed, the selection was based on the overall priorities of the mission, not on the properties of the Ap targets, so that it is essentially unbiassed with respect to the latter. Therefore, any inference about the longest period Ap stars that we derive from the TESS observations is representative of their actual rate of occurrence. In particular, all the stars are dealt with in the same way regardless of the strength of their magnetic field. This ensures that weakly magnetic stars are duly included in the statistics. On the other hand, identifications of new long-period, strongly magnetic stars that may be achieved as part of this study are also valuable, since as previously mentioned, simple models of the geometrical structure of the fields of such stars can be derived in a particularly straightforward manner.

\subsection{TESS data}
\label{sec:tess}

The TESS (Transiting Exoplanet Survey Satellite) Mission \citep{2015JATIS...1a4003R}, launched in 2018 April, is providing a bonanza of photometric data on the Ap stars. It observed over 1000 Ap stars in the southern ecliptic hemisphere in its first year of operations, and these have led to many discoveries of new roAp stars and of rotation periods for Ap stars from their spot variations \citep{2019MNRAS.487.3523C,2019MNRAS.487.2117B,2019MNRAS.487..304D}. 

Decades of studies of magnetic Ap stars have shown that most, and possibly even all, have spots that cause photometric variations with rotation. It is thus apparent that Ap stars observed with TESS that do {\it not} show evidence of rotational variations must either be nearly pole-on so that there is no change of aspect of the spots, or must have the magnetic obliquity to the rotation axis very small so there is little change in the aspect of the spots with rotation (see Sect.~\ref{sec:disc}), or the rotation period of the star must be longer than the TESS data set for that star, at least one Sector of 27\,d. 

The spots in Ap stars are known to be stable over decades, or more, hence we expect to see rotational variation caused by the varying aspect of the spots as long as the geometry of the observation is such that substantially different regions of the stellar surface come into view at different rotation phases. A study of HD~142070  \citep{2017A&A...601A..14M} showed a large amplitude magnetic curve with a deduced inclination of $i = 8^\circ$. Spot and magnetic variations are typically seen together in Ap stars, hence we deduce that TESS data should detect spot variations when $i > 5^\circ$, where $i$ is the inclination of the rotation axis to the line of sight. The probability of nearly pole-on viewing angle goes as $1 - \cos i$, thus if we define `nearly pole-on' as $i \le 5^\circ$, the probability of this is only 0.4 per~cent for random inclinations. With a study of about 1000 Ap stars with TESS, we  expect only about 4 to show no indication of rotation because of their inclination. The fraction of Ap stars with low magnetic obliquity is more difficult to estimate, but it is not expected to be large, and a high-resolution spectrum is sufficient to identify such false positives, i.e., stars with no photometric spot variations because their magnetic and rotation axes are nearly aligned.
Ap stars without spots are not expected, so all other Ap stars that show no rotational variation in the TESS data are likely to have rotation periods longer than 27\,d, or more where there are data for more than one sector. We have searched Sectors $1 - 13$ of the first year of TESS data (the southern ecliptic hemisphere) and found 60 such stars. They are listed in Table\,\ref{table:1}, along with many of their attributes, references and notes. 

Since all but one of the stars have an HD number, and  these stars are known in the literature by their HD numbers, we have chosen to order the table by those in Column~1. Column~2 gives the TIC number. Column~3 gives spectral classifications; all of those are some type of Ap star, since that was a selection criterion. Columns 4, 5 and 6 give $V$, $T_{\rm eff}$ and $\log g$ where there are estimates of those available. The uncertainties of $T_{\rm eff}$ and $\log g$ are likely to be 100s K and up to 0.3\,dex, respectively; we have not rounded the values taken from the compilations to reflect these uncertainties. The data in Columns $3 - 6$  were taken from the TIC (version 8.0) or SIMBAD compilations.  Column~7 indicates whether a star is a known roAp star. Column~8 gives information about the measured magnetic field strengths. All three values correspond to the mean over a rotation period (if known) or over the existing observations (otherwise) of the range spanned by each of the following magnetic field moments: $\langle B_z\rangle_{\rm rms}$ is the root-mean-square longitudinal field, as defined by \citet{1993A&A...269..355B}. It is essentially the quadratic mean of the mean longitudinal magnetic field. Since the latter is a signed quantity, and may reverse its sign over a rotation cycle, taking the quadratic mean is the way to avoid cancellation and to define a parameter that characterises the actual strength of the longitudinal field. $B_0$ is the average value over a rotation cycle of the mean magnetic field modulus $\langle B\rangle$; this is the same quantity as listed in Column~3 of Table~13 of \citet{2017A&A...601A..14M}. $Q_0$ is the average value over a rotation cycle of the mean quadratic magnetic field $\langle B_{\rm q}\rangle$; this is the same quantity as listed in Column~10 of Table~13 of \citet{2017A&A...601A..14M}. Column~10 gives measurements for the stellar rotation periods, with the following column giving the references for those; the references are for compilations, where the original source reference can be found. Column~12 gives a comment on the width and the magnetic resolution of the spectral lines, from the source identified in Col.~13 (see below).  Column~14, the penultimate column, gives various notes, including variable star names, primarily for the roAp stars, since the stars studied here, by definition, do not show rotational variation and hence are not $\alpha^2$\,CVn stars; that column also notes that some of these stars show $\delta$~Sct variations. The final column indicates in which TESS 27-d sectors the star was observed.

The line profile information appearing in Col.~12 comes in part from the literature, and in part from the visual inspection of high-resolution spectra that we obtained within the framework of previous projects, such as, for instance, the systematic search for Ap stars with resolved magnetically split lines \citep{1997A&AS..123..353M,2008MNRAS.389..441F,2012MNRAS.420.2727E,2017A&A...601A..14M}. The 19 stars of Table~\ref{table:1} that show such magnetically resolved lines were listed in Tables~1 and 2 of \citet{2017A&A...601A..14M}. In addition, in a spectrum of HD~217704 recorded at a resolving power $R=10^5$ with the Coud\'e Echelle Spectrograph (CES) of the European Southern Observatory (ESO), the Fe~{\sc ii} $\lambda\,6149.2$ line appears marginally resolved into its magnetic components. The ten stars listed in Table~\ref{table:1} as having sharp, or very sharp, unresolved lines were all visually identified in spectra obtained with the ESO CES ($R=10^5$, 6 stars), the Fibre-fed Extended Range Optical Spectrograph (FEROS) at ESO ($R=4.8\times10^4$, 3 stars: HD~77438, HD~85284, and HD~151860), and the High Resolution Spectrograph (HRS; \citealt{2010SPIE.7735E..4FB}) on the Southern African Large Telescope (SALT), which was used to acquire high-resolution ($R=6.5\times10^4$) spectra of J06514218-6325495 at five different epochs spread over $\sim$5 months. The star shows no hint of variability over the considered time interval. The sharp, unresolved lines of some of these ten stars have also been noted by other authors, but the present characterisation of their line profiles is based exclusively on our spectra listed above.

The rotation periods of six of the stars of Table~\ref{table:1} have been accurately determined; all are longer than 150\,d. Five of these stars show resolved magnetically split lines. We are not aware of the existence of high-resolution spectra of the sixth one (HD~66821), whose period has been obtained from photometry. In addition, the periods of two more stars with magnetically resolved lines (HD~29578 and HD~92499) must be longer than 5\,yr. In summary, we definitely know that 31 of the 60 stars of Table~\ref{table:1} have either long periods or low projected equatorial velocity (or both).

\begin{figure*}
\centering
\includegraphics[width=0.48\linewidth,angle=0]{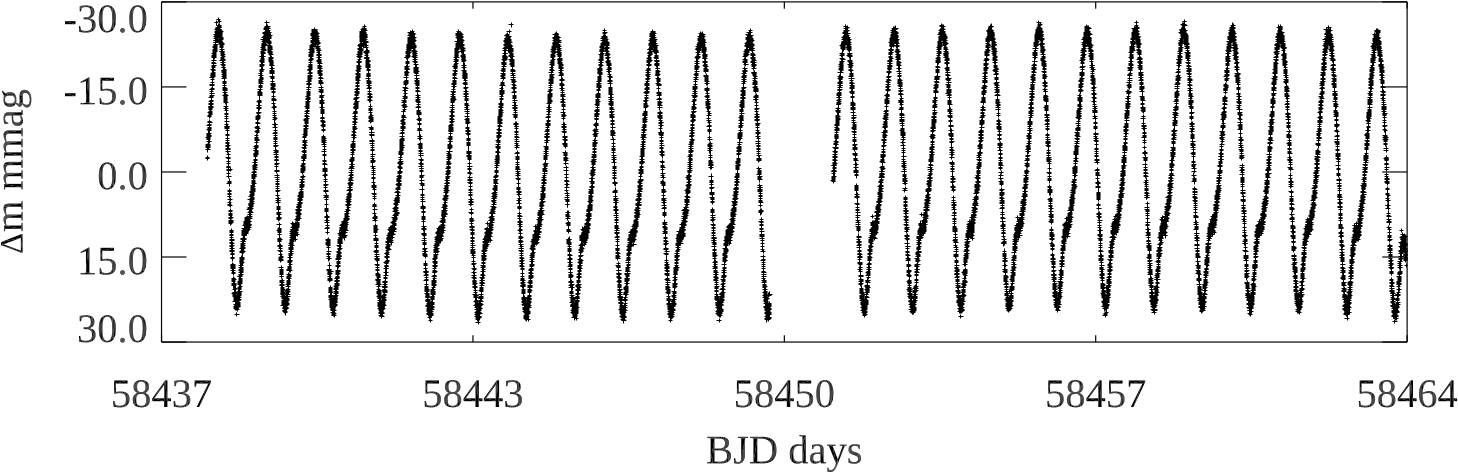}	
\includegraphics[width=0.48\linewidth,angle=0]{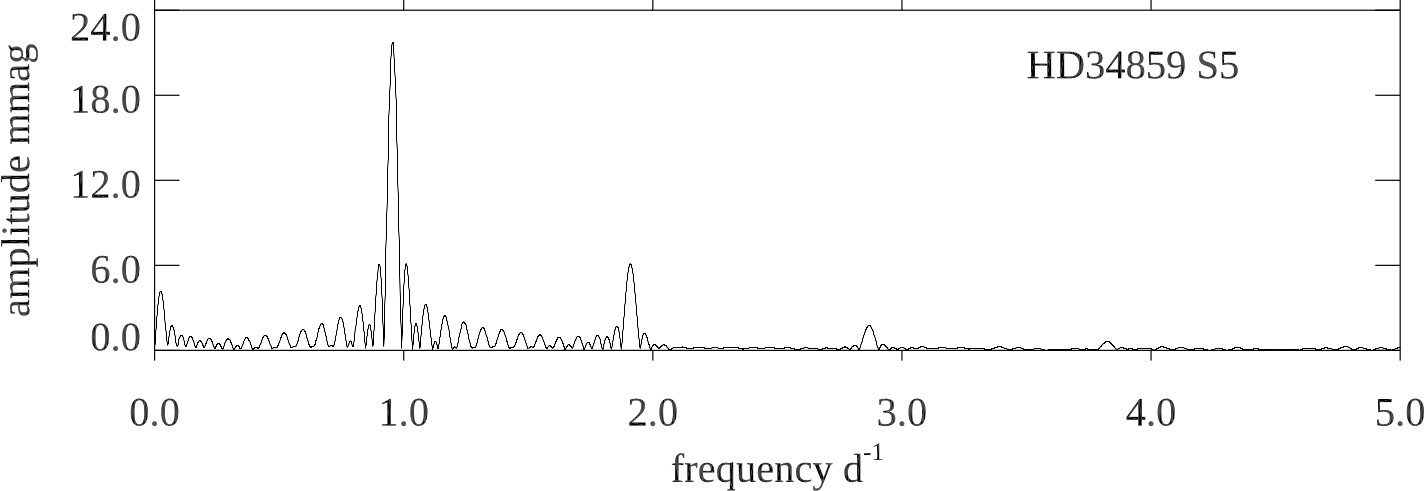}	
\includegraphics[width=0.48\linewidth,angle=0]{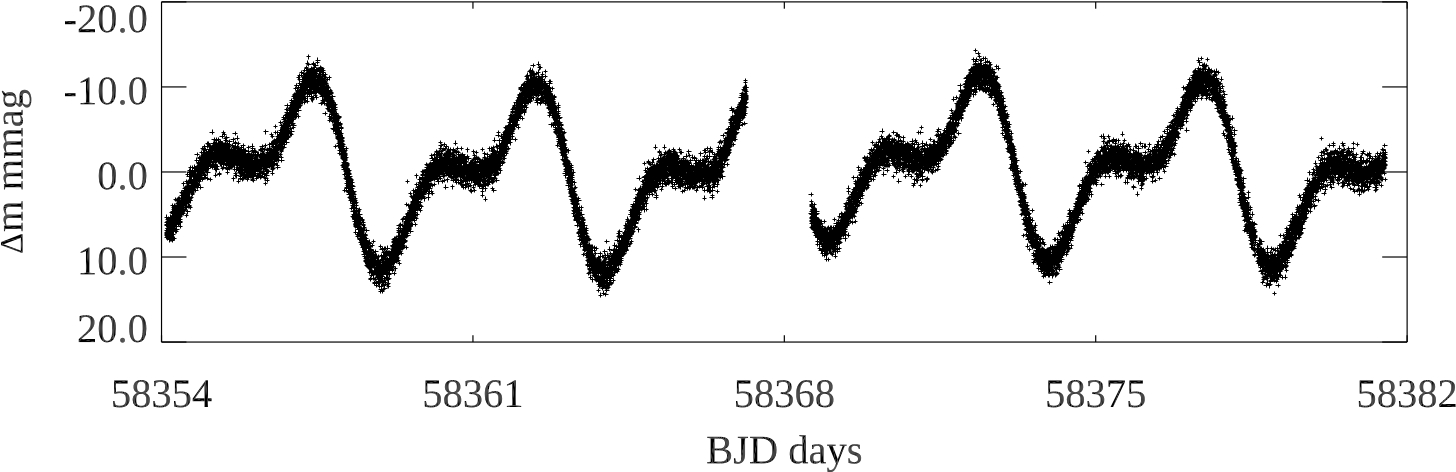}	
\includegraphics[width=0.48\linewidth,angle=0]{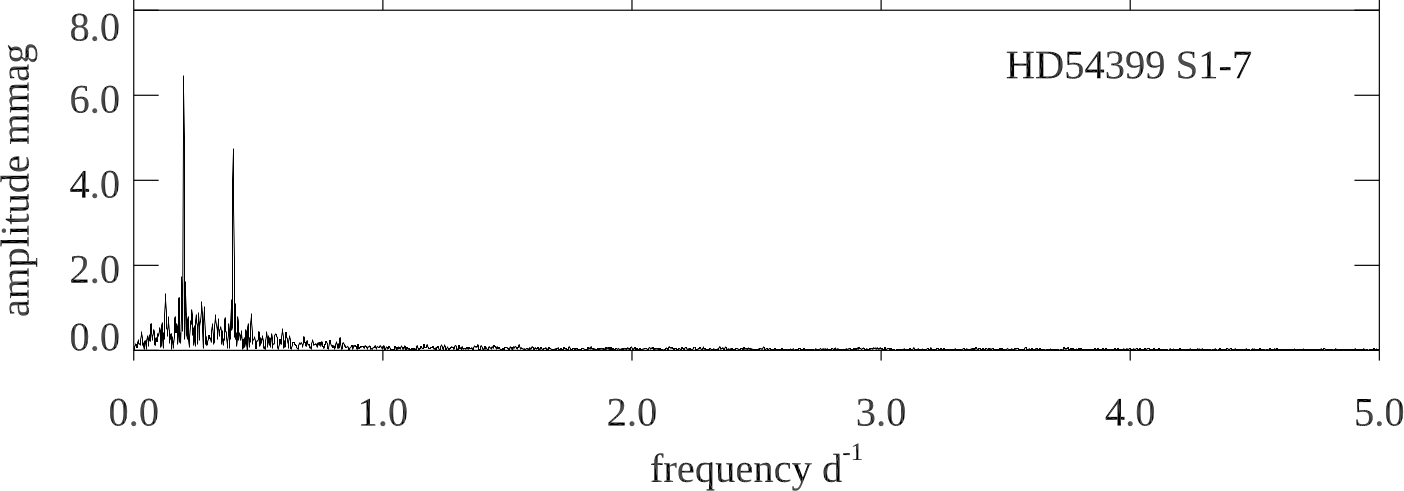}	
\includegraphics[width=0.48\linewidth,angle=0]{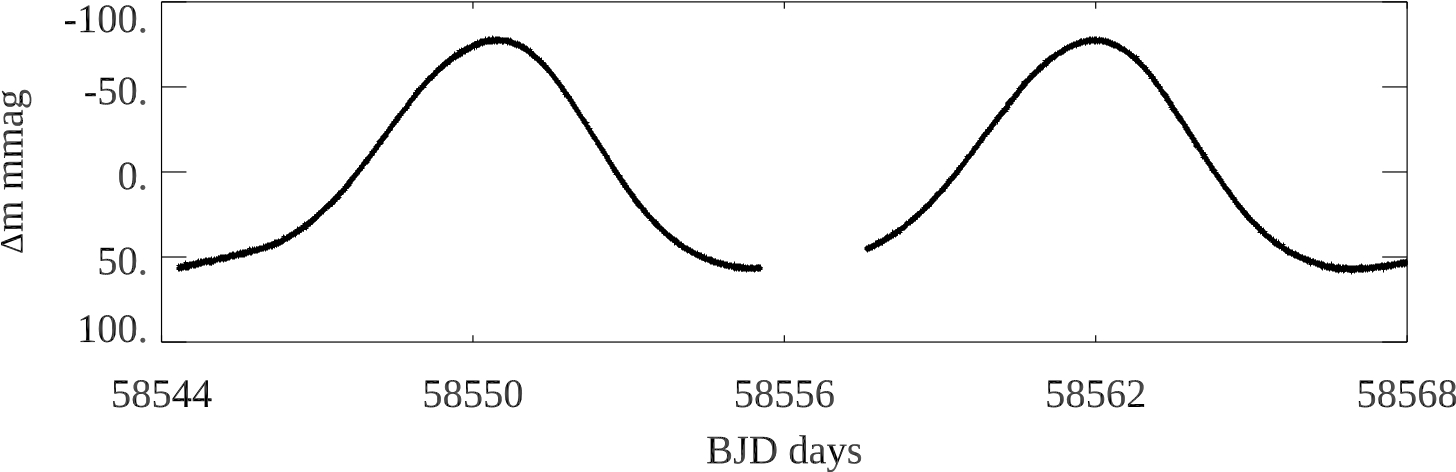}	
\includegraphics[width=0.48\linewidth,angle=0]{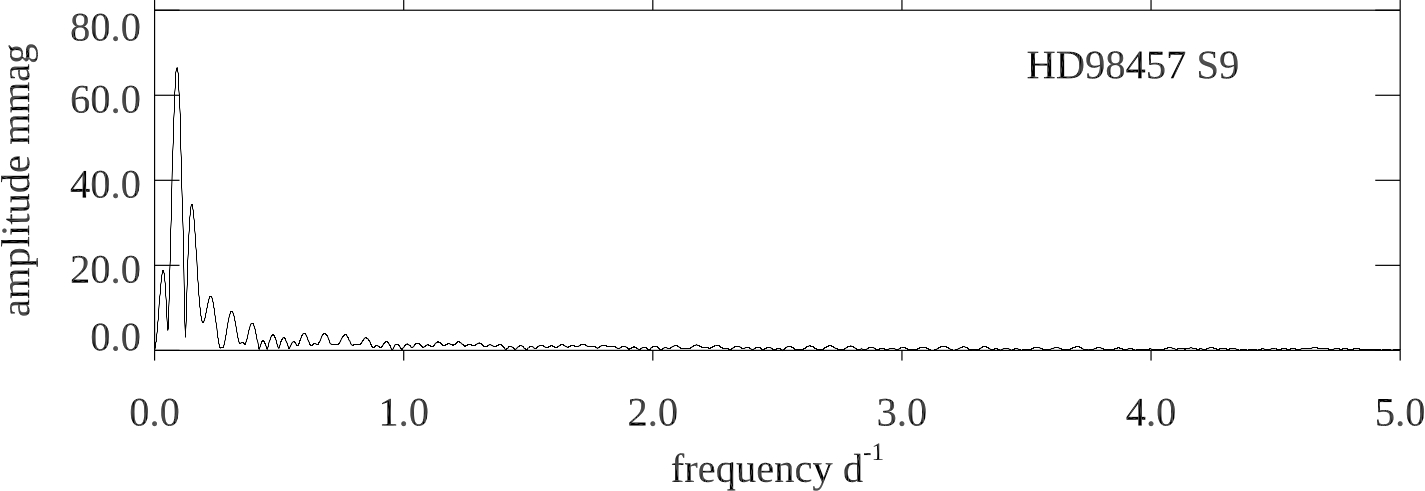}	
\caption{Top left: The Sector 5 light curve of HD~34859 (TIC~4070682), an ApSi star, showing clear rotational variations with a period of $P_{\rm rot} = 1.046 \pm 0.001$\,d. This is a relatively short rotation period Ap star. Top right: An amplitude spectrum for the Sector 5 data set for  HD~34859 showing the rotational harmonics. Middle left: The Sector 2 light curve of  HD~54399 (TIC~348898673), an ApSr(CrEu) star, showing the clear rotational variations with a period of $P_{\rm rot} = 5.015 \pm 0.001$\,d. This is a typical rotation period Ap star of a few days. Middle right: An amplitude spectrum for the Sector $1-7$ data set for HD~54399 showing the rotational harmonics. Note the differences in shape and amplitude of the light curves; these are a function of the size of the spots and their geometry with respect to the line of sight. Bottom left: The Sector 9 light curve of HD~98457  (TIC~57065604), an ApSi star, showing the clear rotational variations with a period of $P_{\rm rot} = 11.5394 \pm 0.0003$\,d. This is still a relatively short rotation period Ap star. Bottom right: An amplitude spectrum for the Sector 9 data set for HD~98457 showing the rotational harmonics. These rotational light curves are typical for magnetic Ap stars, hence set a standard for the Ap stars to follow that do not show rotational variations. Note that the ordinate scales are different for each star, and note the scales themselves for comparison with the stars to come that do not show rotational variability. The units for the abscissa in the left-hand panels is in BJD$-$2\,400\,000.0.} 
\label{fig:lc1}
\end{figure*}

At least some magnetic field measurements exist for 24 of the 60 stars of Table~\ref{table:1}. Nineteen of them have resolved magnetically split lines, and the other five have sharp or very sharp, unresolved lines. Nothing is known about the spectral line shapes of 24 of the 36 stars for which magnetic data have not been obtained yet. Of the remaining twelve, five have sharp spectral lines and six have broad or very broad lines. The latter definitely are not ``super-slowly rotating Ap (ssrAp) stars'' \citep{2019arXiv191206107M}; they will be further discussed in Sect.~\ref{sec:disc}. Finally, in the spectrum of HD~217704, the marginal resolution of the Fe~{\sc ii} $\lambda\,6149.2$ line is insufficient to allow the mean magnetic field modulus to be reliably derived by measuring the wavelength separation of its split components.

Our primary goal for the stars of Table~\ref{table:1} is to begin long-term monitoring of their magnetic field strengths to increase the small number of known slowly rotating magnetic Ap stars for the purposes of understanding the mechanism of magnetic braking, and of establishing if very slow rotation occurs more frequently in strongly magnetic than in weakly magnetic stars. 

We have a secondary goal to examine the fraction of such stars that are roAp pulsators. This incidence appears to be higher for more slowly rotating Ap stars, although we do not know if this correlation is causal. In Table~1 there are 12 roAp stars out of 54 ssrAp stars (22 per~cent). This is greater that the fraction of roAp stars found in a study of all magnetic Ap stars in Sectors 1 and 2 by \citealt{2019MNRAS.487.3523C}, which was 12 roAp stars out of 83 Ap stars, or 14~per~cent. This latter sample was somewhat biassed towards a larger fraction of roAp stars, since known roAp stars were particularly targeted for observations by TESS. A more general sample comes from a study of the entire southern ecliptic hemisphere observations in Sectors~1--13 (Holdsworth et al., in preparation). In that sample there are 41 known roAp stars (\citealt{2019MNRAS.487.3523C}; \citealt{2019MNRAS.487.2117B}) out of 1014 stars with spectral classification Ap, a fraction of 4~per~cent. If we look only at the sample that are not ssrAp, then there are 29 roAp stars out of those 960 Ap stars, or 3~per~cent. 

The temperature range of the currently known roAp stars is $6400 < T_{\rm eff} < 8800$\,K. Neither the ssrAp sample in Table~1, nor  the $>$1000 Ap stars selected for TESS observations were based on $T_{\rm eff}$; the former were selected for lack of rotation signal, and the latter solely by spectral classification Ap with no restriction by sub-type (e.g., Si or SrCrEu or permutations of those). It thus appears that the incidence of roAp stars among ssrAp stars is higher than for Ap stars in general. Whether that is directly a function of rotation, or is rather dependent on some other parameter associated with roAp pulsation, such as magnetic field strength or $T_{\rm eff}$, is not yet known.

\citet{2019MNRAS.485.2380M} studied 15\,000 {\it Kepler} A and F stars to find the pulsator fraction in the instability strip where it crosses the main-sequence. They found that the incidence of $\delta$~Sct stars peaks at 70~per~cent in the middle of the strip, and is about 50~per~cent across the whole strip. Many other stars show low frequency variations from g~modes, as in the $\gamma$~Dor stars, and r~modes \citep{2018MNRAS.474.2774S} so that A stars that are truly photometrically constant are in the minority. The reasons for the non-variability of those stars in the $\gamma$~Dor and $\delta$~Sct instability strips are generally not known. Besides the study of long-period Ap stars we are presenting in this paper, those constant stars are also useful for probing the instrumental characteristics of the TESS cameras, and other stellar photometric missions, such as CHEOPS\footnote{https://cheops.unibe.ch} and PLATO\footnote{https://platomission.com}.

\section{Searching for long period Ap stars}
\label{sec:search}
We have used TESS 2-min cadence SPOC (Science Processing Operations Center; \citealt{2016SPIE.9913E..3EJ})
data for 1014 stars that were classified as, or suspected to be Ap stars. These data were taken with the higher 2-min cadence to search for, and to study, the rapidly oscillating Ap (roAp) stars, in particular \citep{2019MNRAS.487.3523C}. But they are also very useful for studying the rotational variations of Ap stars, although eventually the 30-min cadence Full Frame Images (FFIs) of the TESS main mission, and the upcoming 10-min cadence FFIs for the TESS extended missions, will provide much larger data sets for this purpose. Until those are available, we take advantage of the 2-min cadence data in our search for astrophysically interesting, and relatively rare, long period Ap stars, to initiate the project and as a proof-of-concept. 

\subsection{Stars showing rotational modulation}
\label{sec:spots}

To set a benchmark for comparison, we show here typical examples of rotational variations in some Ap stars. These can then be compared with the non-variable Ap stars we have found. Figure\,\ref{fig:lc1} shows example light curves and amplitude spectra for three typical rotationally variable Ap stars, also known as $\alpha^2$\,CVn stars. 

\subsection{Stars not showing rotational modulation}
\label{sec:withoutspots}

For each of the 1014 stars in our sample we calculated an amplitude spectrum in the frequency range $0 - 10$\,d$^{-1}$ of the 2-min cadence {\sc{pdc\_sap}} light curves generated by the SPOC. We fitted, by linear least squares, the highest amplitude peak in this frequency range to the data to obtain an estimate of the error on the amplitude. This error was then used to calculate the signal to noise ratio (S/N) of the peak. If the S/N was found to be less than 10, an amplitude spectrum was calculated from zero to the Nyquist frequency of 360\,d$^{-1}$ and stored for human inspection. This enabled us to reject most stars where variability is present, but the process also rejects stars where the pipeline has injected signal at low frequency, or where orbit-to-orbit variations of the TESS satellite have not been corrected optimally. In future work, we will reassess this method, especially when we have fully processed FFI data.

\begin{figure*}
\centering
\includegraphics[width=0.48\linewidth,angle=0]{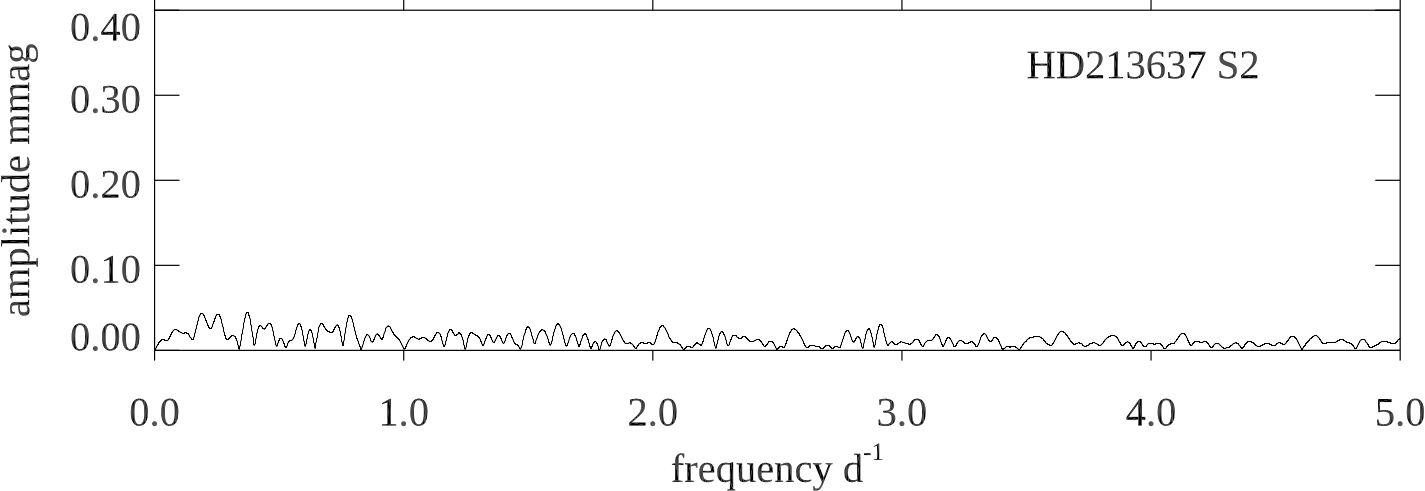}	
\includegraphics[width=0.48\linewidth,angle=0]{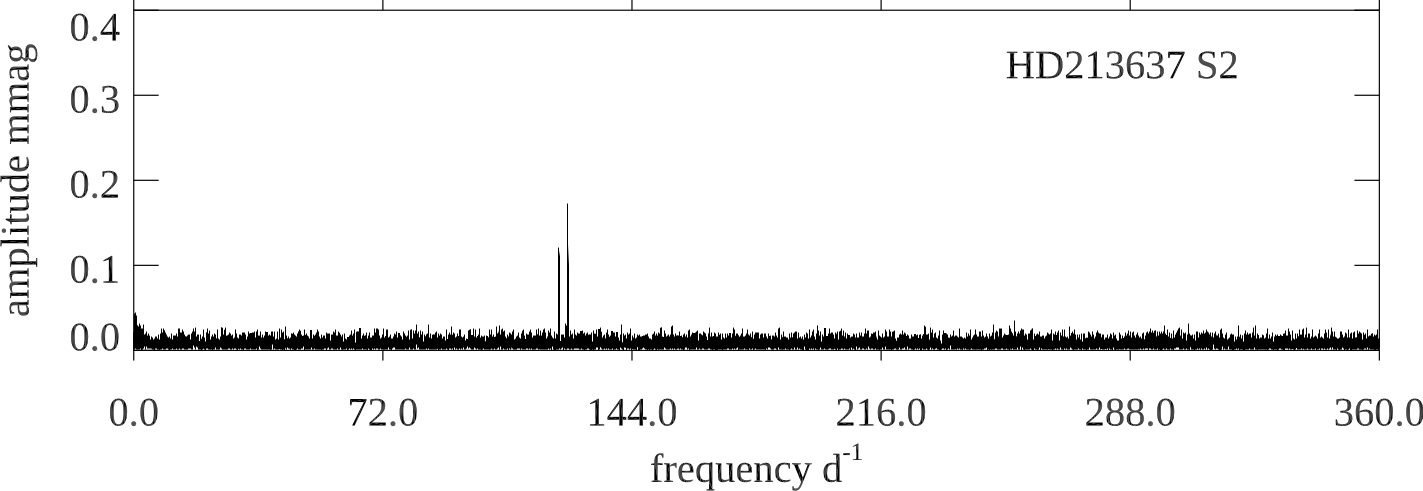}	
\includegraphics[width=0.48\linewidth,angle=0]{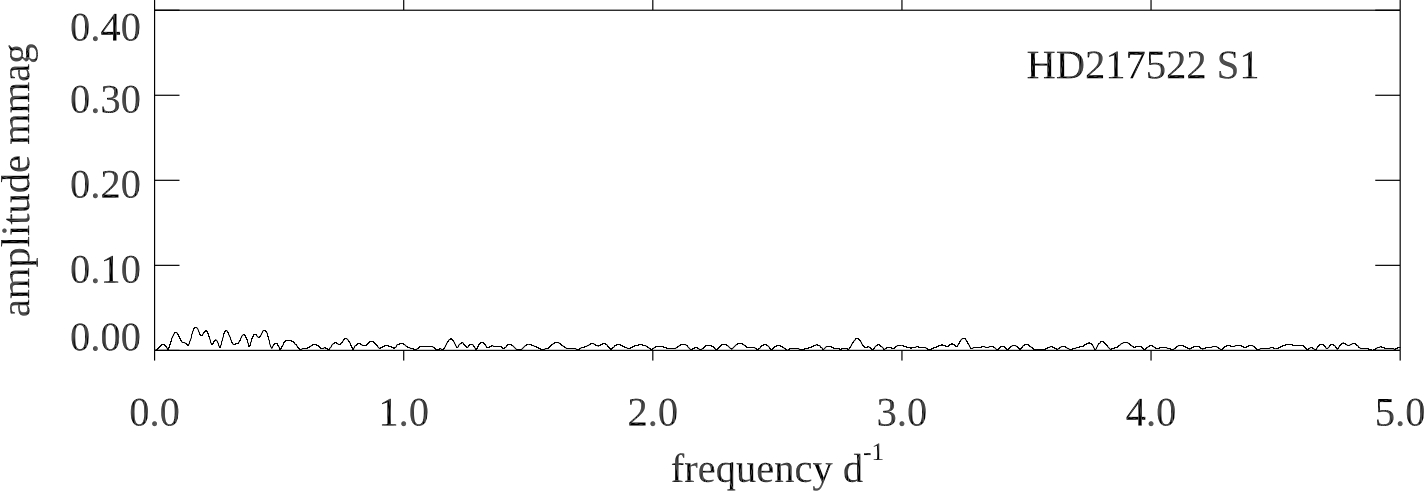}	
\includegraphics[width=0.48\linewidth,angle=0]{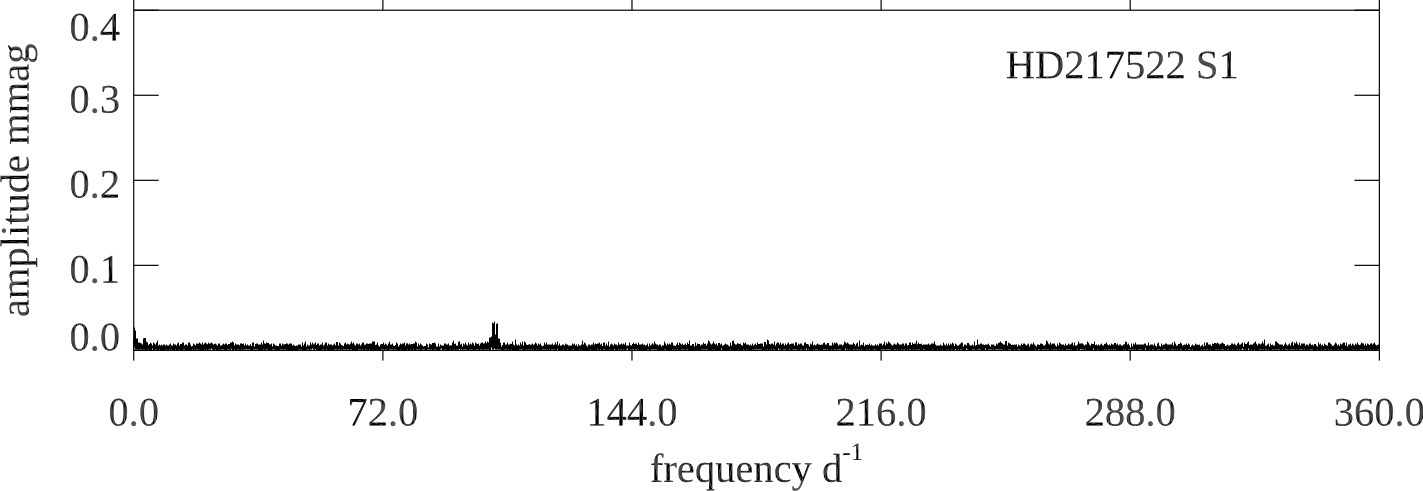}	
\includegraphics[width=0.48\linewidth,angle=0]{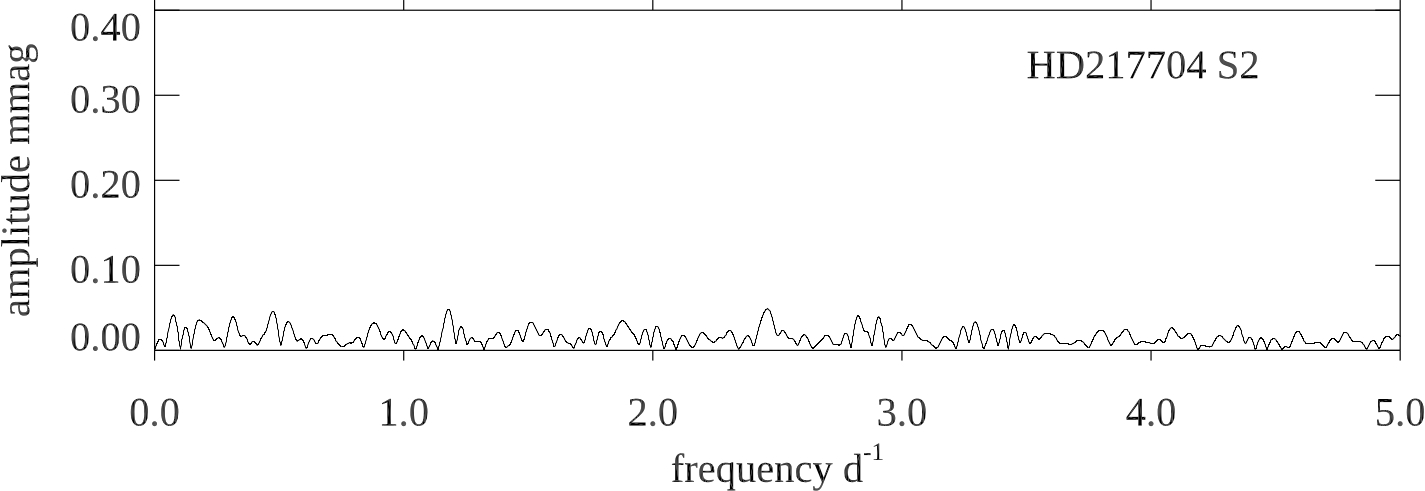}	
\includegraphics[width=0.48\linewidth,angle=0]{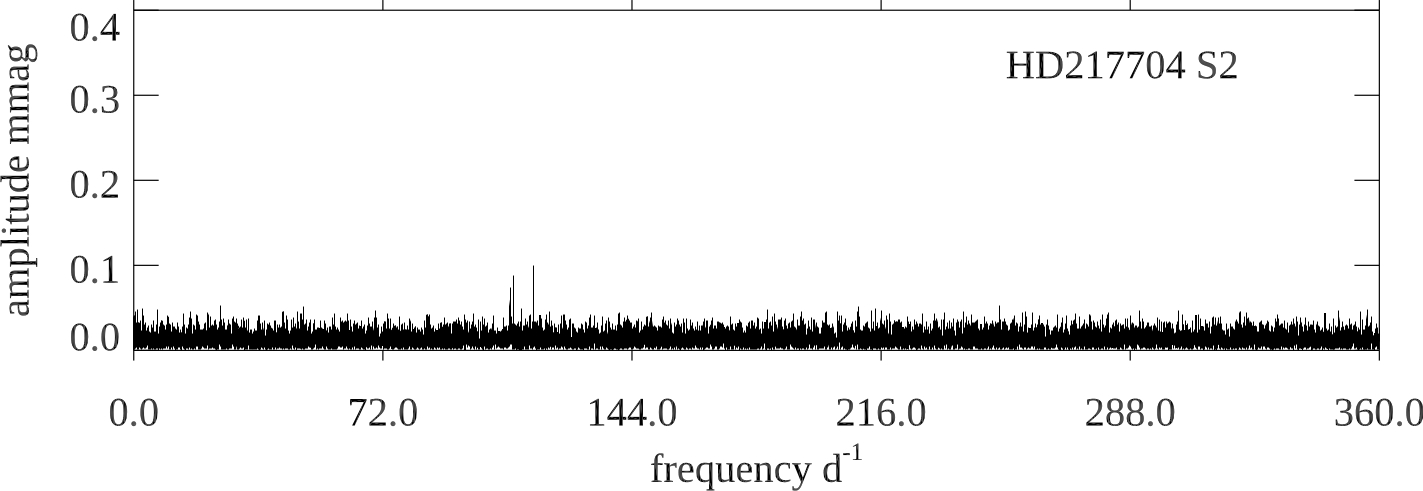}	
\includegraphics[width=0.48\linewidth,angle=0]{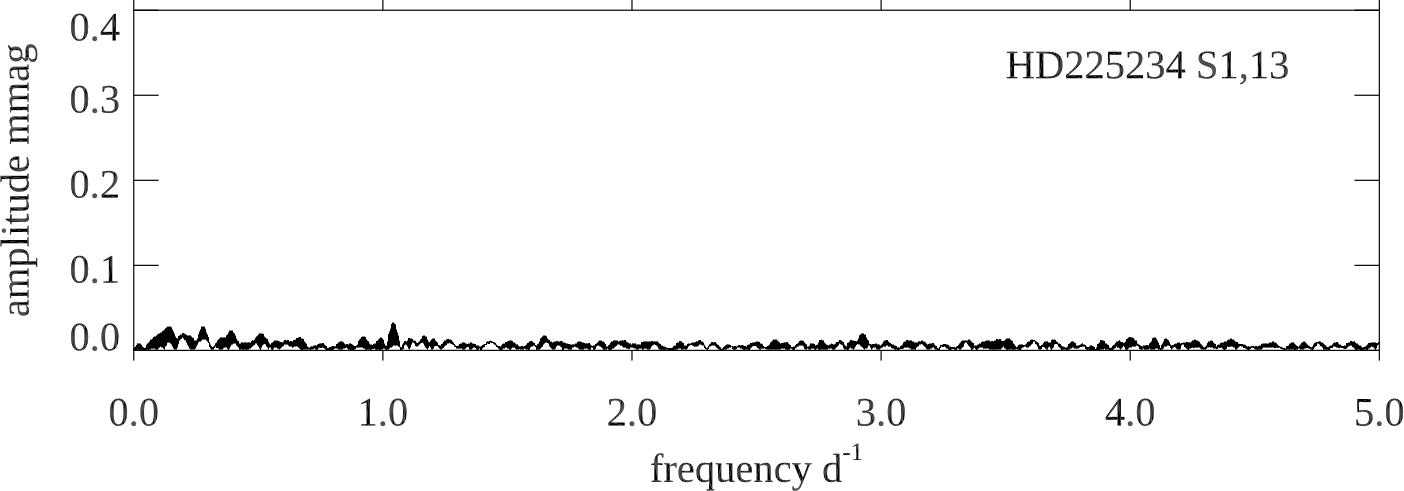}	
\includegraphics[width=0.48\linewidth,angle=0]{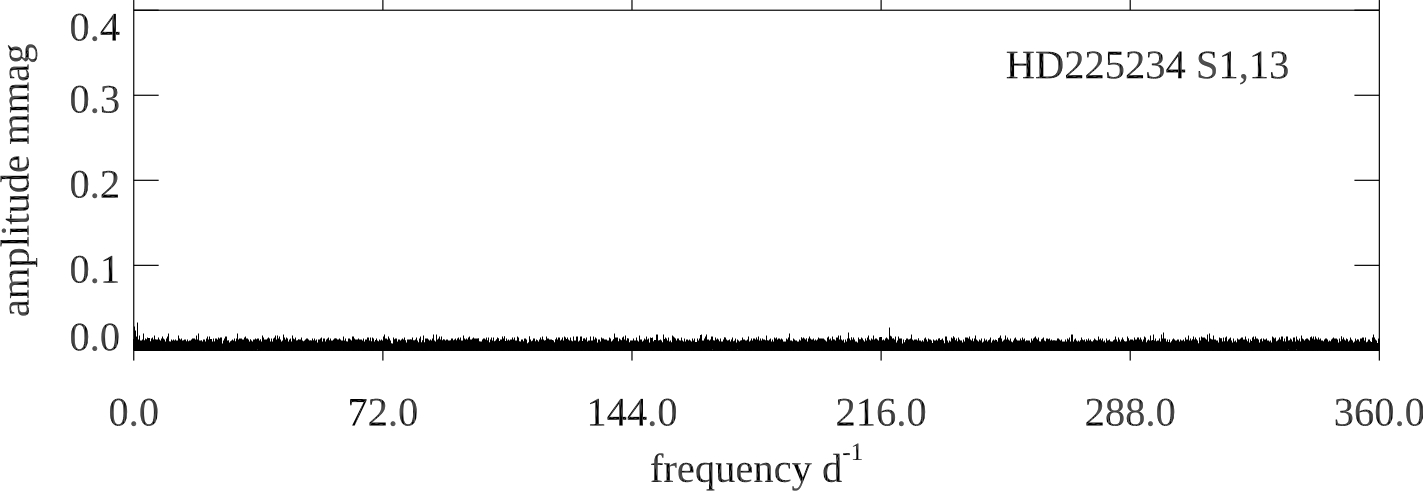}	
\includegraphics[width=0.48\linewidth,angle=0]{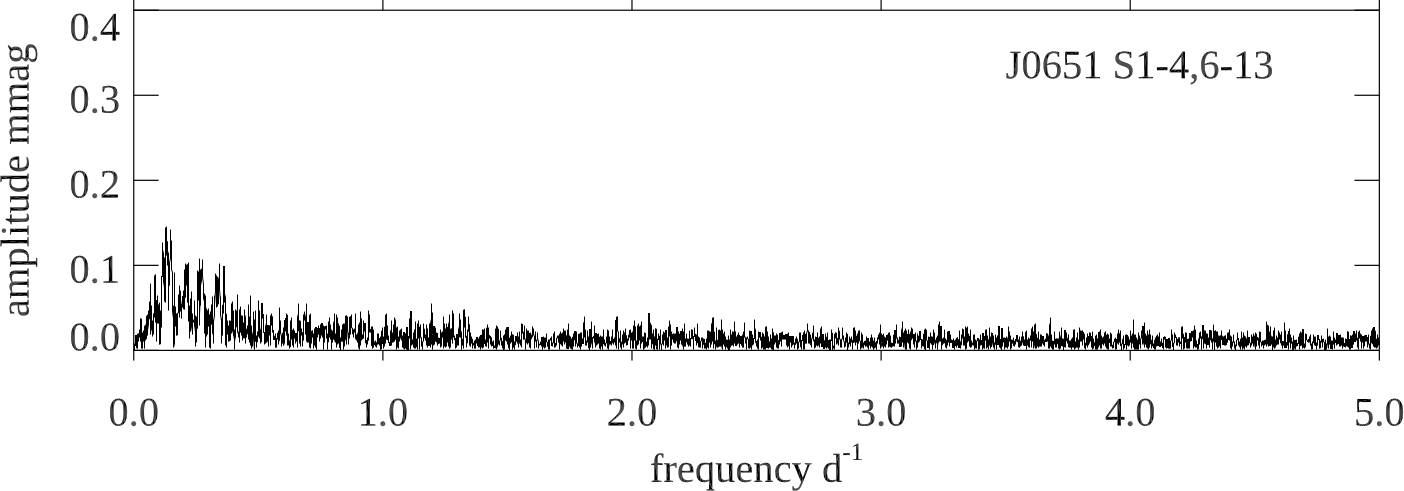}	
\includegraphics[width=0.48\linewidth,angle=0]{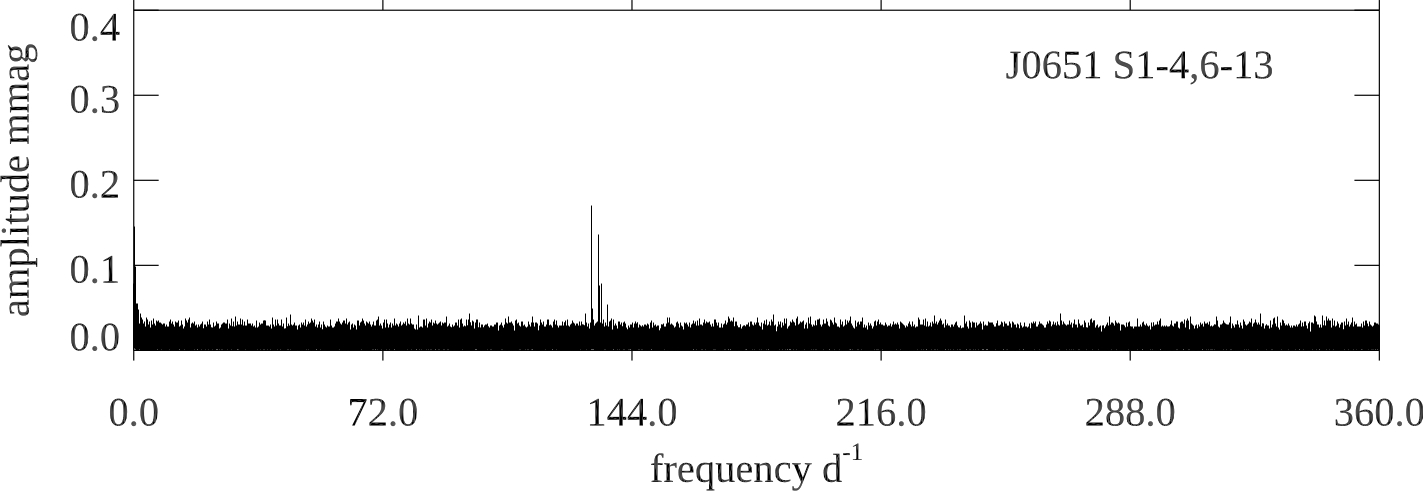}	
\caption{Amplitude spectra for the long-period Ap stars. Each row presents a low-frequency amplitude spectrum showing no rotational variation in the left panel, and a full amplitude spectrum to the Nyquist frequency of 360\,d$^{-1}$ in the right panel, which sometimes indicates pulsation, usually of the roAp variety. Note the large reduction in the ordinate scale compared to Fig.\,\ref{fig:lc1}, and other changes in that scale from panel-to-panel. HD~213637 has no rotational variation, but is a known roAp star (see, e.g.,  \citealt{1998A&A...334..606M}; \citealt{2015MNRAS.446.4126E}; \citealt{2019MNRAS.487.3523C}) with two roAp peaks at $122.8998$\,d$^{-1}$ ($A = 120$\,$\mu$mag) and $125.4853$\,d$^{-1}$ ($A = 172$\,$\mu$mag); these two frequencies have a separation of 30\,$\mu$Hz, which is plausibly half the large separation. 
HD~217522 has no clear rotation variation, but obvious roAp variations around 104\,d$^{-1}$, which are well-known (see, e.g., \citealt{2015MNRAS.446.1347M}). 
HD~217704 is a known roAp star (\citealt{2019MNRAS.487.3523C}) with peaks in the range $108 - 116$\,d$^{-1}$.
J06514218-6325495 is a known roAp star discovered by \citet{2014MNRAS.439.2078H} with a study of the TESS data by \citet{2019MNRAS.487.3523C} showing frequencies in the range $132  - 135$\,d$^{-1}$. }
\label{fig:fts_1}
\end{figure*}

Figure\,\ref{fig:fts_1} shows amplitude spectra for some of the stars we have identified as long rotation period Ap stars, meaning periods longer than the 27-d TESS sector time span for stars with one sector of data, and proportionally longer for those with more than one sector of data. The amplitude spectra of the other candidate long-period Ap stars are shown in the Appendix. The left column plots show the amplitude spectrum out to 5\,d$^{-1}$, a range that encompasses the rotational frequencies of Ap stars, the shortest of which is around 2\,d$^{-1}$. The ordinate scale has been chosen to have a maximum of only 0.4\,mmag, which should be compared with the much higher amplitude variations typical of shorter period Ap stars illustrated in Figure\,\ref{fig:lc1}. The right column shows the amplitude spectrum out to the Nyquist frequency of 360\,d$^{-1}$ for the 2-min cadence data. This is because some of the stars are constant at all frequencies, while a few others are either $\delta$~Sct or roAp stars. The ordinate scale has been chosen appropriately for each star when a maximum amplitude of 0.4\,mmag was smaller than the pulsation amplitude, thus note the ordinate scales for these plots. The stars are illustrated in order of HD number, hence increasing right ascension (for one fainter star for which an HD number is not available, the TYC number is given). The figure captions give some comments for the variable stars.

\section{Discussion and conclusions}
\label{sec:disc}
The long-period Ap stars are of special interest because their sharp spectral lines allow the Zeeman components to be seen and the surface magnetic field to be measured directly. Most of the known long-period Ap stars have been studied over decades by \citet{2017A&A...601A..14M}, and others. These studies have shown that some of the long-period Ap stars have rotation periods of years, decades, and even centuries. This raises important astrophysical questions: Why do they rotate so slowly? Typical rotation periods for non-peculiar A stars are of the order of 1\,d. How did Ap stars lose their angular momentum? If magnetic braking is the mechanism for angular momentum loss, then how strong is the correlation between magnetic field strength and rotation period? Finally, with the TESS data, we have an unbiased method to find these long rotation period Ap stars, and we have a large data set for shorter rotation period Ap stars among the 1014 Ap stars observed in short (2-min) cadence in the first year of TESS observations, i.e. the stars in the southern ecliptic hemisphere. With this large, unbiased data set we can begin the long-term magnetic observations of the newly discovered long-period stars to address these questions.

As a group, the long rotation period Ap stars include some of the most peculiar of the Ap stars, and it can be seen from Table~\ref{table:1} that they are more likely to be roAp stars than more rapidly rotating Ap stars. Six of the stars in Table~\ref{table:1} have broad lines, hence short rotation periods. They are constant as a consequence of low magnetic obliquity (see below), and none is an roAp star. After rejecting those 6, we found 12 out of 54 stars to be roAp stars, or 22~per~cent. That can be compared to only 4~per~cent for the $>$1000 Ap stars observed by TESS in Sectors~1--13. Hence, roAp stars are more likely to be found among the most slowly rotating Ap stars, even though some roAp stars have rotation periods as short as 2\,d, such as HD~6532 \citep{1996MNRAS.281..883K}. This statistic possibly addresses the question of the driving mechanism in roAp stars \citep{2001MNRAS.323..362B}. There is much that we can learn from these stars that are chemically peculiar, pulsationally rare, and rotationally rare. 

\begin{figure}
\resizebox{\hsize}{!}{\includegraphics{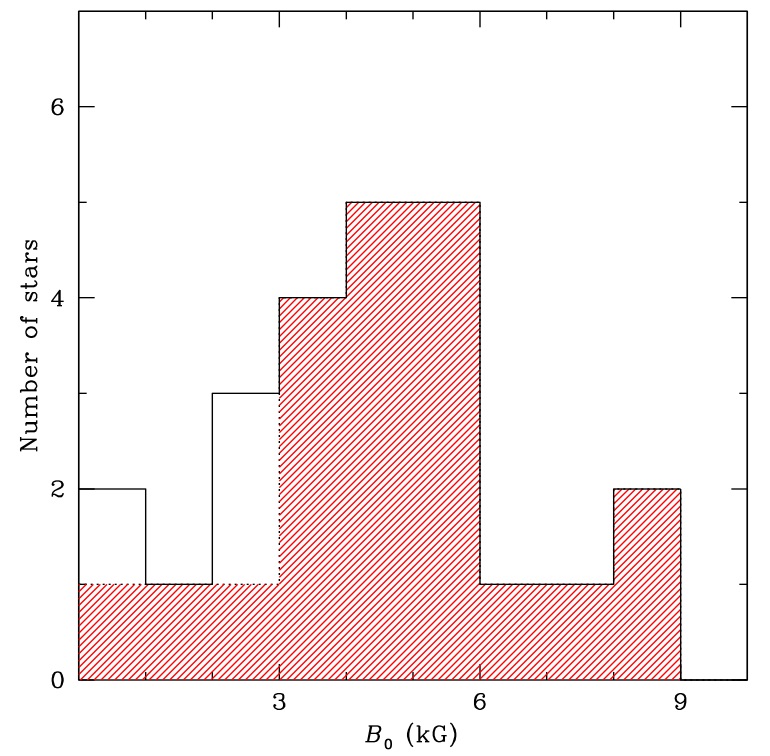}}
\caption{Distribution of the phase-averaged magnetic field strength $B_0$ for the long-period stars of Table~\ref{table:1}. The shaded part of the histogram corresponds to the stars for which measurements of the mean magnetic field modulus or of the mean quadratic magnetic field are available; for the remaining stars, a lower limit of $B_0$ was inferred from the existing mean longitudinal magnetic field measurements.} 
\label{fig:bavhist}
\end{figure}

Figure~\ref{fig:bavhist} shows the distribution of the phase-averaged mean magnetic field modulus $B_0$ for those long period stars of Table~\ref{table:1} for which magnetic measurements have been obtained. For those stars for which no mean magnetic field modulus determinations are available, an estimate of $B_0$ was computed by dividing the phase-averaged value of the mean quadratic field $Q_0$ by 1.28 \citep{2017A&A...601A..14M}. When only mean longitudinal magnetic field measurements had been obtained, we used $B_0=3\,\langle B_z\rangle_{\mathrm{rms}}$. This latter value is actually a lower limit \citep{2017A&A...601A..14M}. The corresponding parts of the histogram were left unshaded to reflect this difference.

To first order, the distribution illustrated in Fig.~\ref{fig:bavhist} is similar to the distribution of the phase-averaged magnetic field strengths in the entire population of known Ap stars with resolved magnetically split lines \citep[see Figure~1 of][]{2017A&A...601A..14M}, except for the absence of the strong-field tail ($B_0>9$\,kG) of the latter. This absence is consistent with the statistically strong conclusion that the strongest magnetic fields occur only in those Ap stars that rotate sufficiently fast \citep{1997A&AS..123..353M,2017A&A...601A..14M}.

At the low-field end of the distribution, there are five stars from Table~\ref{table:1} that have sharp or very sharp unresolved spectral lines, whose rotation periods have not been determined yet and for which no magnetic field measurements have been obtained yet. The magnetic field of HD~217704, in which the Fe~{\sc ii} $\lambda\,6149.2$ line is marginally resolved and whose rotation period is also unconstrained, has not been determined either. These six stars must all have mean magnetic field moduli well below 3\,kG. They should populate the first three bins of the histogram, making the distribution look mostly flat over the 0--6\,kG range. While the distribution of the magnetic field strengths in the 23 stars of Table~\ref{table:1} for which no high-resolution spectra or magnetic data are available until now is unknown, a priori there is no reason to suspect that it should be significantly different. Thus, a nearly flat distribution of the magnetic field strengths from 0 to 6\,kG in the longest-period Ap stars does not seem implausible. Such a distribution contrasts with that of the root-mean-square longitudinal magnetic fields among all Ap stars, which sharply peaks at the lowest values \citep{2009MNRAS.394.1338B}. This suggests that there may be a deficiency of weakly magnetic stars among the slowly rotating Ap stars. Seeking confirmation of this conclusion represents a strong motivation for a systematic follow-up study aimed at characterising the magnetic fields of all the long period Ap stars. 

 For six of the 60 stars of Table~\ref{table:1}, accurate values of the rotation periods have been determined: they all exceed 150\,d. In addition, two stars whose magnetic variations have been followed over a time base of five years, HD~29578 and HD~92499, have not completed a full rotation cycle yet: their rotation periods must be longer than 5\,yr. We could not find any information in the literature about the rotation periods of the remaining 52 stars. Thus, all the stars of Table~\ref{table:1} for which meaningful constraints on the rotation period are available rotate extremely slowly.

Among the 52 stars with unconstrained rotation periods, 13 have (marginally) resolved magnetically split lines (identified by ``r'' or ``mr'' in Col.~12 of Table~\ref{table:1}) and ten have (very) sharp unresolved lines whose profiles hardly differ from the instrumental profile of the spectrograph (identified by ``s'' or ``vs'' in Col.~12 of Table~\ref{table:1}). The spectral lines of these 23 stars show very little or no rotational broadening. On the other hand, there are six stars with spectral lines that are either broad ($v\,\sin i\ga50$\,km\,s$^{-1}$) or very broad (with a much greater $v\,\sin i$). This leaves 23 stars in which TESS data do not show any rotational modulation and for which no other information about the period or \vsi\ is available.

Even if all Ap stars have spotted surfaces, we should expect some of them that have a high projected equatorial velocity to show no photometric variability. Indeed, we can reasonably assume that, to first order, the brightness distribution over the surface of an Ap star is symmetric about the axis of its predominantly dipole-like magnetic field, which is inclined at an angle $\beta$ to its rotation axis. If $\beta$ is too small, that is, if the magnetic and rotation axes are nearly aligned with each other, the star will not show any photometric variability, regardless of its rotation period.

If we define `nearly aligned' as $\beta\leq5^\circ$, consistently with the definition adopted in Section~\ref{sec:tess} for `nearly pole-on' ($i\leq5^\circ$), and if we assume a random distribution of $\beta$ (the magnetic obliquity), we expect about 0.4 per cent of the Ap stars to show no detectable photometric variation on account of the near alignment of their magnetic and rotation axes. Within the statistical uncertainties, this fraction is not inconsistent with the identification of six non-variable fast-rotating stars among the 1014 stars of our survey.

Admittedly, the distribution of $\beta$ has never been definitely characterised. The results of early studies \citep{1967ApJ...150..547P,1970ApJ...159.1001L}, which were compatible with random magnetic obliquities, have not been seriously challenged, and a more recent study seems to support this conclusion \citep{2019MNRAS.483.3127S}, but this distribution is not uniquely constrained and others (e.g., bimodal) could plausibly be consistent with the observations \citep{2007AN....328..475H}. Furthermore, one cannot rule out the possibility that the distribution of the magnetic obliquity may be different in groups of stars that have different rotation rates, as advocated in particular for the super-slowly rotating stars, albeit with discrepant conclusions from different studies \citep{2000A&A...359..213L,2020A&A...636A...6M}. Within this context, the consistency of the fraction of non-variable fast-rotating stars with the prediction for a random distribution of $\beta$ can be regarded as supporting the plausibility of the random distribution, albeit not in a conclusive and unambiguous manner. 

The interpretation that the six non-variable fast-rotating stars of Table~\ref{table:1} have their rotation and magnetic axes nearly aligned can be confirmed by obtaining measurements of their mean longitudinal magnetic fields at various epochs sampling the longest rotation period compatible with their $v\,\sin i$ (a few days, at most). If $\beta$ is small, the longitudinal field  should show no significant variations, or at most, variations of very low amplitude. We are planning to obtain the required spectropolarimetric observations to carry out this test. The opportunity to identify unambiguously Ap stars in which the magnetic and orientation axes are nearly aligned is particularly valuable because in most cases, the angles $i$ and $\beta$ are interchangeable in magnetic models of such stars. 

As mentioned above, 23 stars whose rotation periods have not been constrained until now have a projected equatorial velocity $v\,\sin i$ of the order of the spectral resolution of the spectra (typically, $\sim$3\,km\,s$^{-1}$), or lower. Under the assumption that the inclinations of their rotation axes to the line of sight are random, which is borne out by the results of the systematic study of the Ap stars with resolved magnetically split lines \citep{2017A&A...601A..14M}, the vast majority of these stars must be genuine (very) slow rotators, with rotation periods significantly exceeding one month, and possibly reaching up to years or decades.

Therefore, we also expect most of the 23 stars of Table~\ref{table:1} whose period is unknown and for which we have not yet obtained any high-resolution spectra to have similarly long periods. As a next step, we shall obtain such spectroscopic observations to confirm that the projected equatorial velocities of all these stars do not significantly exceed 3\,km\,s$^{-1}$. This will also enable us to obtain a first characterisation of their magnetic fields. We expect some of these stars to show resolved magnetically split lines, and others to show sharp spectral lines with little or no detectable broadening of either magnetic or rotational origin. 

Thus, most of the constant TESS stars that are identified in our survey belong, with a high probability, to the group of the most slowly rotating Ap stars. Can all the Ap stars with a rotation period longer than a certain threshold (say, a couple of months) be detected through this approach?

The recent critical compilation by \citet{2019arXiv191206107M} of the ssrAp stars lists the 33 Ap stars known to date that have an accurately determined rotation period longer than 50 days. Of these 33 stars, 27 do not appear in Table~\ref{table:1}. Twenty of them were not part of the sample that we analysed because there were no TESS Sectors 1--13 2-min cadence data for them: eleven are northern ecliptic hemisphere stars,  two are in the ecliptic equatorial band, and for seven, no 2-min cadence observations were obtained. 

Thus, only seven stars that were expected to be constant over 27 days were missed by our search. For three of them, HD~103844, HD~149766, and HD~187474, the data seem to suffer from instrumental effects of various types. The resulting apparent variations do not resemble the rotational variations of Ap stars. One star, HD~166473, is in a crowded field, and the low-amplitude variability of its light curve is most likely due to contamination. The light curve of HD~102797 shows eclipses, which may either originate from a contaminating star or from the Ap star itself. If the latter, this would be only the fourth Ap star known in an eclipsing binary \citep{2018MNRAS.478.1749K,2019MNRAS.487.4230S,2019MNRAS.490.4154S}. Finally, HD~94660 and HD~116458 both show variability of the type observed in cool stars. Both are well known spectroscopic binaries \citep[and references therein]{2017A&A...601A..14M}, but it is unclear at present if the observed variability originates from the secondary component or if it should be attributed to a contaminating source. 

In summary, a survey such as the one that we carried out does not yield a complete list of the most slowly rotating Ap stars. While seven missed detections out of thirteen long period stars observed may seem to represent a high failure rate, the significance of this result is limited by the small number statistics, and it is impossible at present to quantify reliably the incompleteness of the outcome of our survey.  Some stars are definitely missed, albeit for reasons that are not related to their intrinsic physical properties. Accordingly, while the identifications achieved with the approach used here are incomplete, they are also unbiased, so that the resulting sample of stars is suitable to study possible relations between slow rotation and other physical properties of Ap stars. For instance, it should lend itself to a comparison of the rates of occurrence of slow rotation in weakly and strongly magnetic Ap stars. This kind of study will require a longer-term observational follow-up to constrain the values of the rotation periods from the study of the magnetic variations. For those stars that show magnetically resolved lines, magnetic field modulus determinations from high-resolution spectra recorded in natural light may be sufficient in most cases, but for the weaker field stars, spectropolarimetric observations will be required to obtain measurements of the mean longitudinal magnetic field. 

On the other hand, none of the TESS light curves of the known slow rotators that were missed resembles a typical Ap star light curve. This suggests that the probability is low that a short rotation period may mistakenly be derived for a slowly rotating Ap star. 

In conclusion, our survey finds the known long-period and/or low $v\,\sin i$ stars, and it has found  23 new long-period candidates, of which only a few, at most, will be pole-on, or have low magnetic obliquity. Hence our false-positive rate is low. We have thus developed a novel technique for finding long-period Ap stars that demonstrably works. We will continue with this study in the TESS northern ecliptic hemisphere short cadence data set, then eventually greatly expand it with the TESS FFIs. 

The emphasis and the excitement in studying the TESS photometric data arises from studying variable stars -- for finding exoplanets, the mission's prime objective; for asteroseismology; for binary stars; for gyrochronology; for flare studies; i.e., for anything astrophysical that can be gleaned from stellar variability. Now, we have shown that there is interesting astrophysical inference to be made from {\it stars that do not vary}. These non-variable stars are the best targets for characterising the instrumental responses of TESS and other stellar photometric space missions, such as CHEOPS\footnote{https://cheops.unibe.ch} and PLATO\footnote{https://platomission.com}.  These are payoffs that the mission planners did not anticipate: We are actually getting ``something for nothing''! 

\begin{acknowledgements}
DWK and DLH acknowledge financial support from the Science and Technology Facilities Council (STFC) via grant ST/M000877/1. This paper includes data collected by the TESS mission. Funding for the TESS mission is provided by the NASA Explorer Program. This work was initiated during a stay of DWK at the European Southern Observatory office in Santiago, within the framework of the ESO Scientific Visitor Programme. DWK thanks ESO for funding this visit. Some of the observations reported in this paper were obtained with the Southern African Large Telescope (SALT). We thank the referee, Margarida Cunha, whose constructive comments helped us to improve the presentation of our results. 
\end{acknowledgements}

\bibliographystyle{aa}
\bibliography{18version_longperiod}

\appendix

\section{Amplitude spectra of candidate long rotation period Ap stars}
\label{sec:amp_sp}
\begin{figure*}
\centering
\includegraphics[width=0.48\linewidth,angle=0]{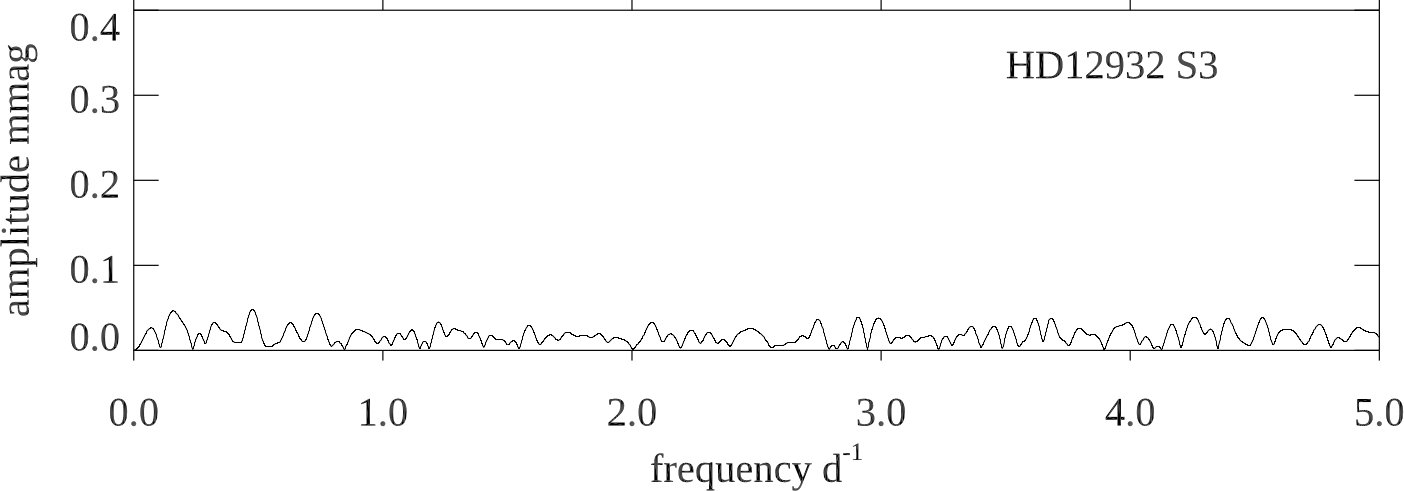}	
\includegraphics[width=0.48\linewidth,angle=0]{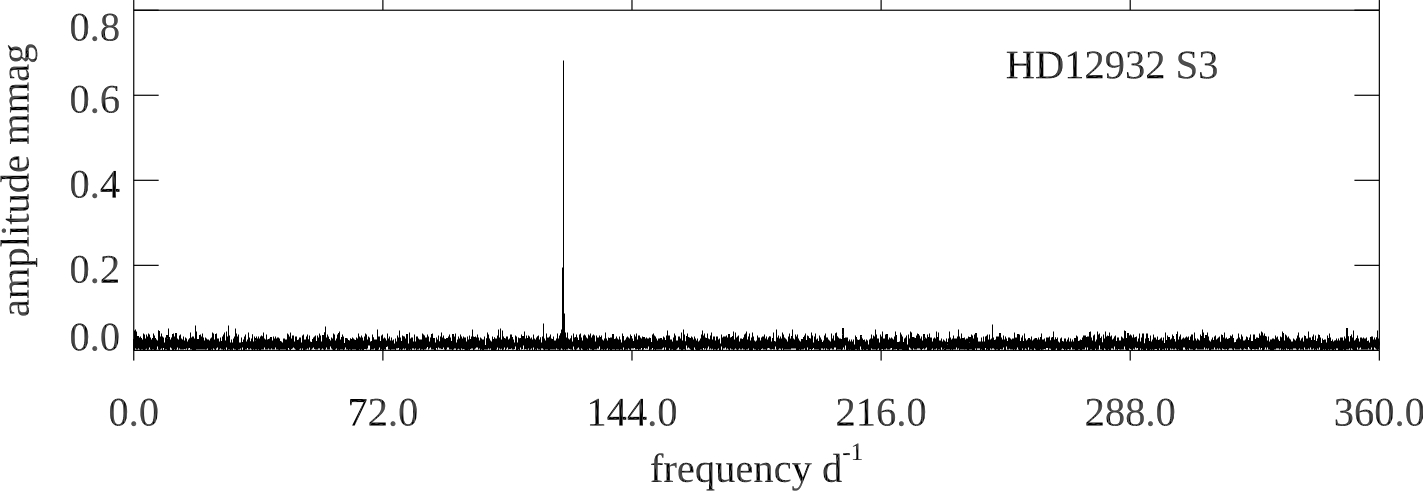}	
\includegraphics[width=0.48\linewidth,angle=0]{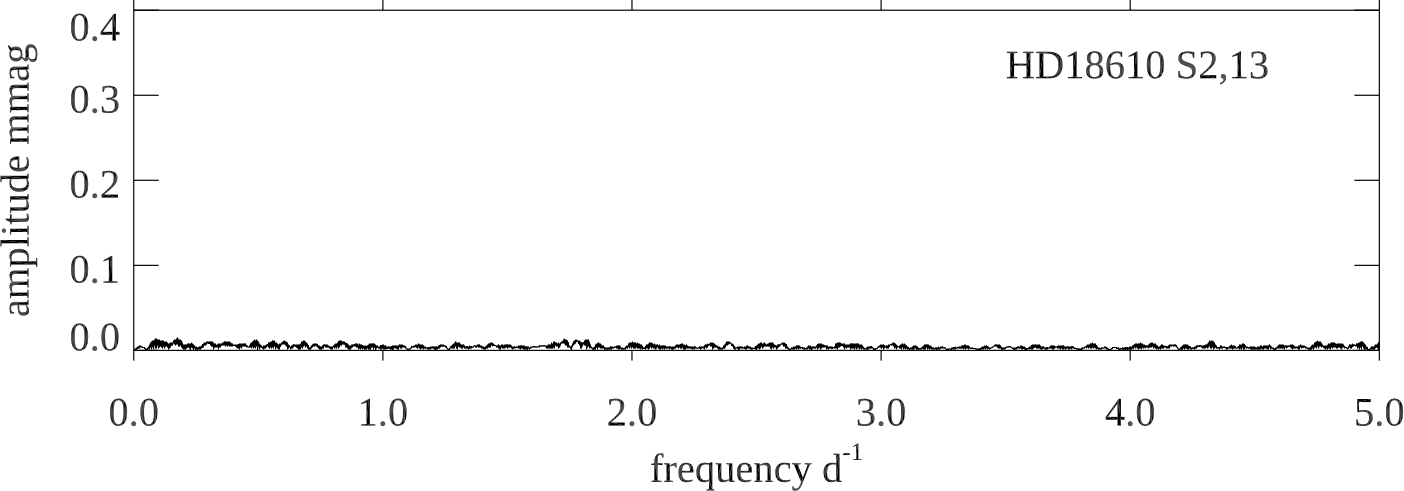}	
\includegraphics[width=0.48\linewidth,angle=0]{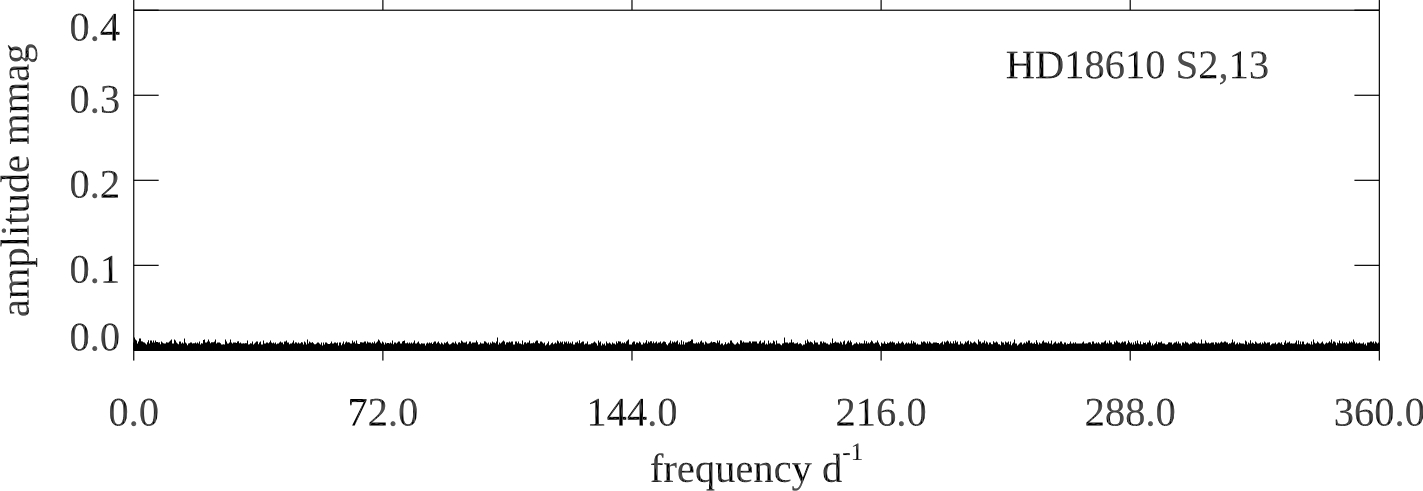}	
\includegraphics[width=0.48\linewidth,angle=0]{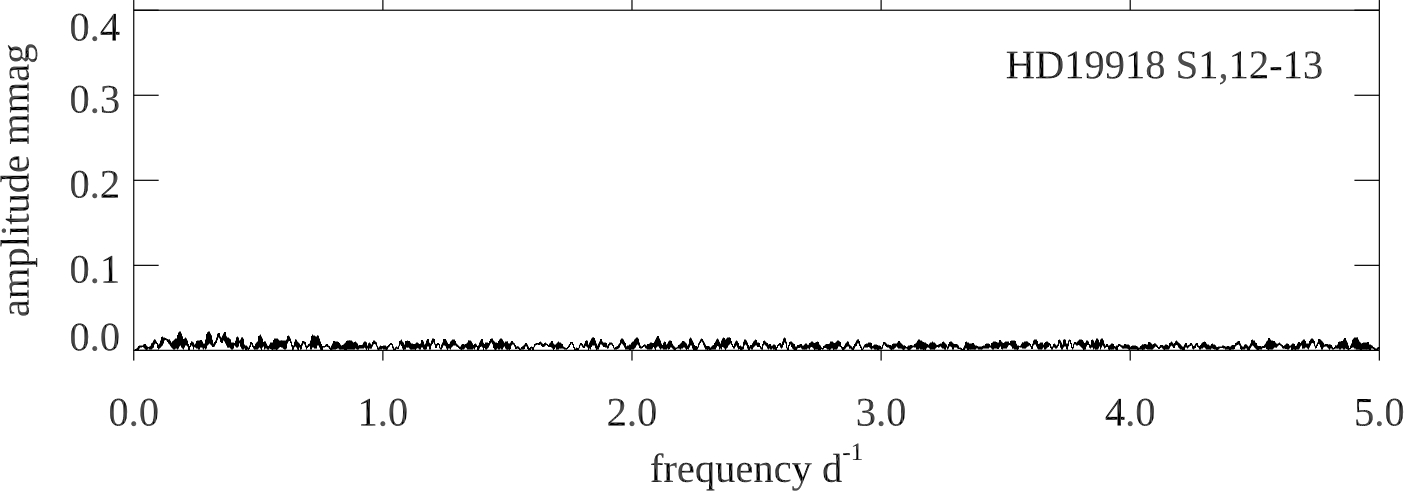}	
\includegraphics[width=0.48\linewidth,angle=0]{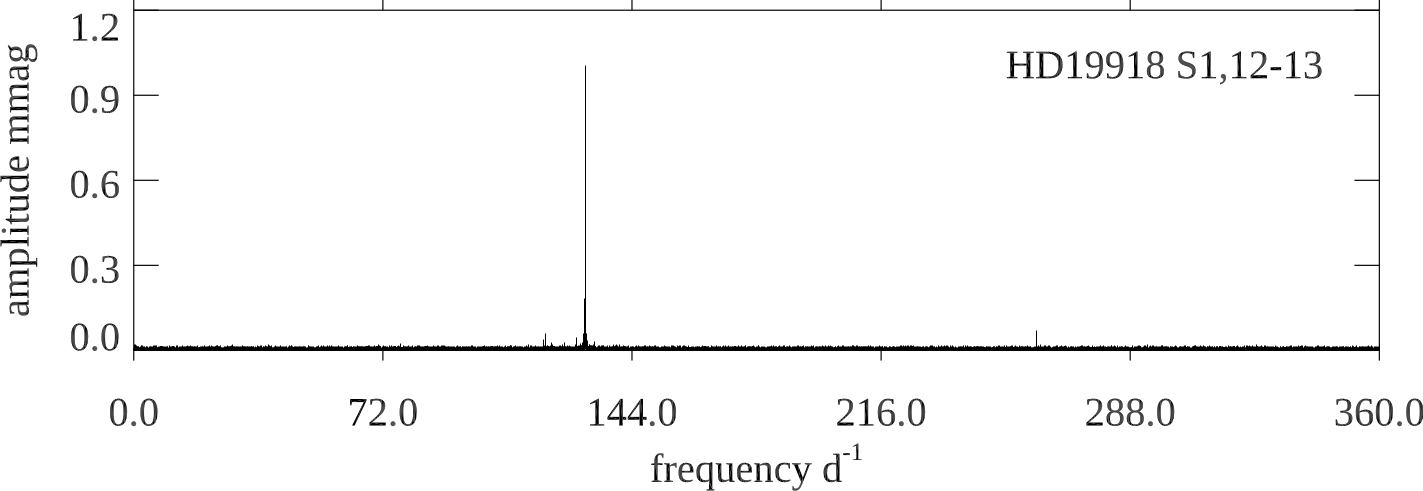}	
\includegraphics[width=0.48\linewidth,angle=0]{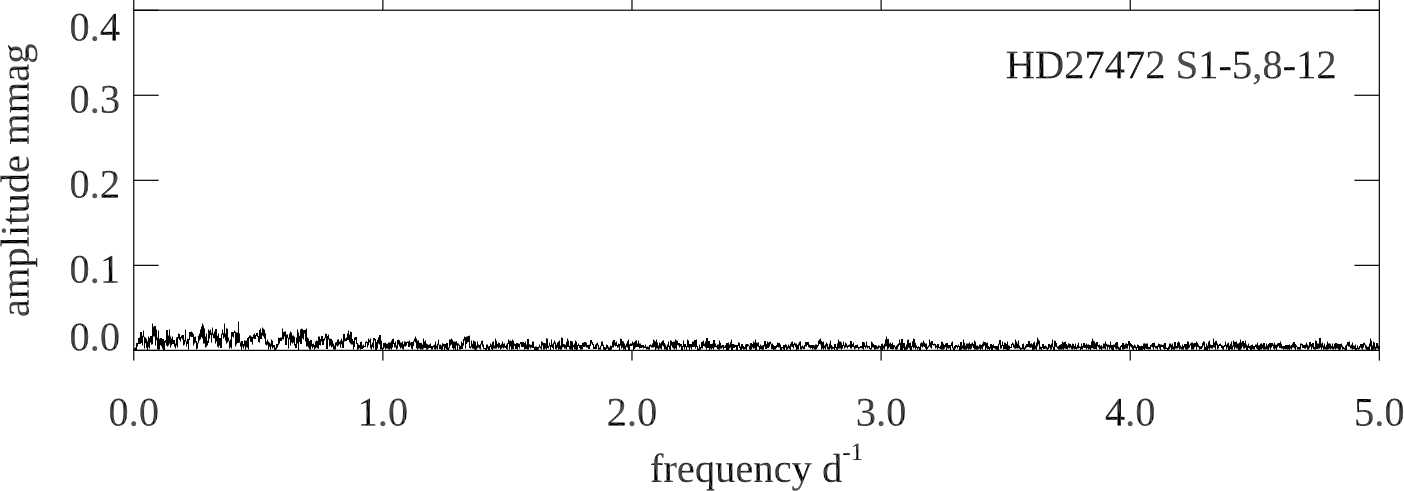}	
\includegraphics[width=0.48\linewidth,angle=0]{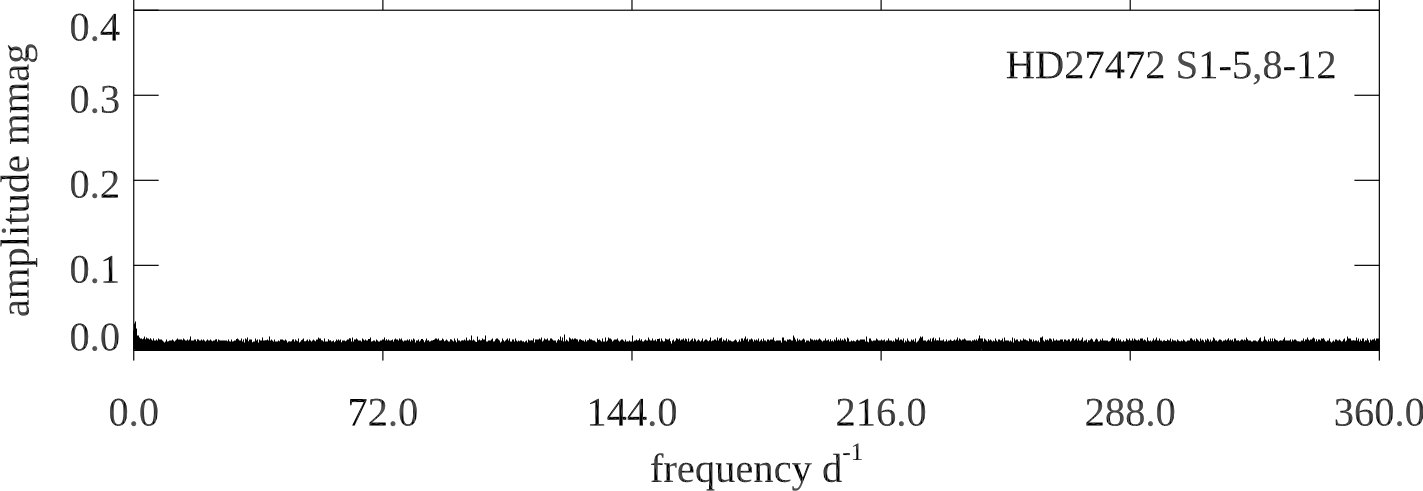}	
\includegraphics[width=0.48\linewidth,angle=0]{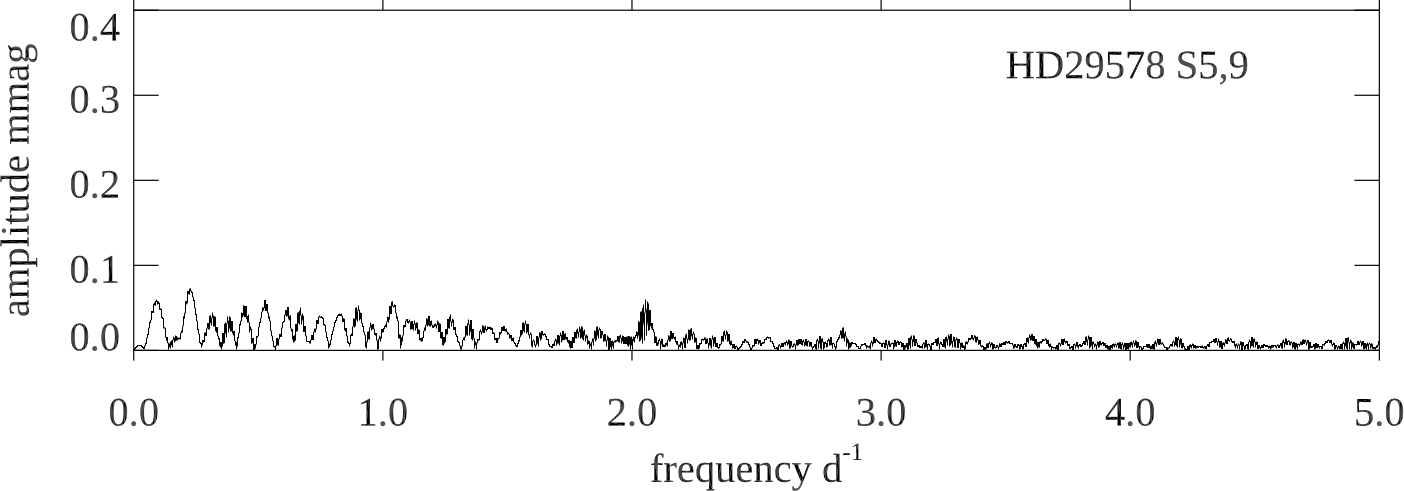}	
\includegraphics[width=0.48\linewidth,angle=0]{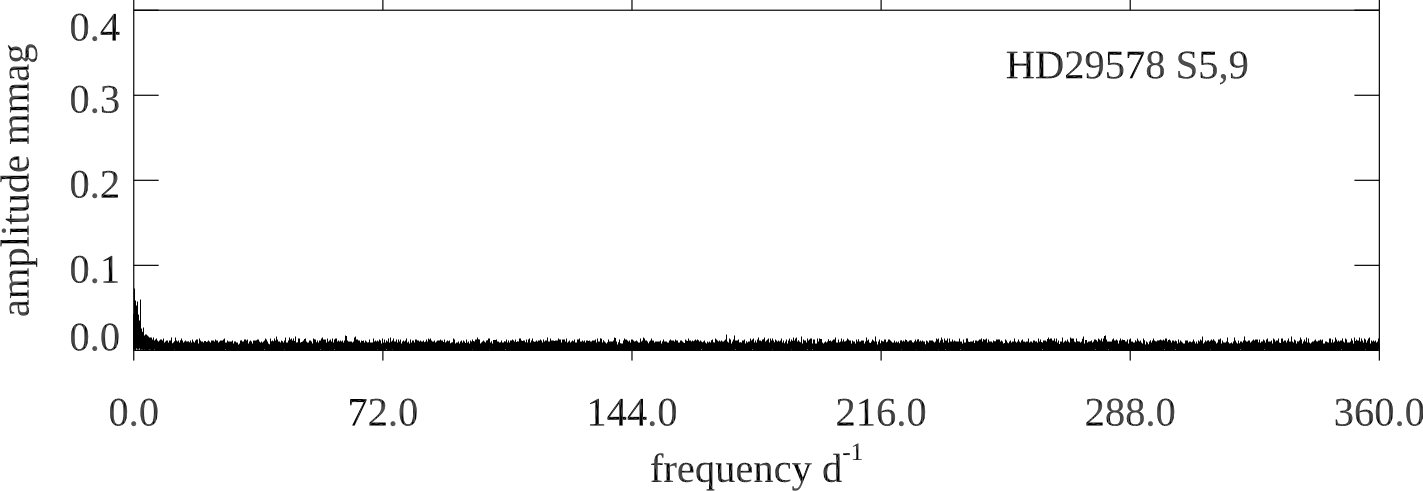}
\includegraphics[width=0.48\linewidth,angle=0]{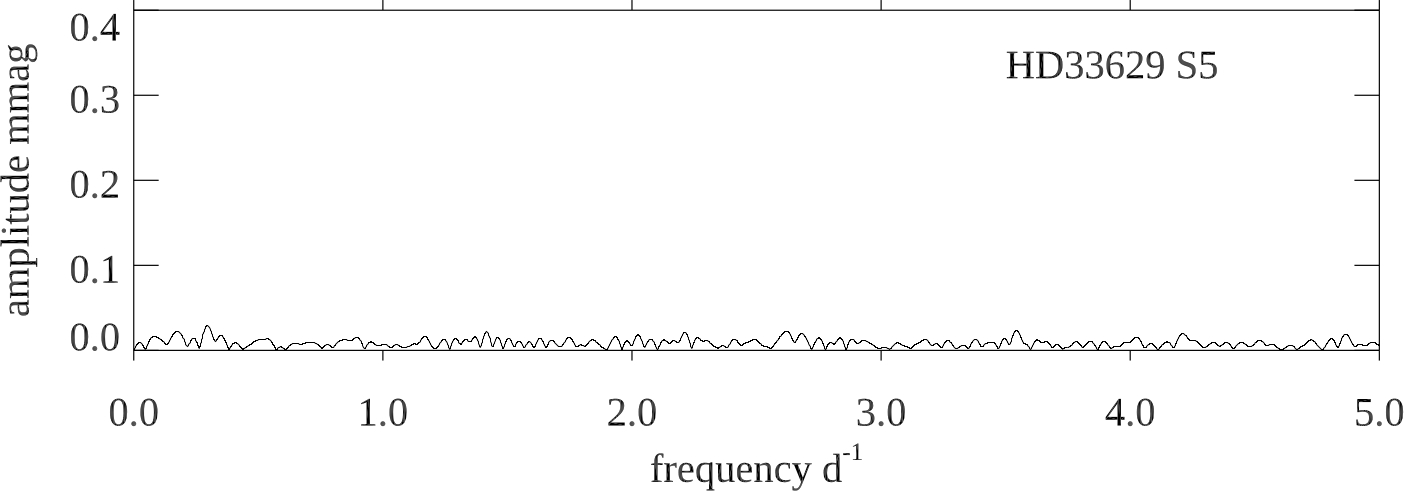}	
\includegraphics[width=0.48\linewidth,angle=0]{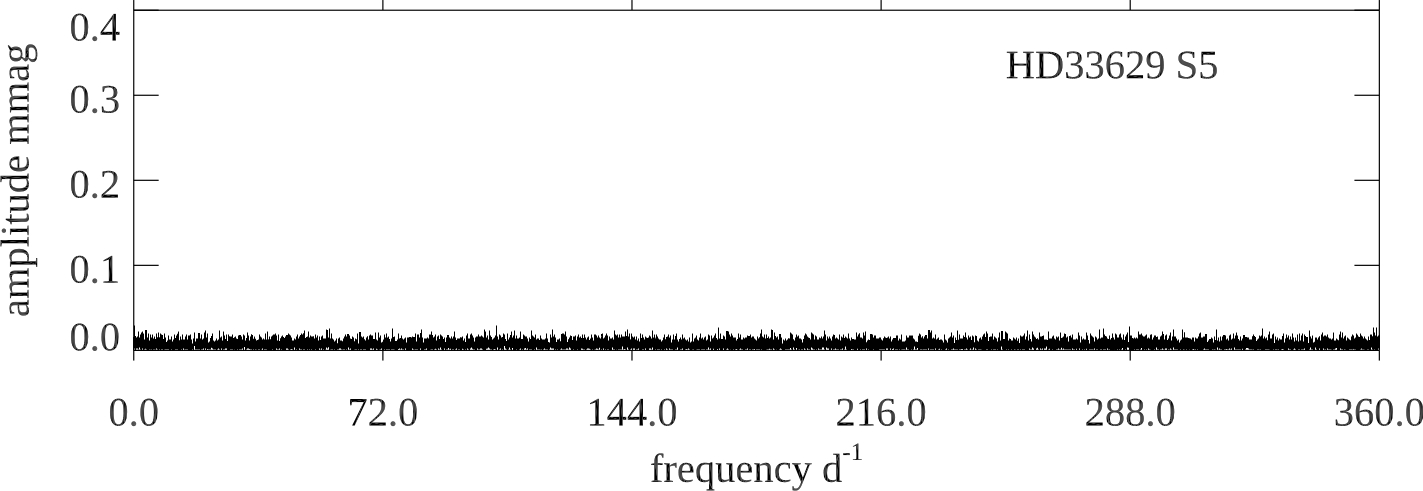}		
\includegraphics[width=0.48\linewidth,angle=0]{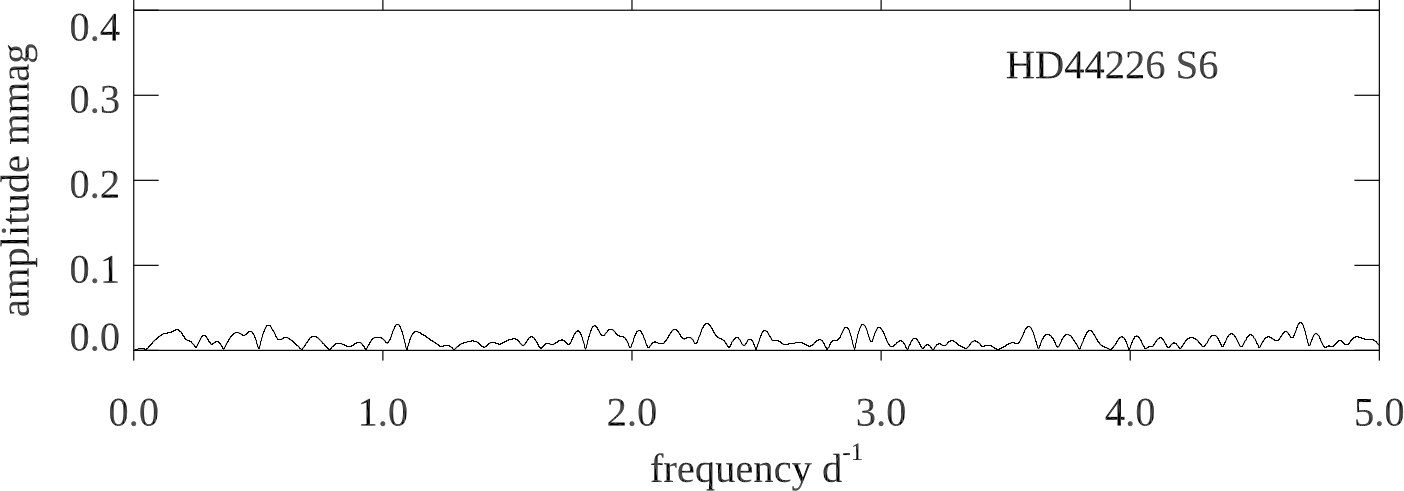}	
\includegraphics[width=0.48\linewidth,angle=0]{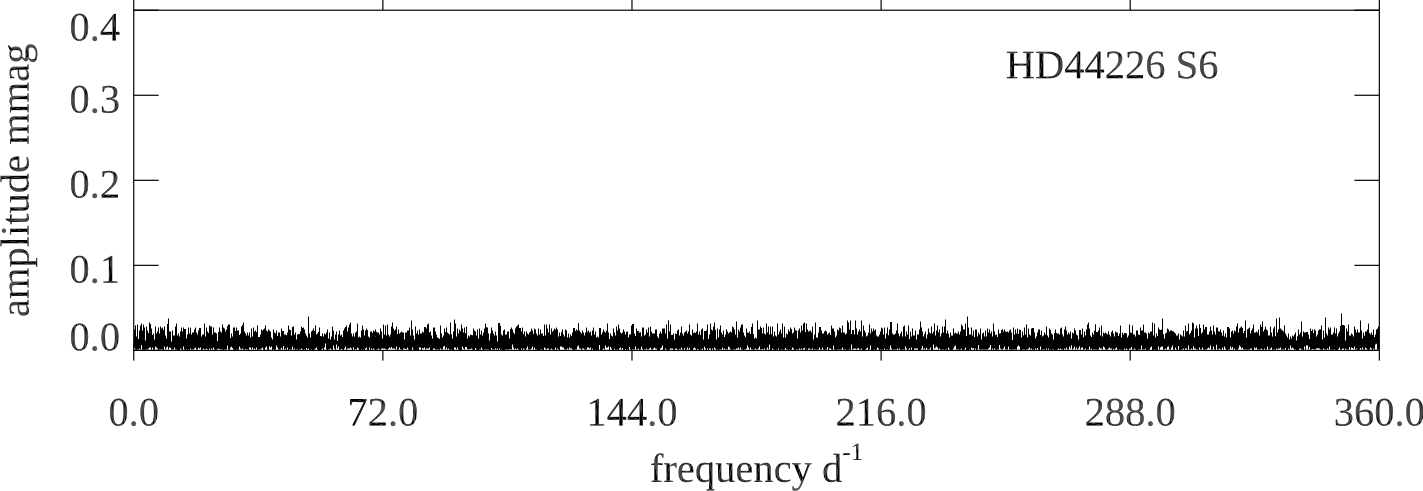}	
\caption{Same as Fig.~\ref{fig:fts_1}. For HD~29578 the small peak at 2 d$^{-1}$ is real, but of unknown origin; the magnetic measurements show a rotation period $> 5$\,yr for this star, so a period of half a day is not plausible as its rotation period.  HD~12932 and HD~19918 are well-known roAp stars (see, e.g., \citealt{2019MNRAS.487.3523C}; \citealt{2019MNRAS.487.2117B}). }
\label{fig:fts_2}
\end{figure*}

\begin{figure*}[p]
\centering
\includegraphics[width=0.48\linewidth,angle=0]{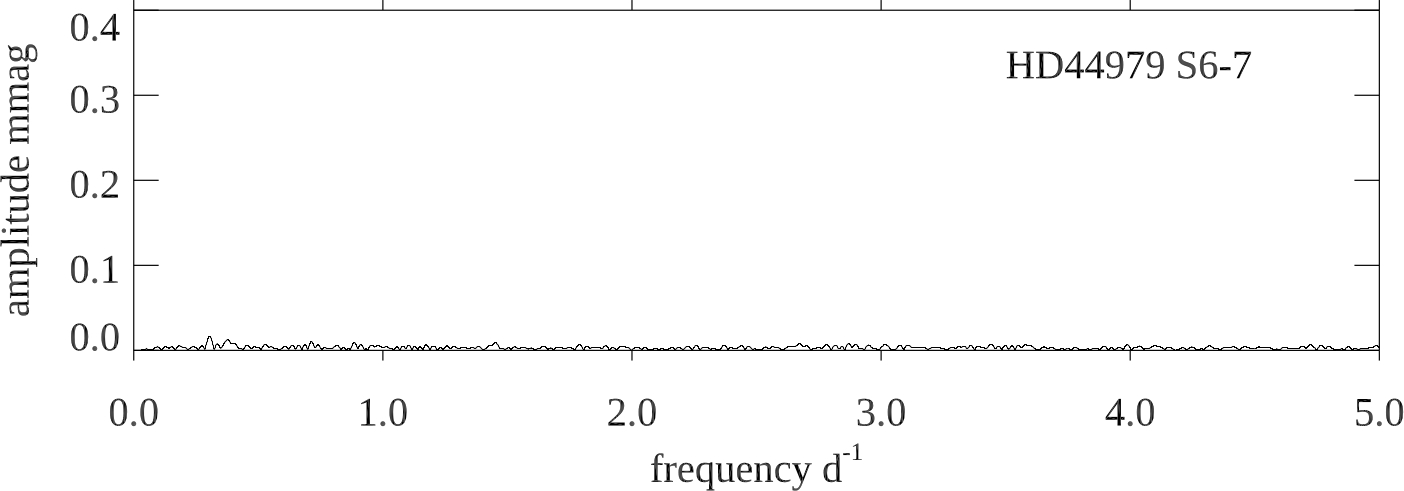}	
\includegraphics[width=0.48\linewidth,angle=0]{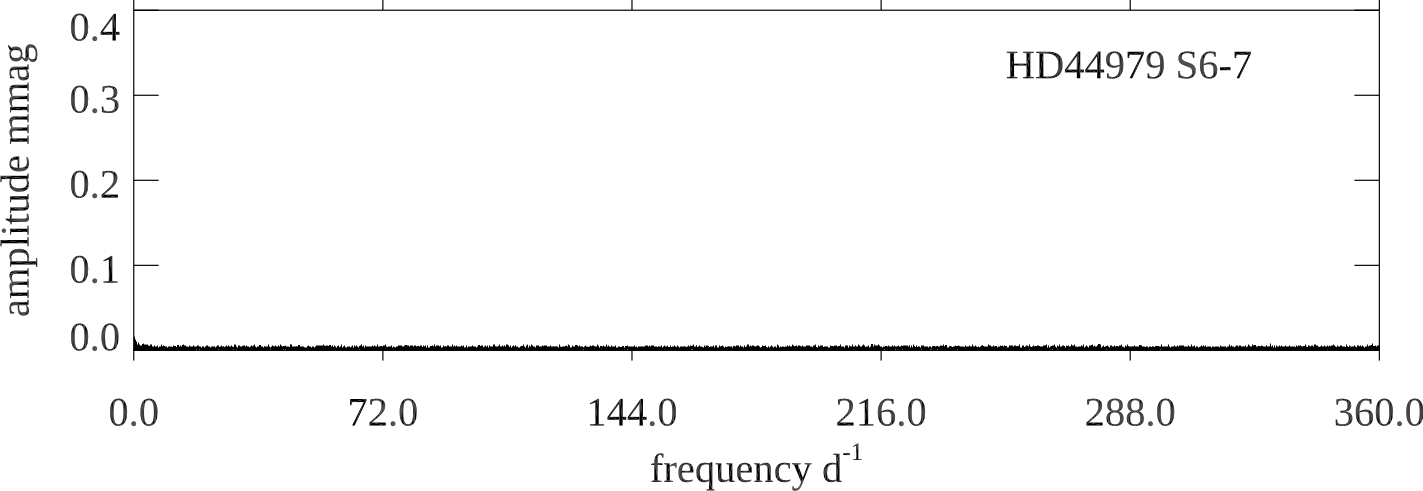}	
\includegraphics[width=0.48\linewidth,angle=0]{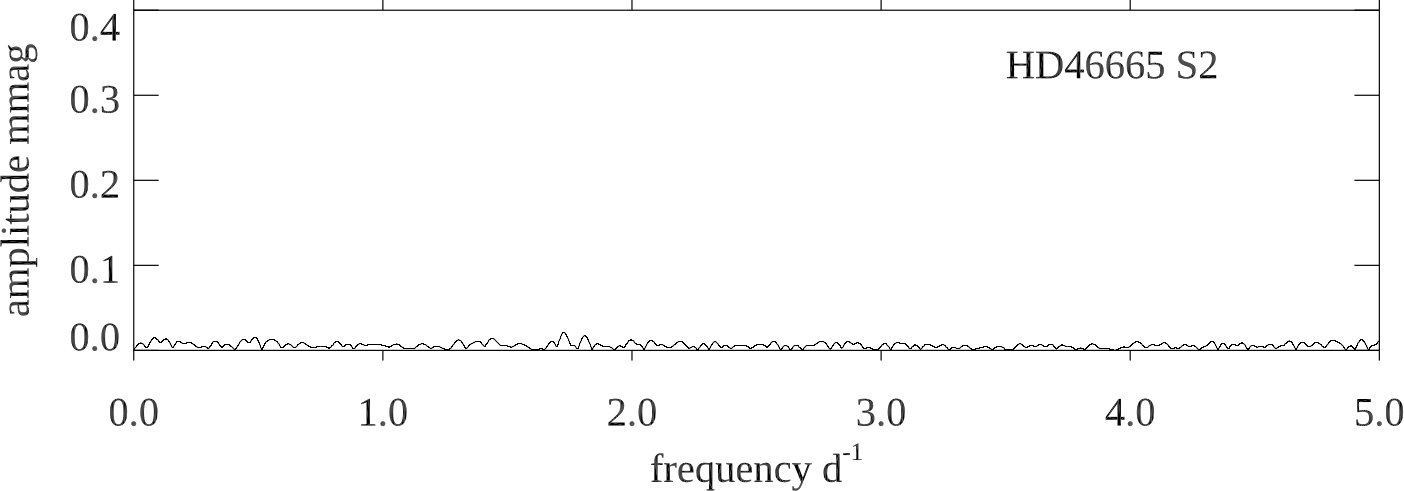}	
\includegraphics[width=0.48\linewidth,angle=0]{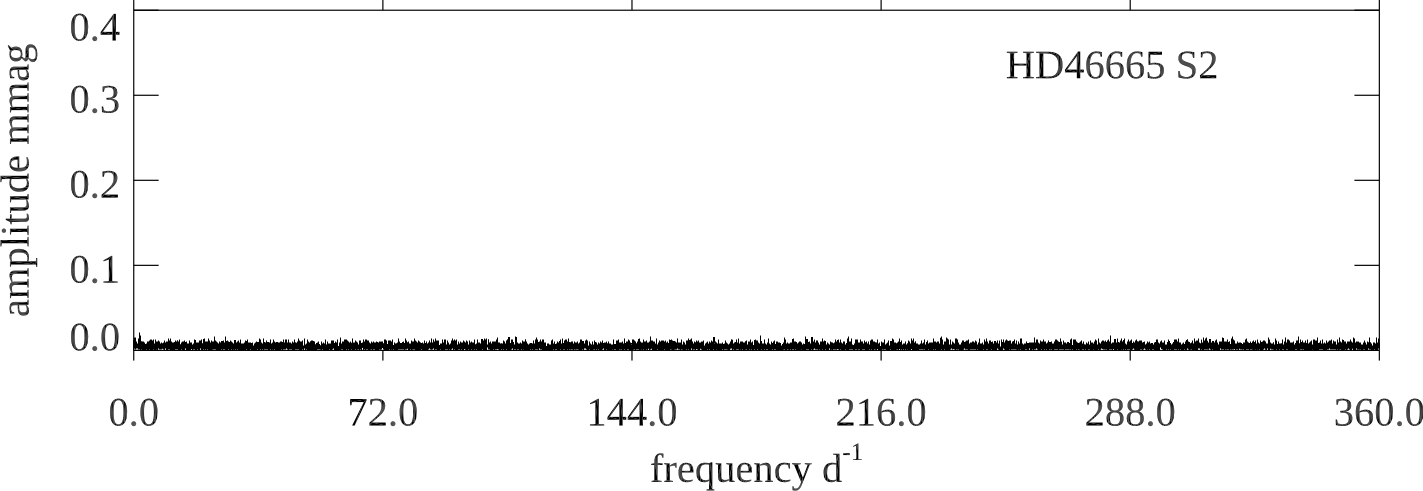}	
\includegraphics[width=0.48\linewidth,angle=0]{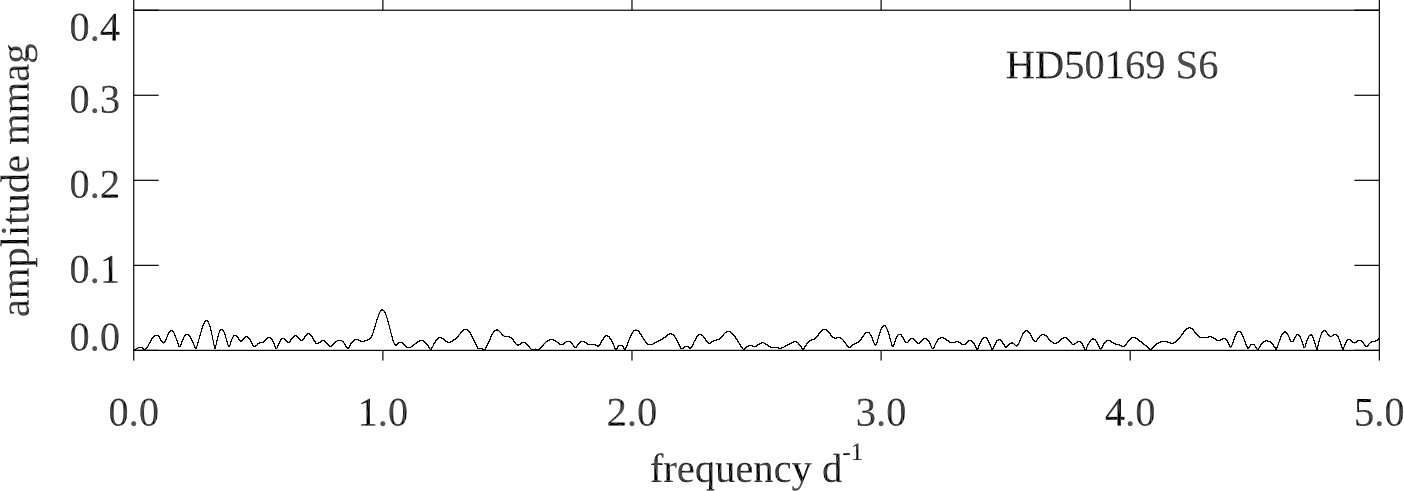}	
\includegraphics[width=0.48\linewidth,angle=0]{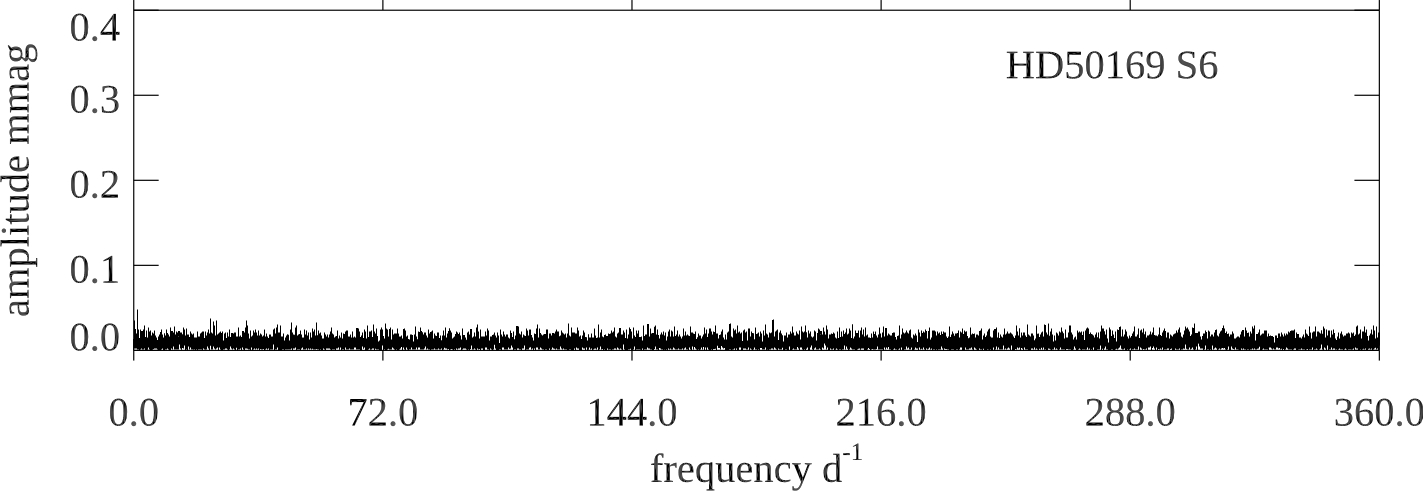}	
\includegraphics[width=0.48\linewidth,angle=0]{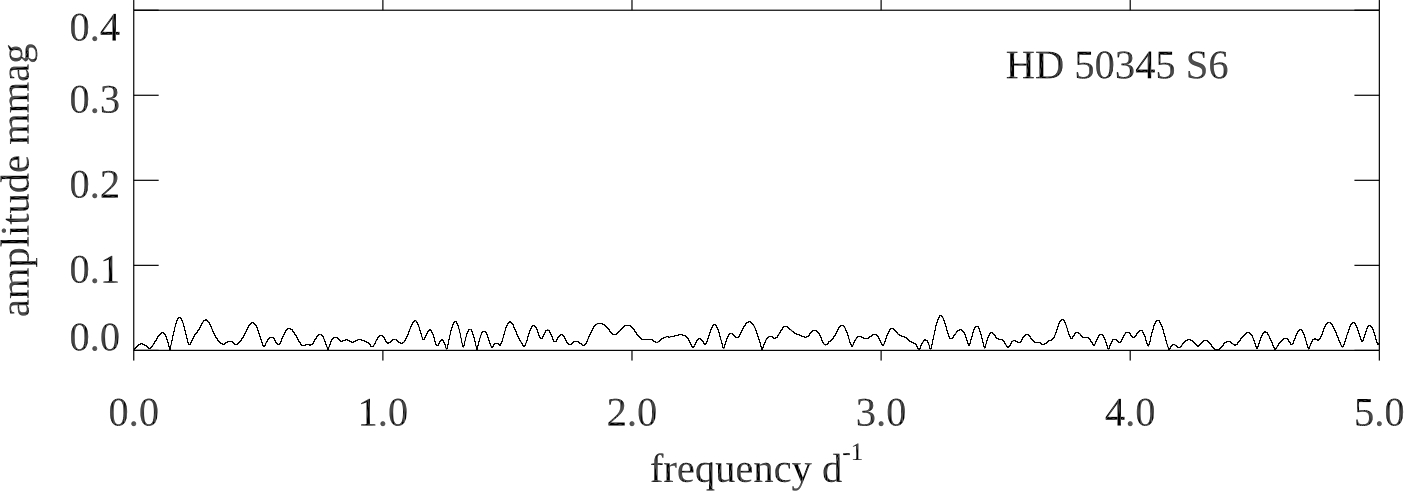}	
\includegraphics[width=0.48\linewidth,angle=0]{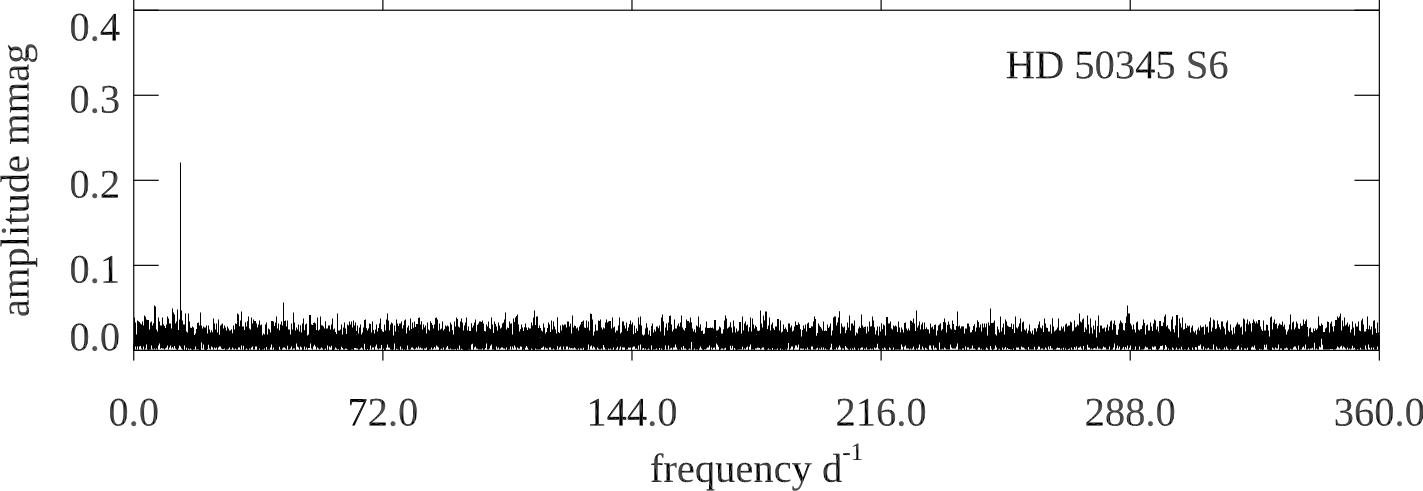}	
\includegraphics[width=0.48\linewidth,angle=0]{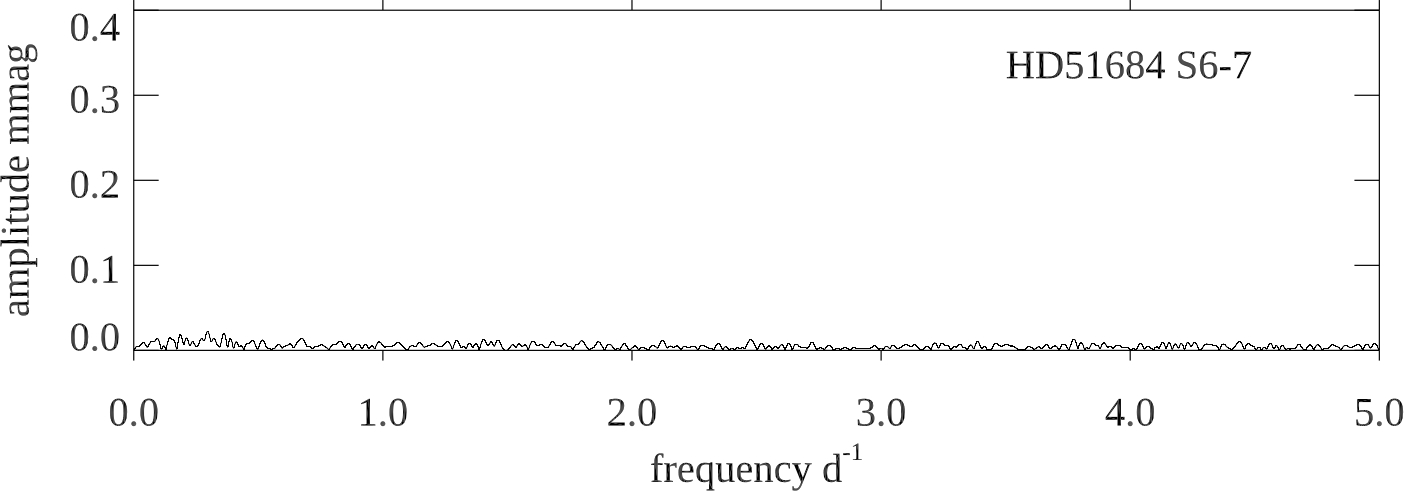}	
\includegraphics[width=0.48\linewidth,angle=0]{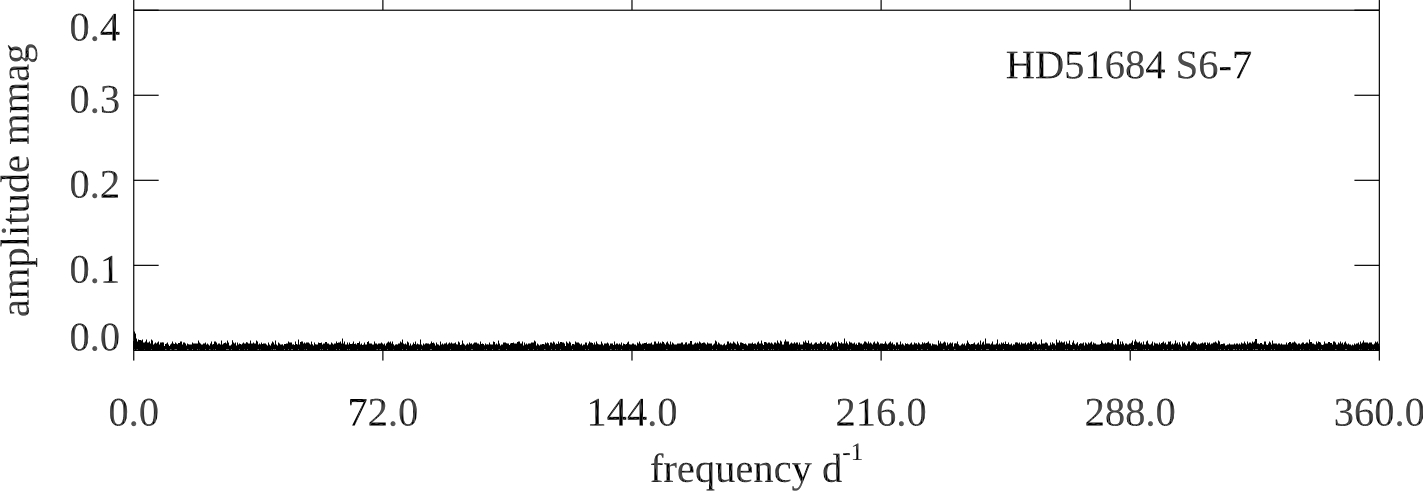}	
\includegraphics[width=0.48\linewidth,angle=0]{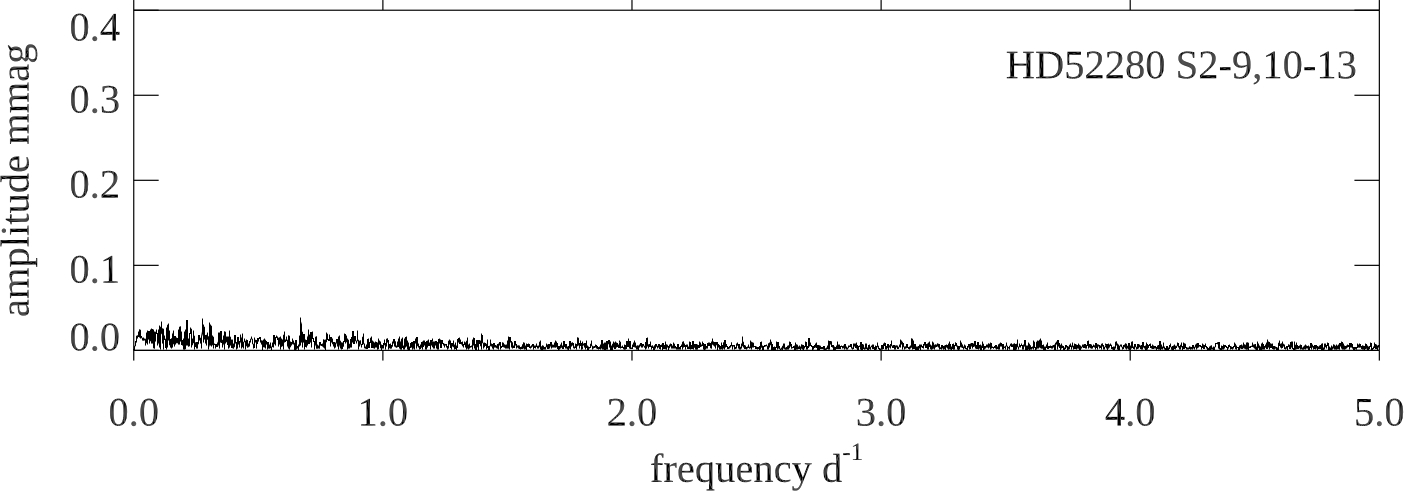}	
\includegraphics[width=0.48\linewidth,angle=0]{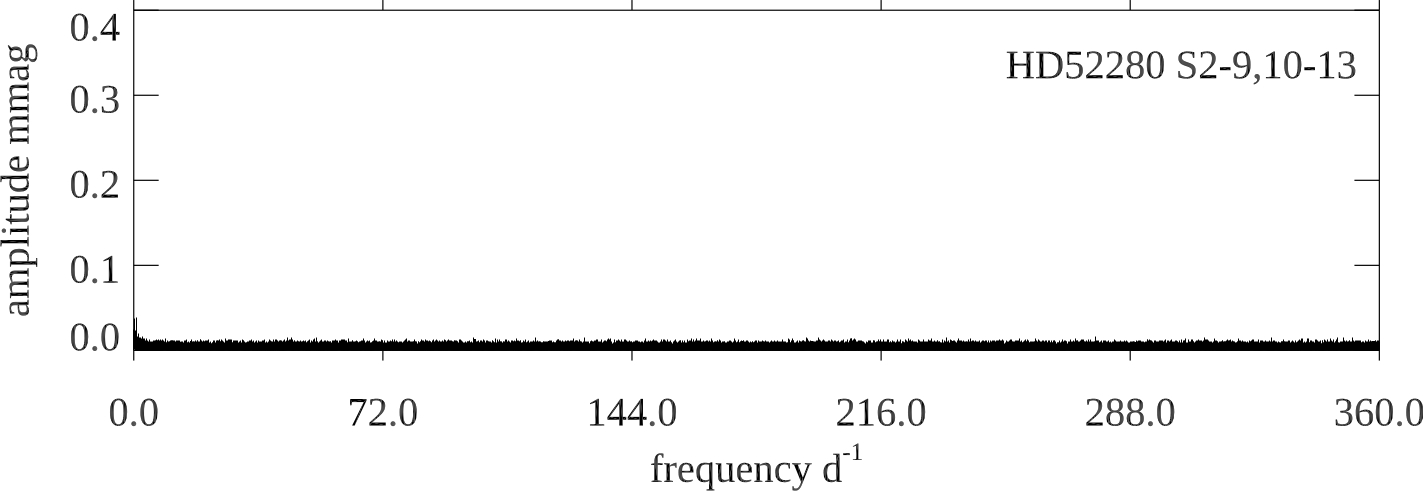}	
\includegraphics[width=0.48\linewidth,angle=0]{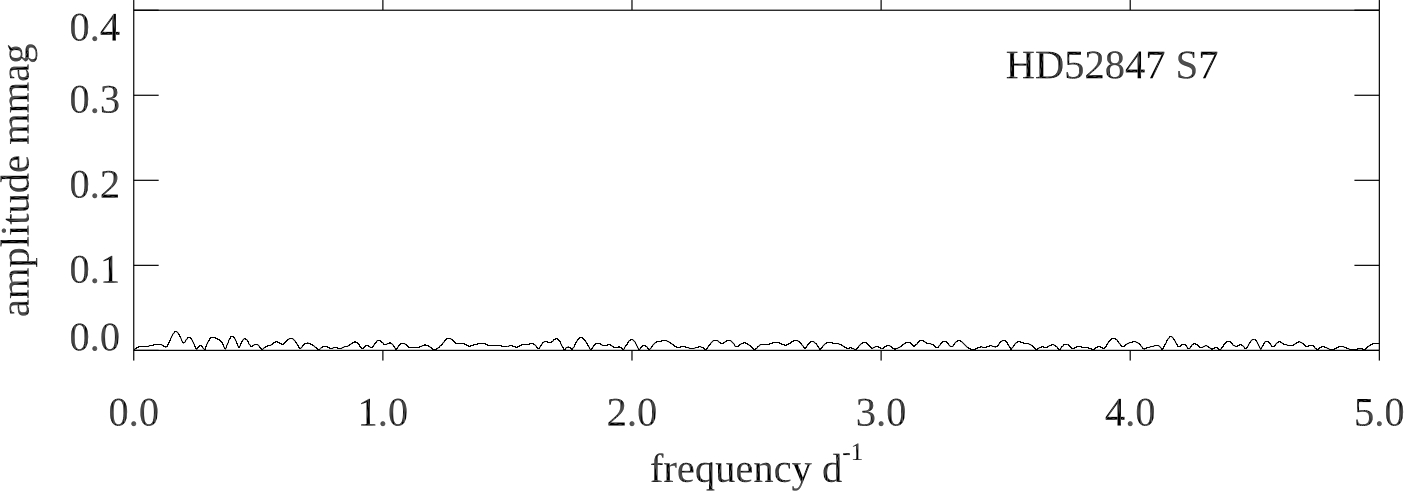}	
\includegraphics[width=0.48\linewidth,angle=0]{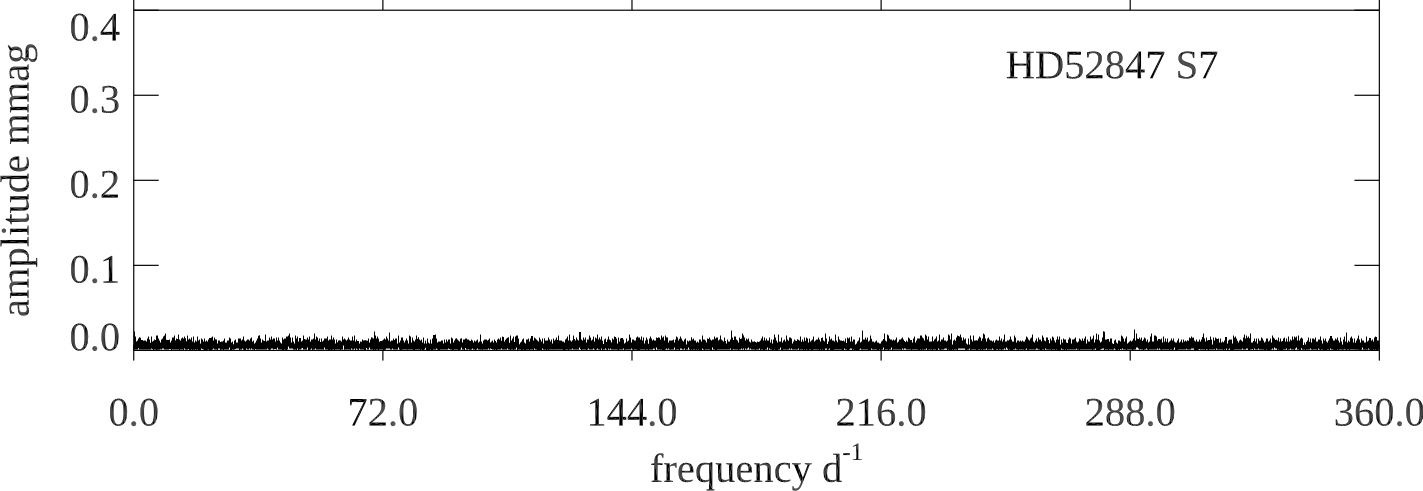}	
\caption{Same as Fig.~\ref{fig:fts_1}. HD~50345 appears to have a single $\delta$~Sct peak at 13.5464\,d$^{-1}$ not previously noted in the literature; $\delta$~Sct pulsations in magnetic Ap stars are rare, so this star is of interest.
}
\label{fig:fts_3}
\end{figure*}

\begin{figure*}[p]
\centering
\includegraphics[width=0.48\linewidth,angle=0]{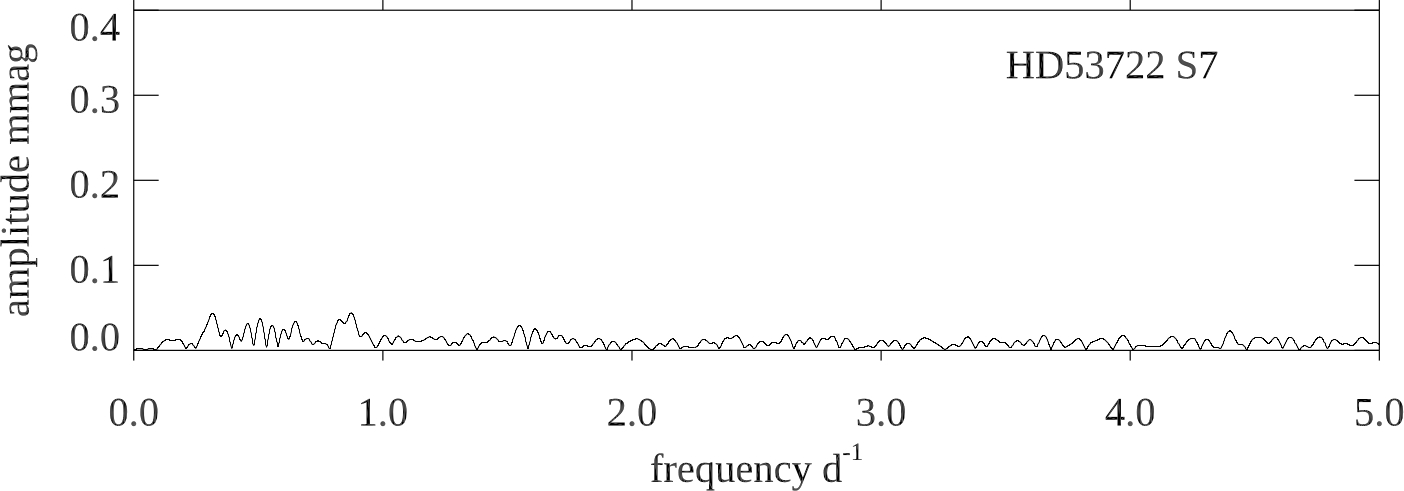}	
\includegraphics[width=0.48\linewidth,angle=0]{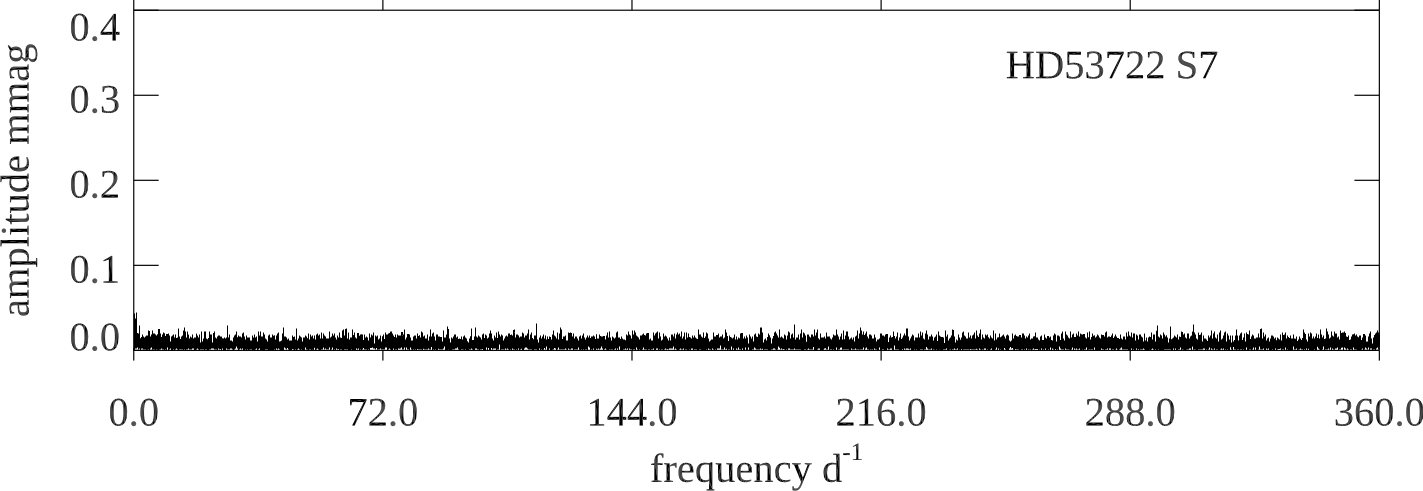}	
\includegraphics[width=0.48\linewidth,angle=0]{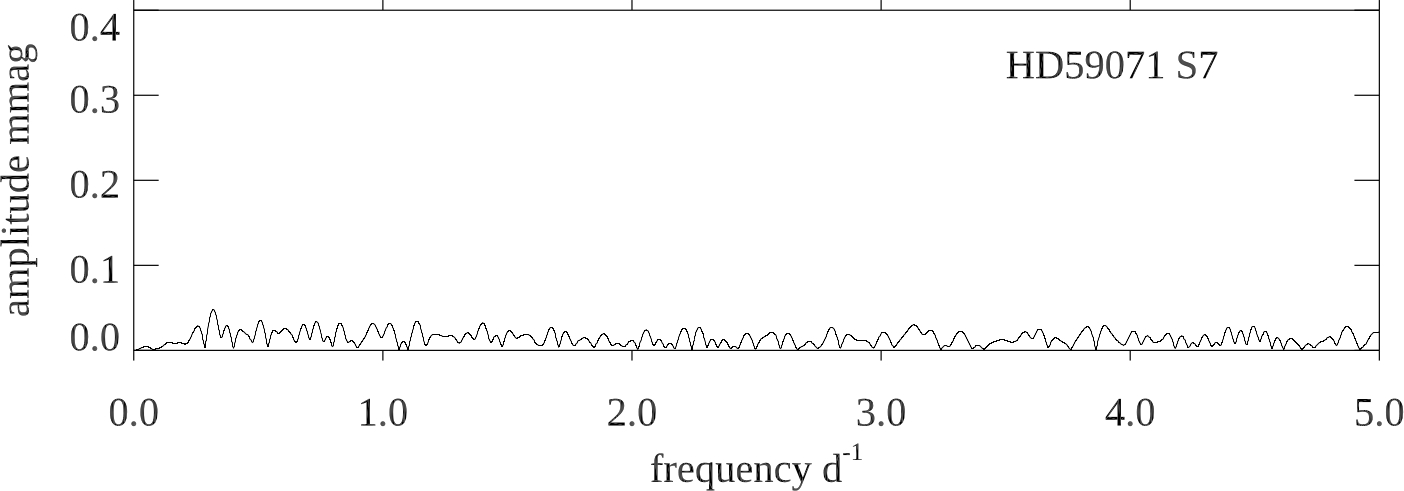}	
\includegraphics[width=0.48\linewidth,angle=0]{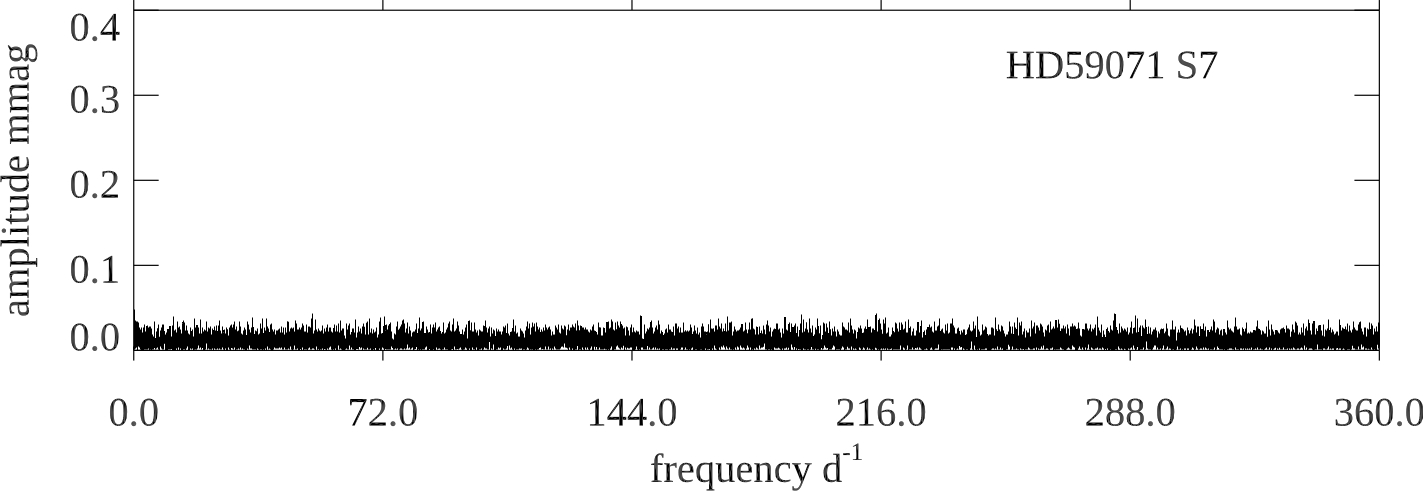}	
\includegraphics[width=0.48\linewidth,angle=0]{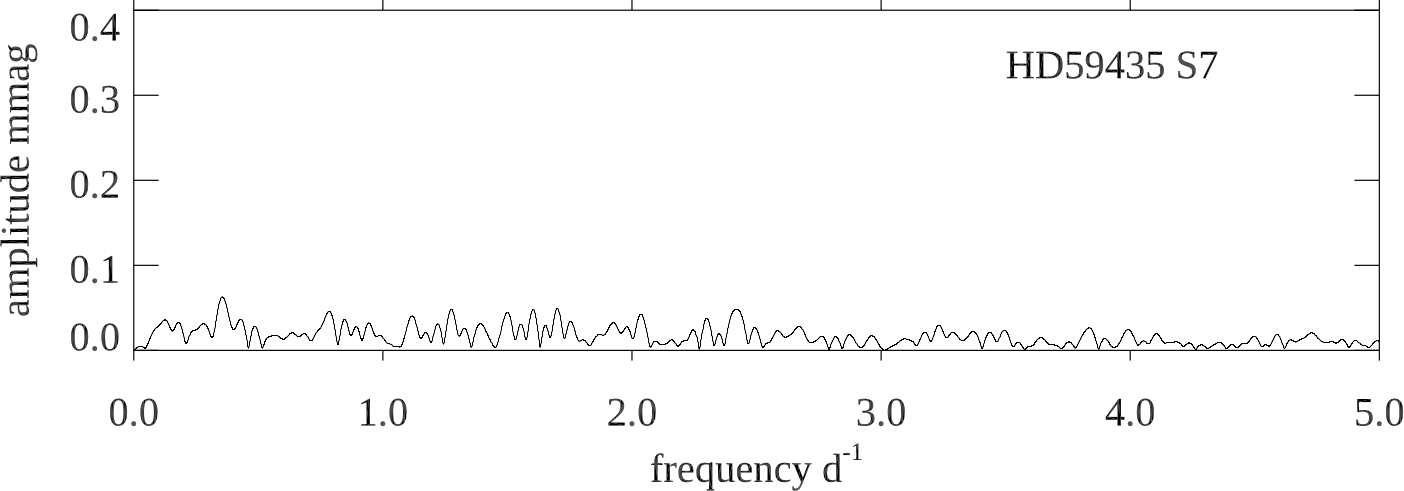}	
\includegraphics[width=0.48\linewidth,angle=0]{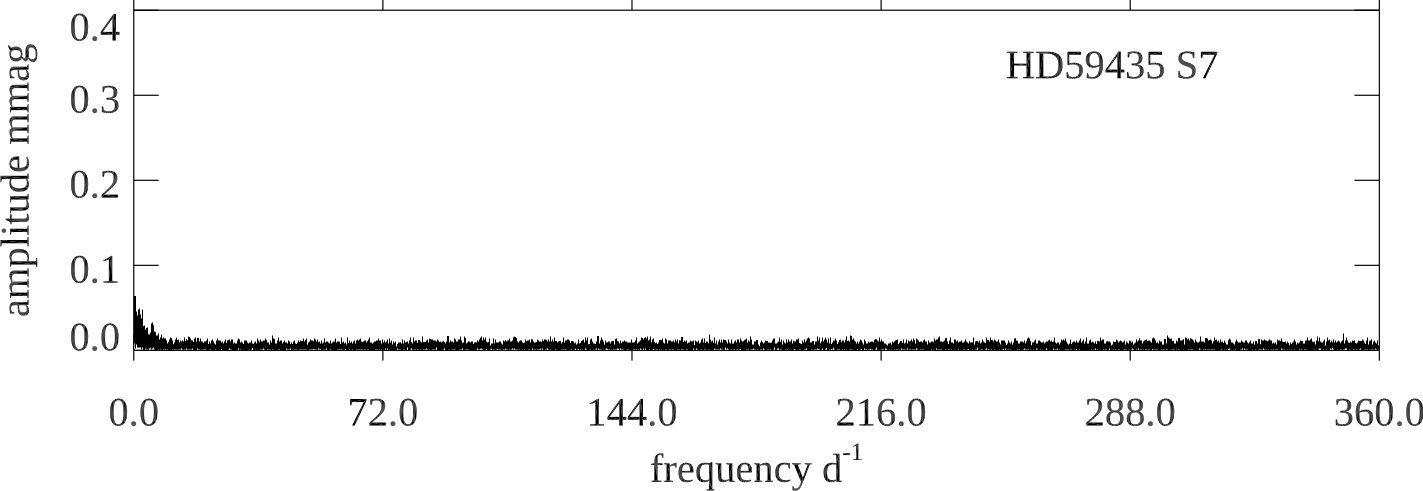}	
\includegraphics[width=0.48\linewidth,angle=0]{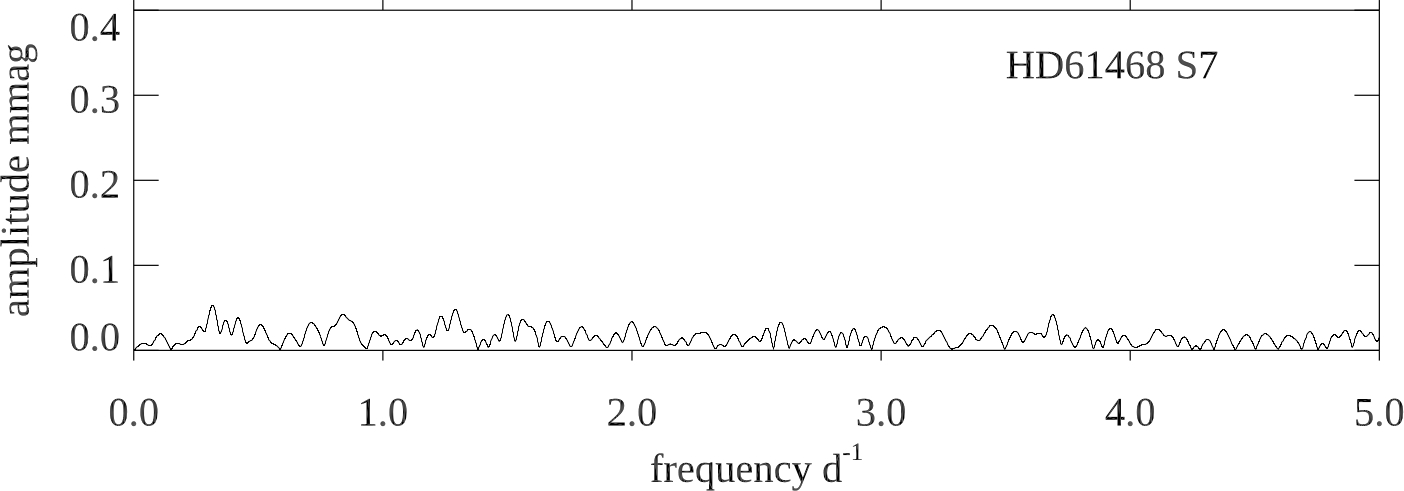}	
\includegraphics[width=0.48\linewidth,angle=0]{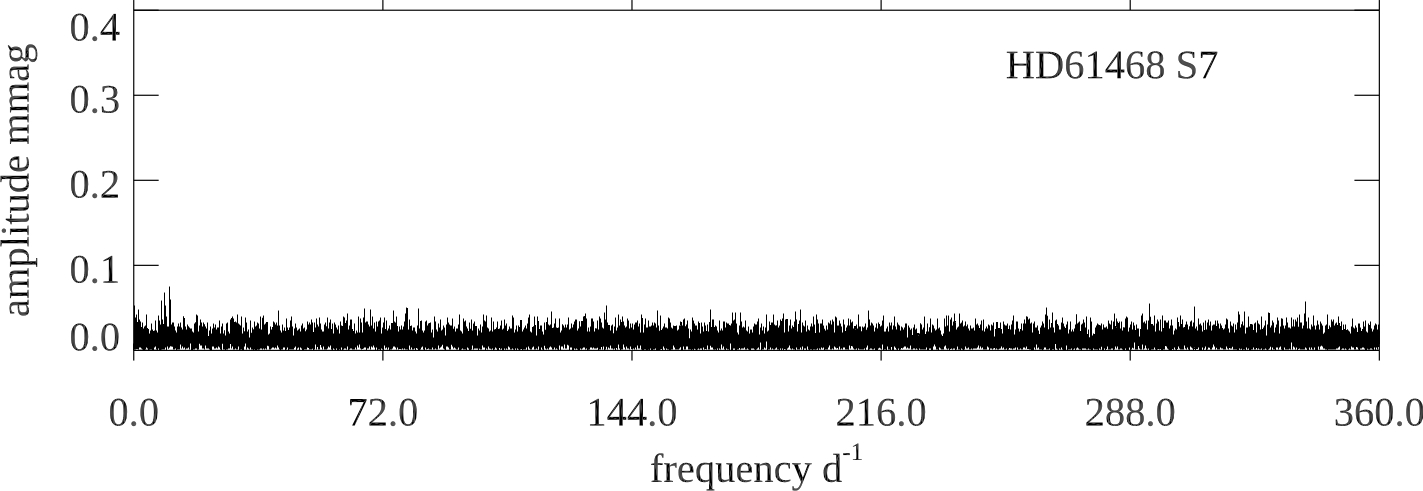}	
\includegraphics[width=0.48\linewidth,angle=0]{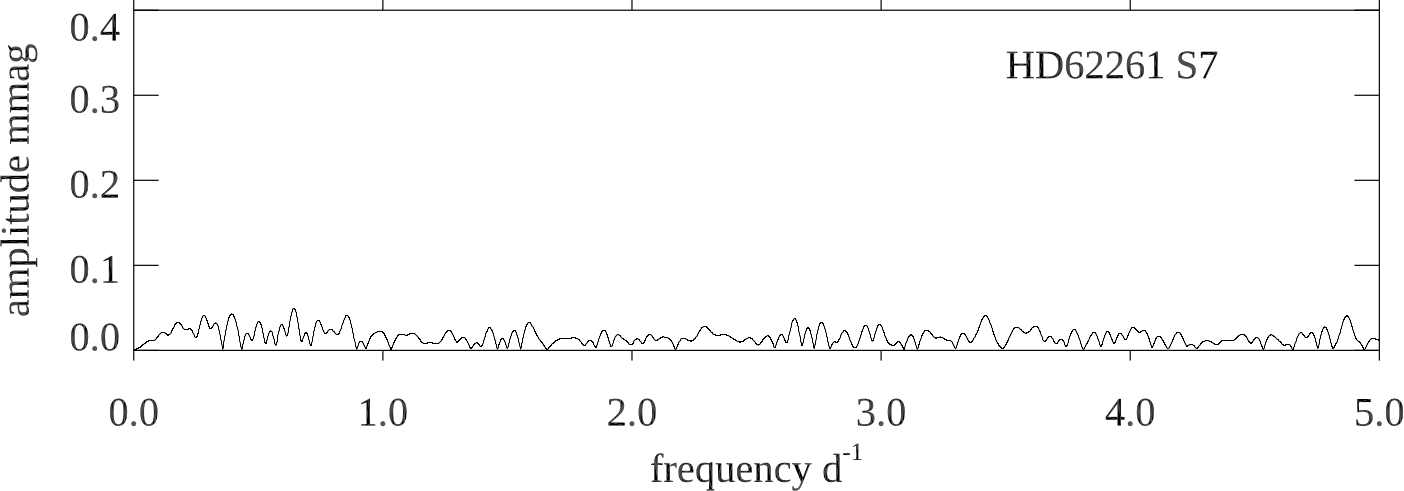}	
\includegraphics[width=0.48\linewidth,angle=0]{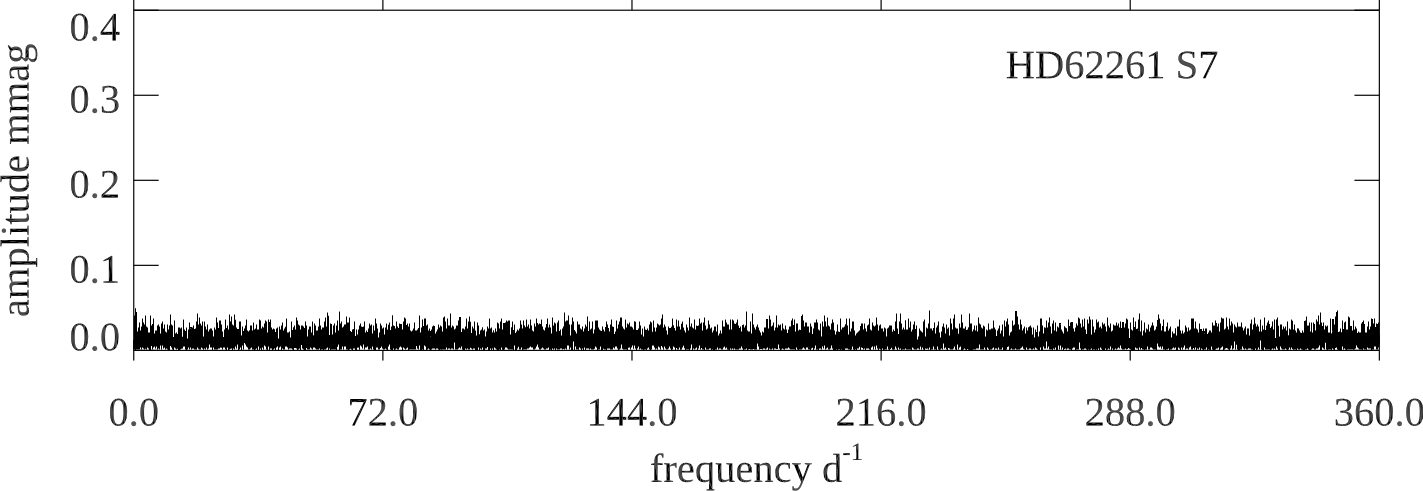}	
\includegraphics[width=0.48\linewidth,angle=0]{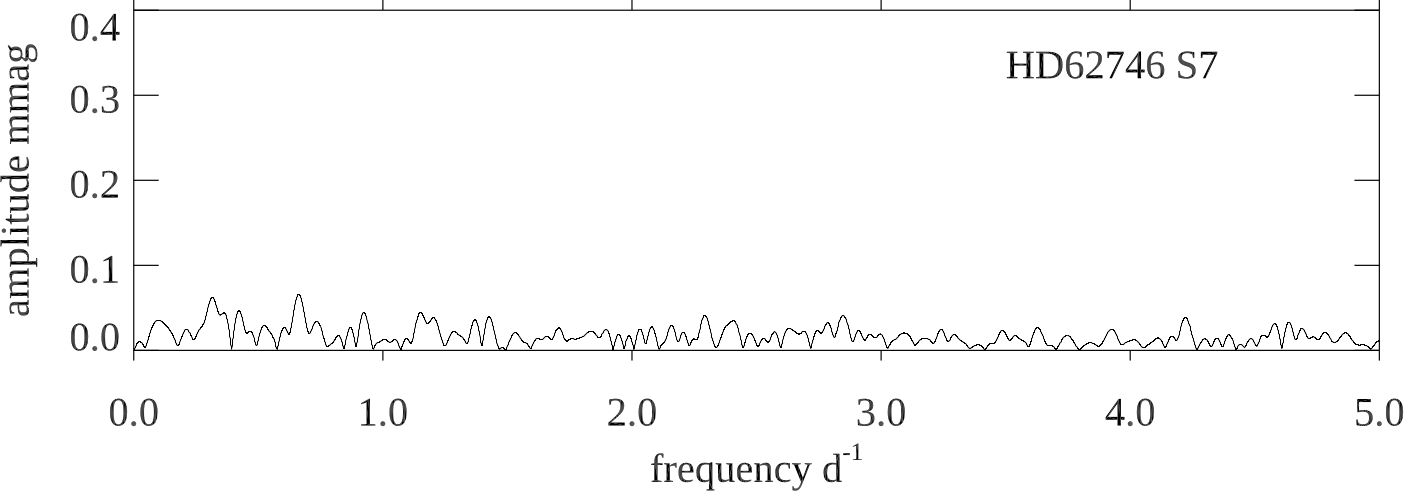}	
\includegraphics[width=0.48\linewidth,angle=0]{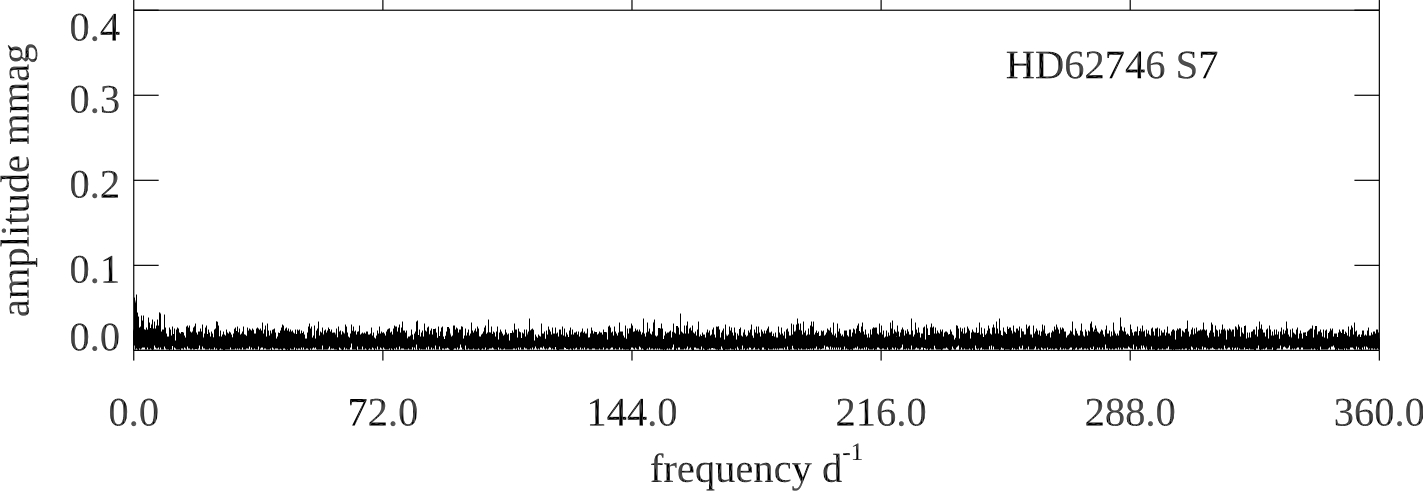}	
\includegraphics[width=0.48\linewidth,angle=0]{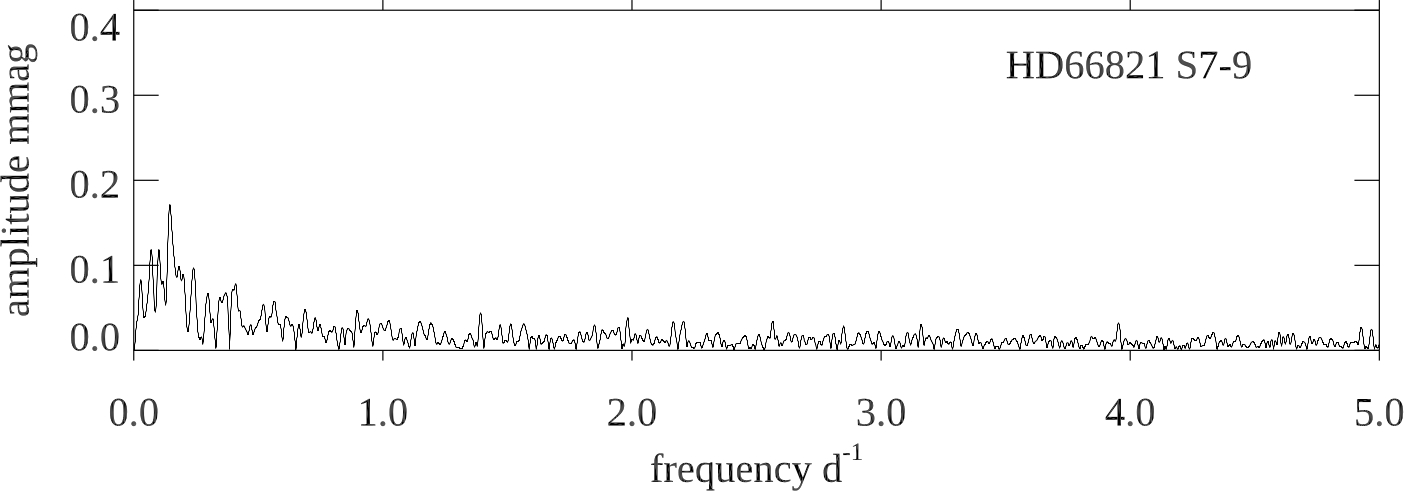}	
\includegraphics[width=0.48\linewidth,angle=0]{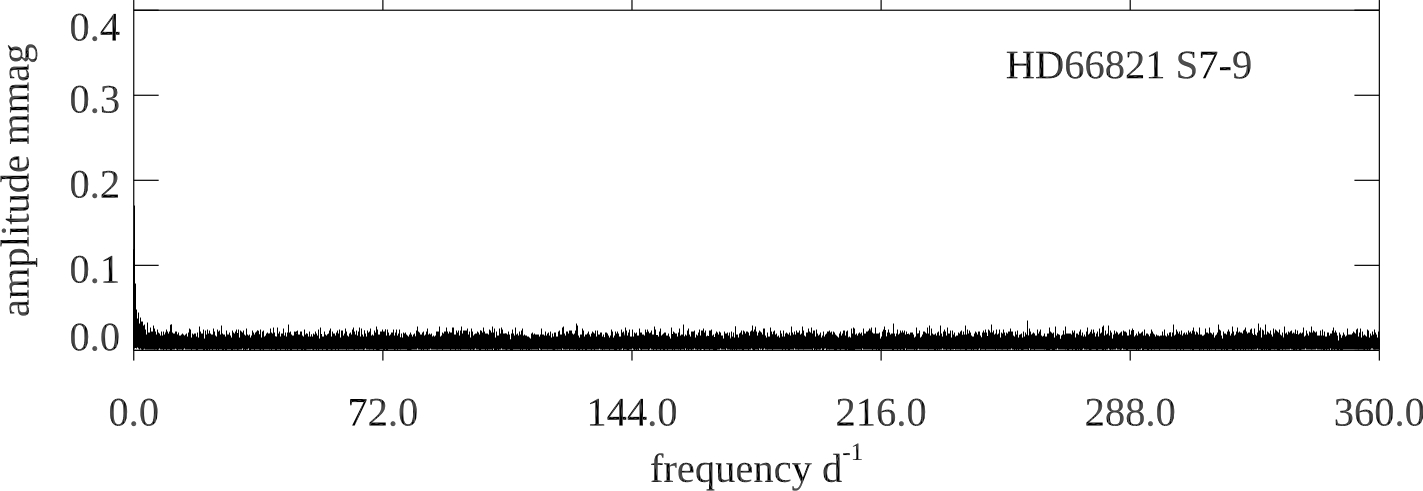}	
\caption{Same as Fig.~\ref{fig:fts_1}. HD~61468 has possible $\delta$~Sct p~mode or mixed mode peaks of amplitude $\sim$70\,$\mu$mag in the frequency range $8-11$\,d$^{-1}$. }
\label{fig:fts_4}
\end{figure*}

\begin{figure*}[p]
\centering
\includegraphics[width=0.48\linewidth,angle=0]{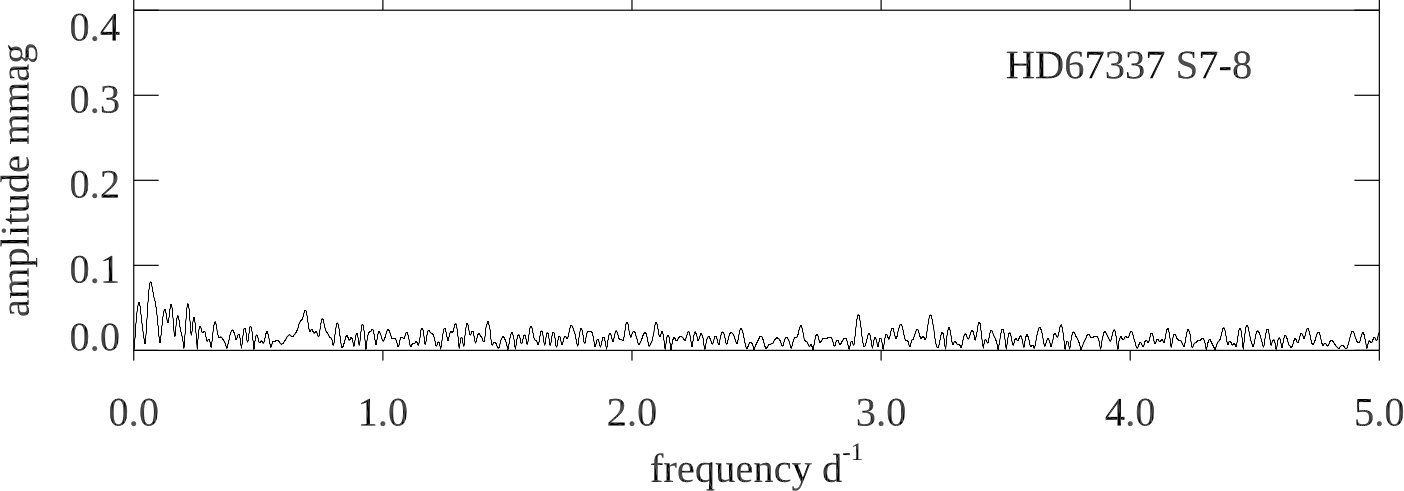}	
\includegraphics[width=0.48\linewidth,angle=0]{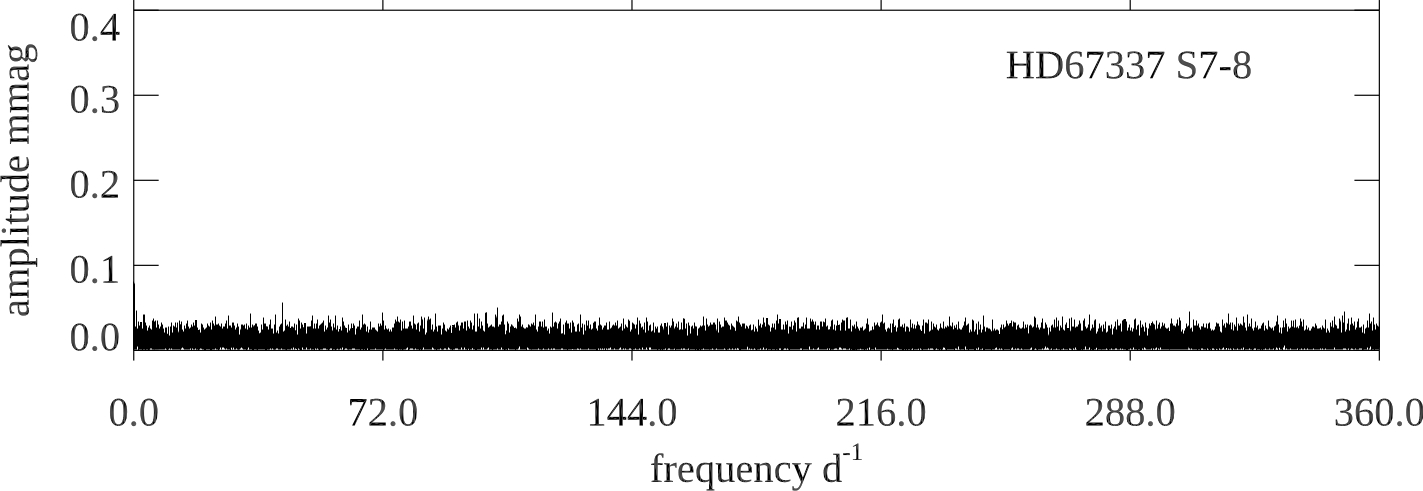}	
\includegraphics[width=0.48\linewidth,angle=0]{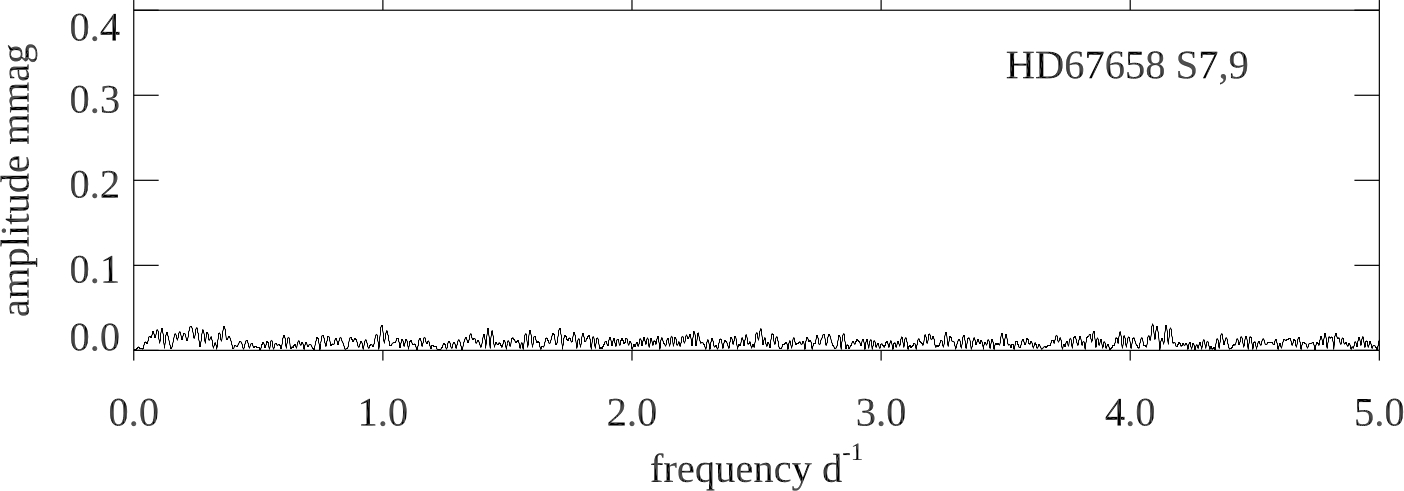}	
\includegraphics[width=0.48\linewidth,angle=0]{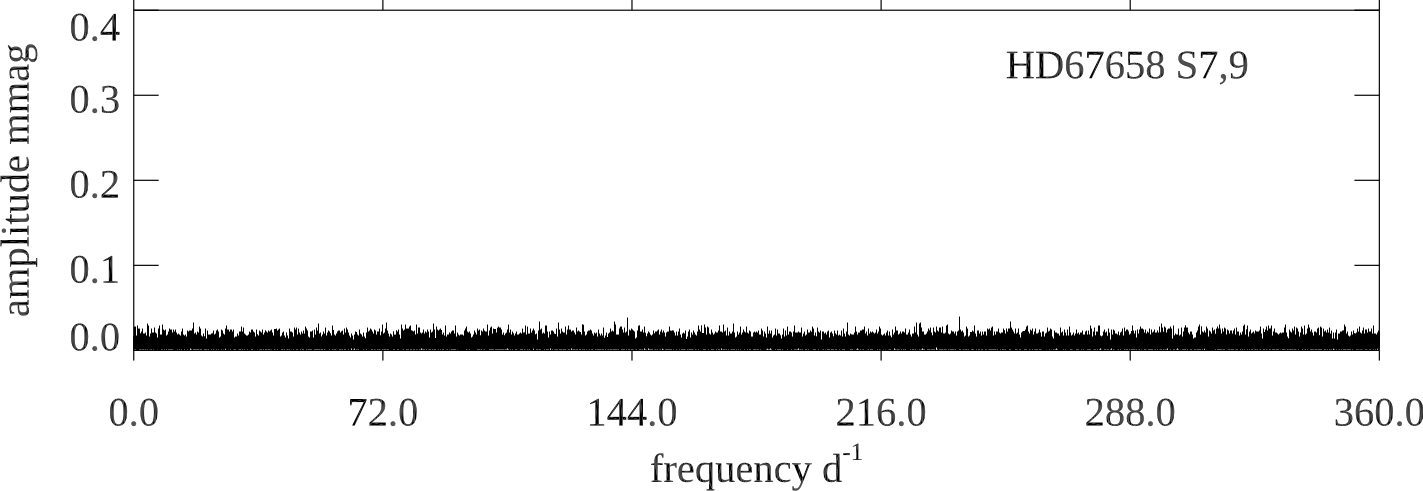}	
\includegraphics[width=0.48\linewidth,angle=0]{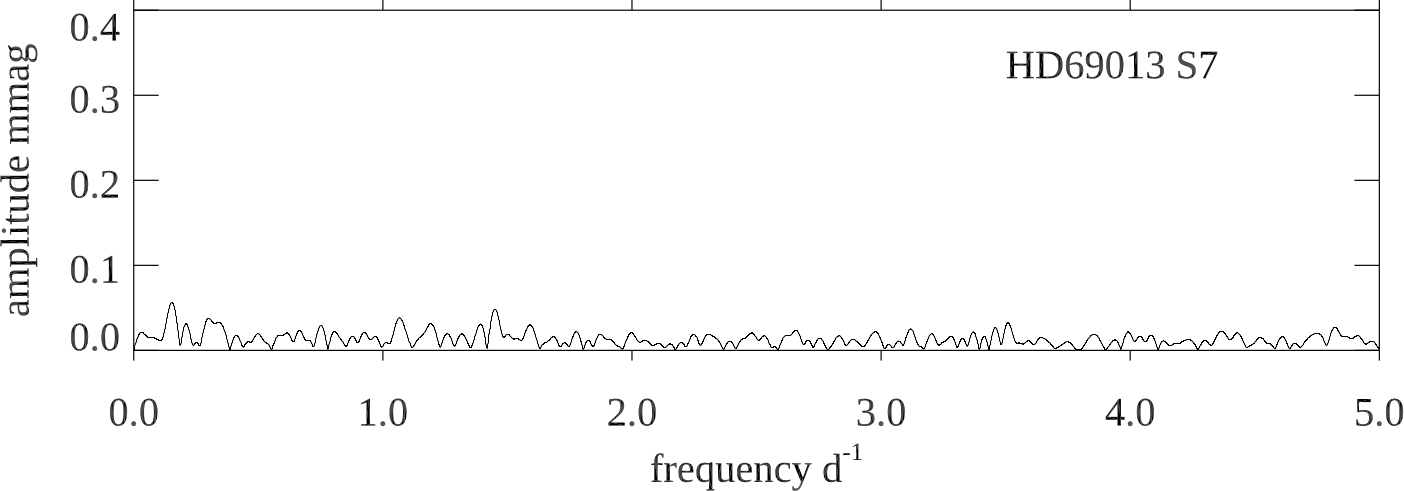}	
\includegraphics[width=0.48\linewidth,angle=0]{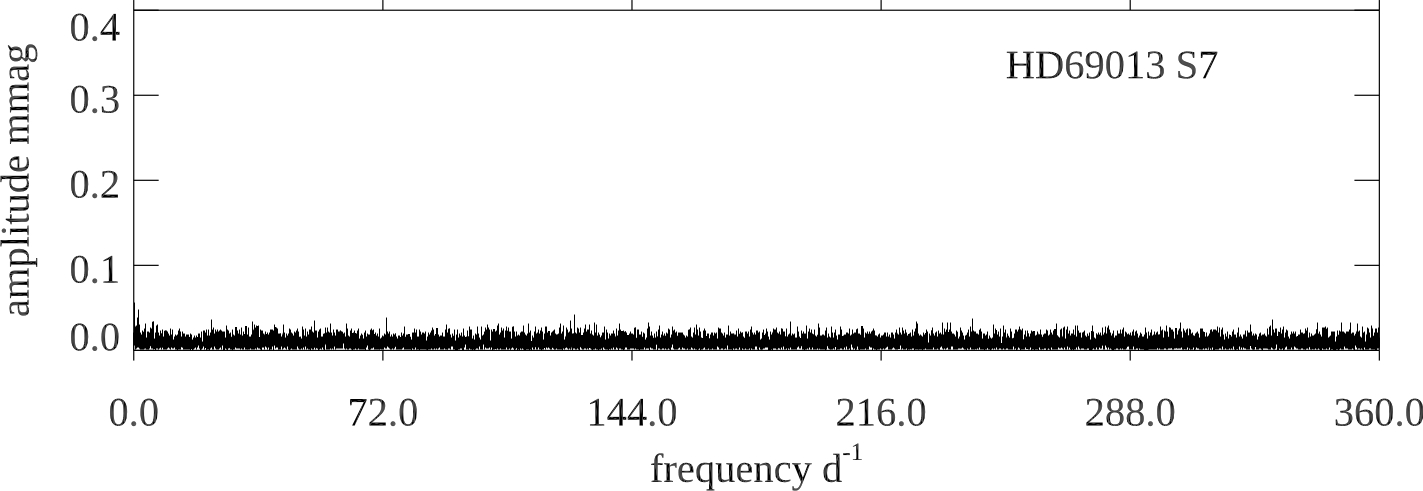}	
\includegraphics[width=0.48\linewidth,angle=0]{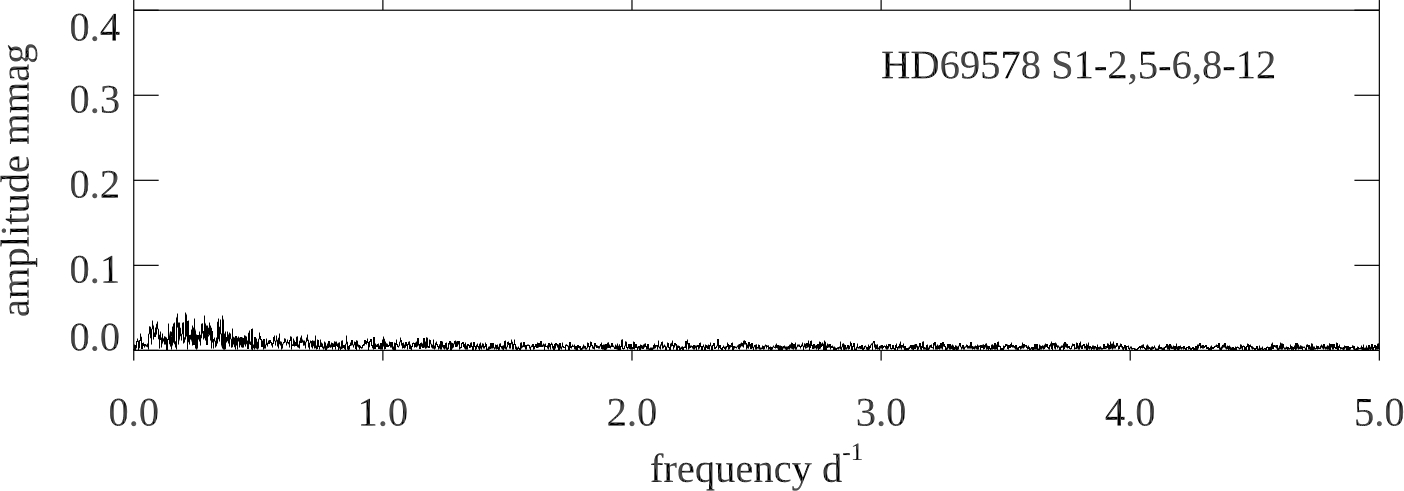}	
\includegraphics[width=0.48\linewidth,angle=0]{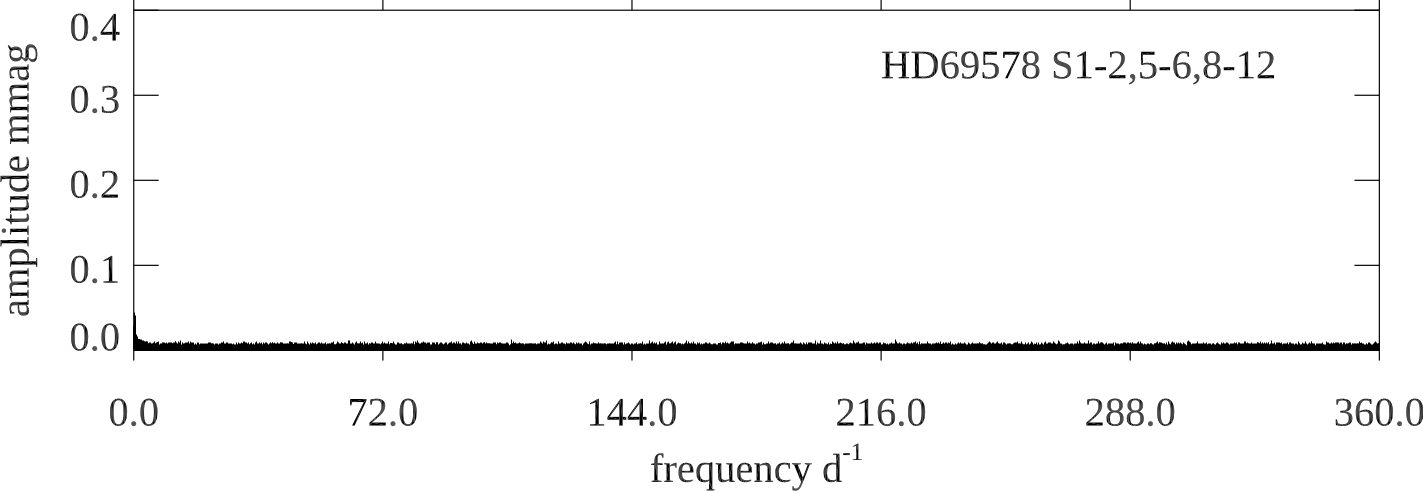}	
\includegraphics[width=0.48\linewidth,angle=0]{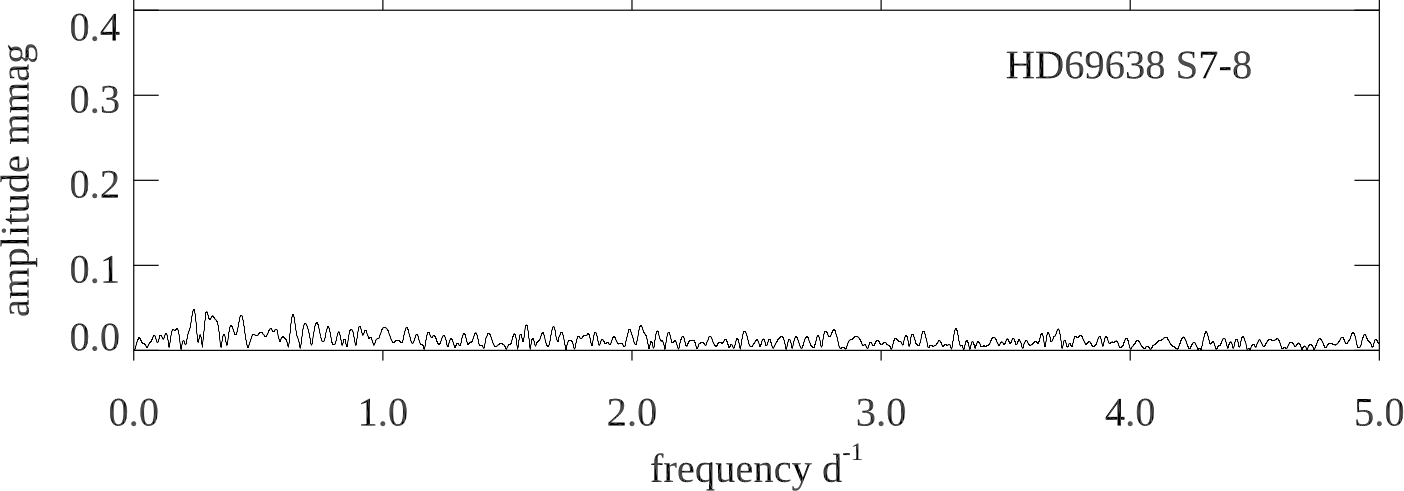}	
\includegraphics[width=0.48\linewidth,angle=0]{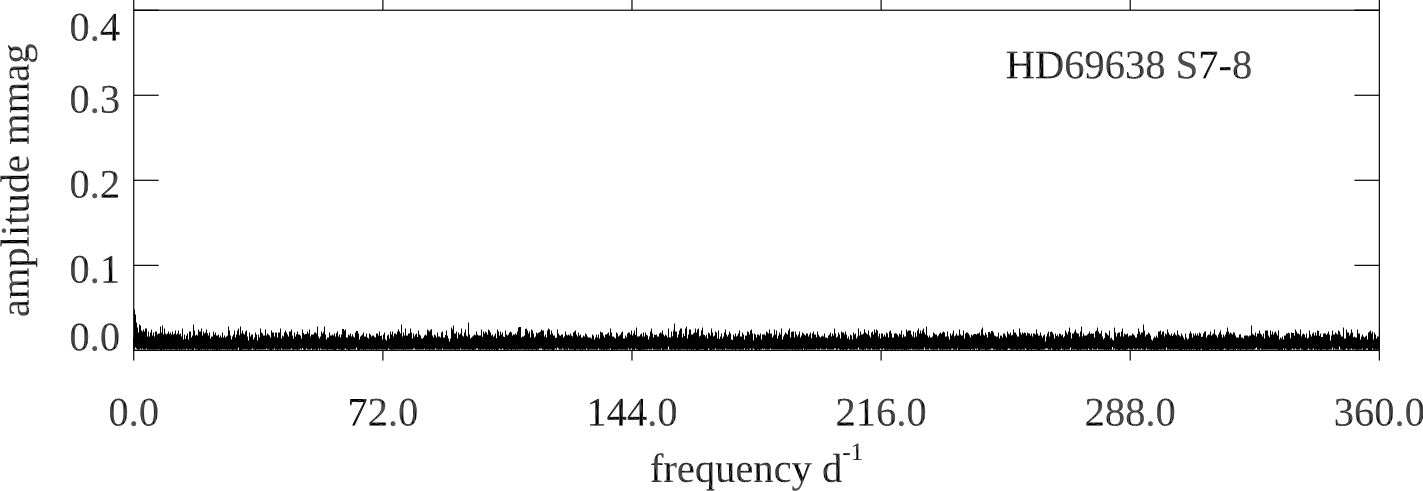}	
\includegraphics[width=0.48\linewidth,angle=0]{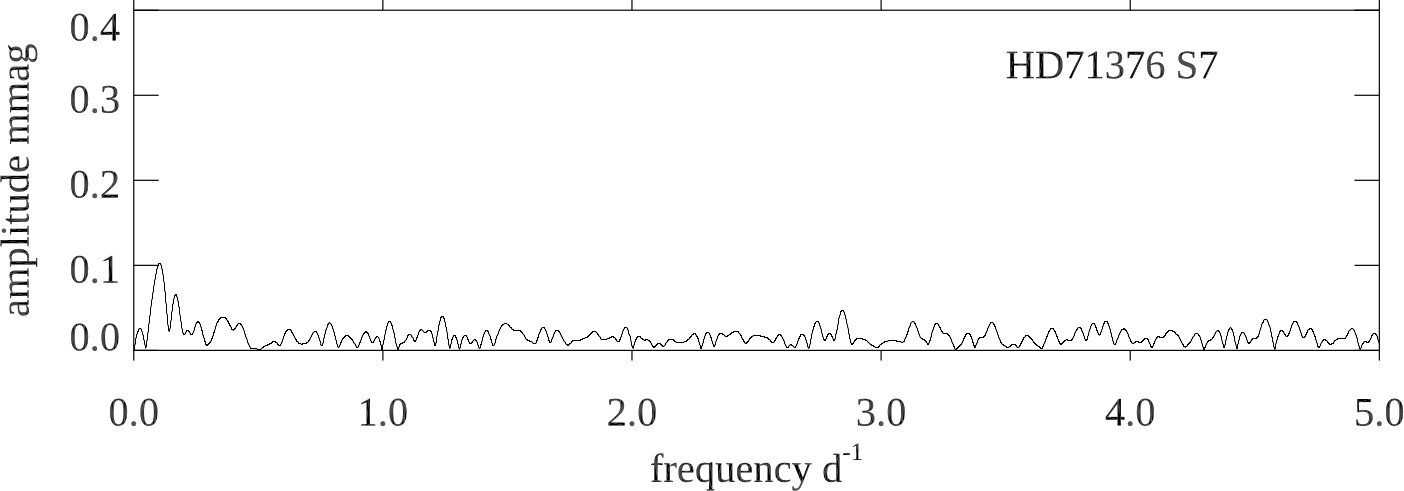}	
\includegraphics[width=0.48\linewidth,angle=0]{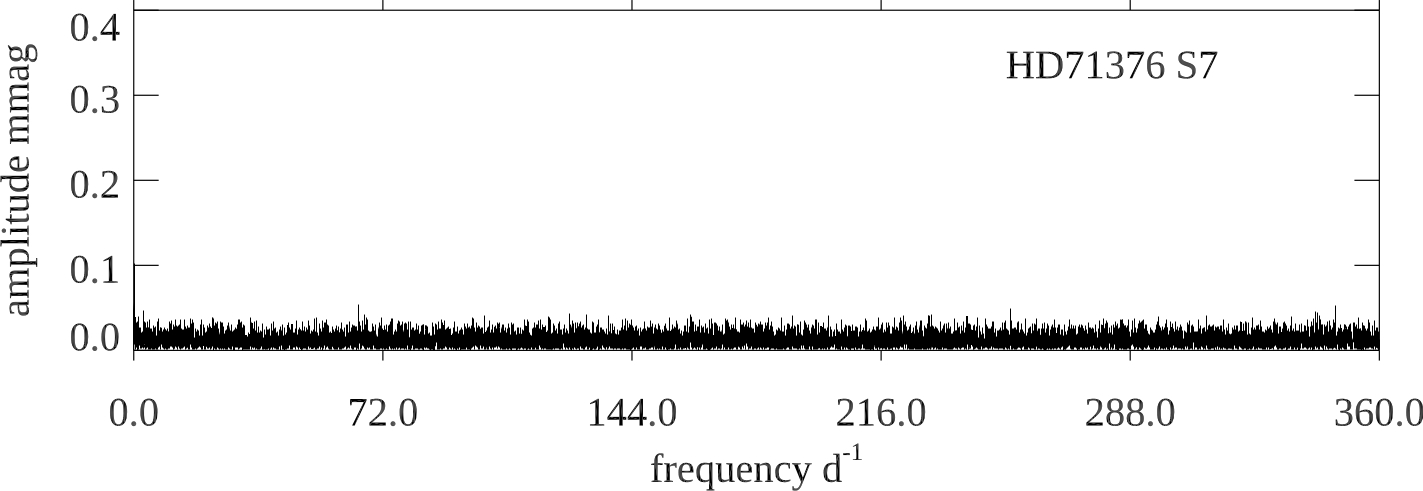}	
\includegraphics[width=0.48\linewidth,angle=0]{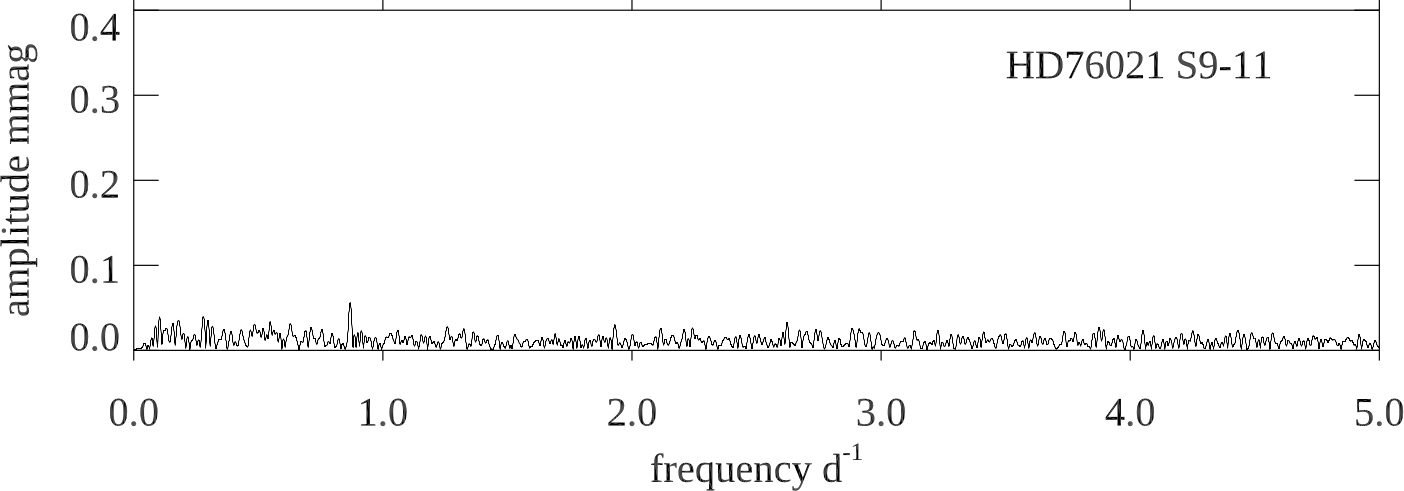}	
\includegraphics[width=0.48\linewidth,angle=0]{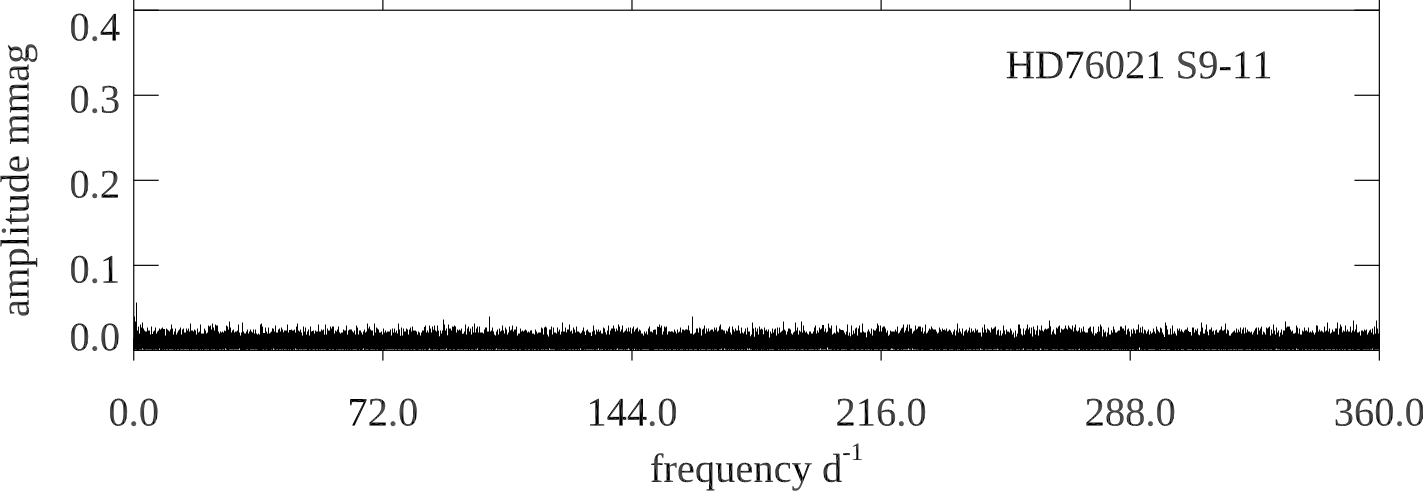}	
\caption{Same as Fig.~\ref{fig:fts_1}. HD~76021 has a low amplitude peak at 0.870\,d$^{-1}$ with an amplitude of 56\,$\mu$mag that could be rotational.
 }
\label{fig:fts_5}
\end{figure*}

\begin{figure*}[p]
\centering
\includegraphics[width=0.48\linewidth,angle=0]{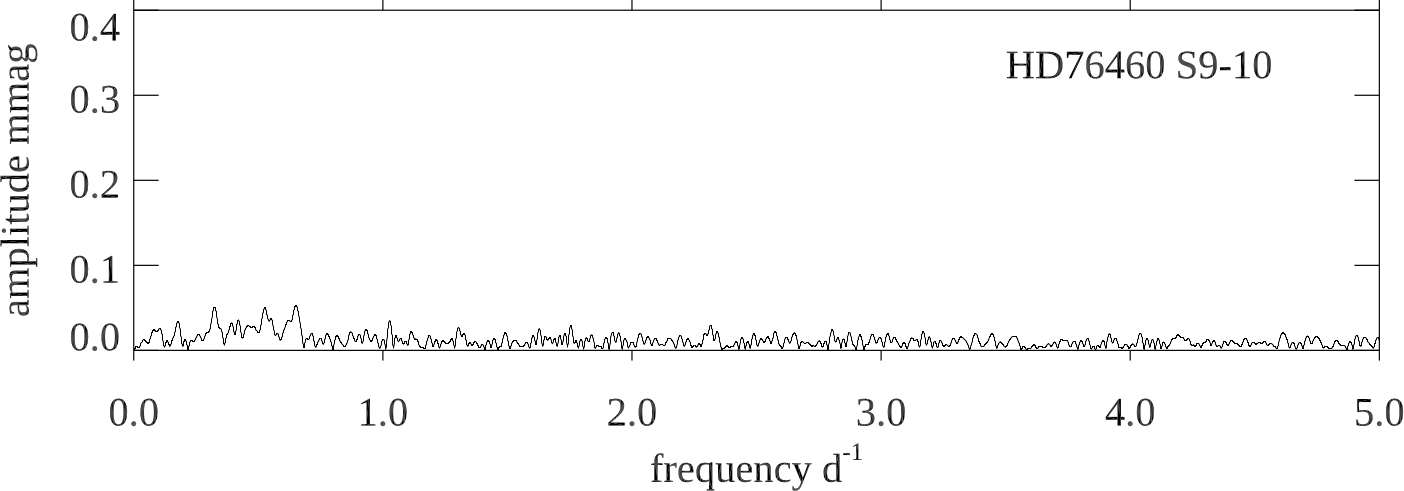}	
\includegraphics[width=0.48\linewidth,angle=0]{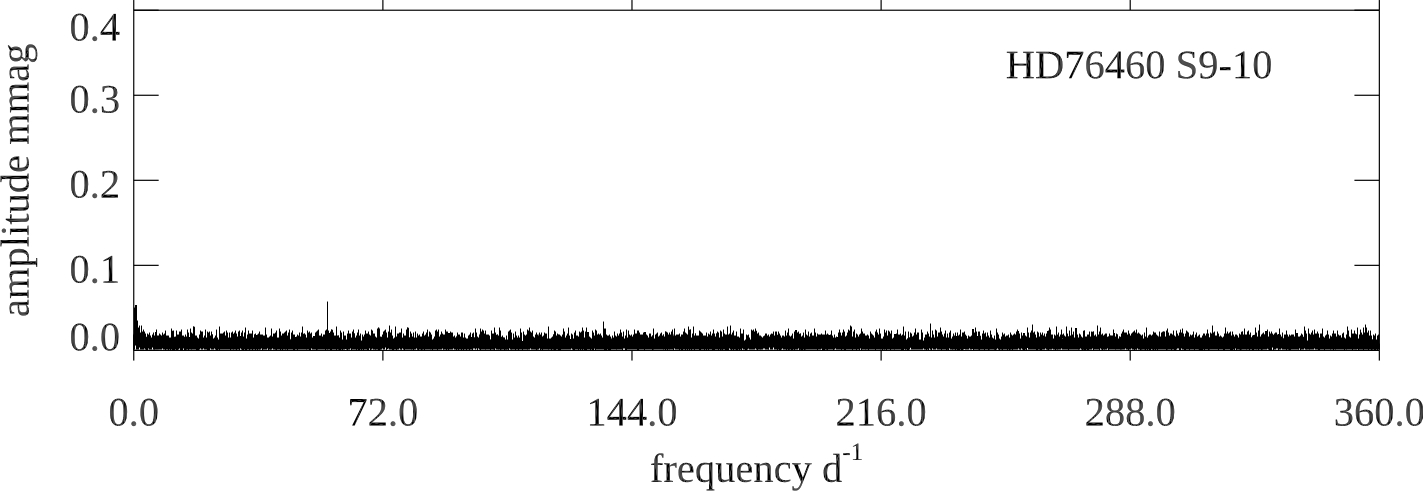}	
\includegraphics[width=0.48\linewidth,angle=0]{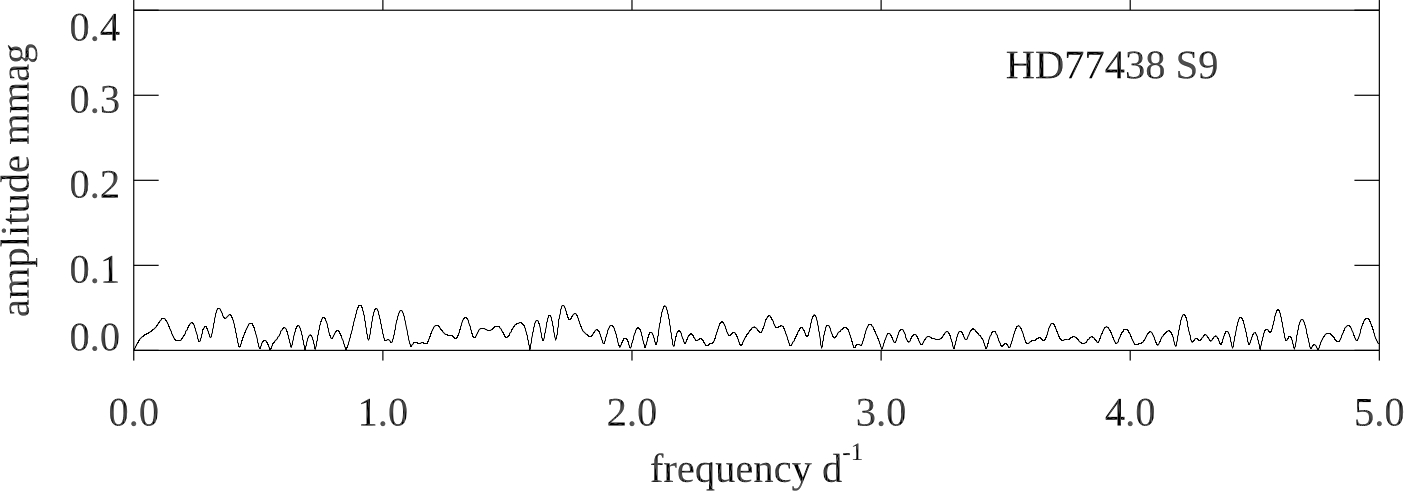}	
\includegraphics[width=0.48\linewidth,angle=0]{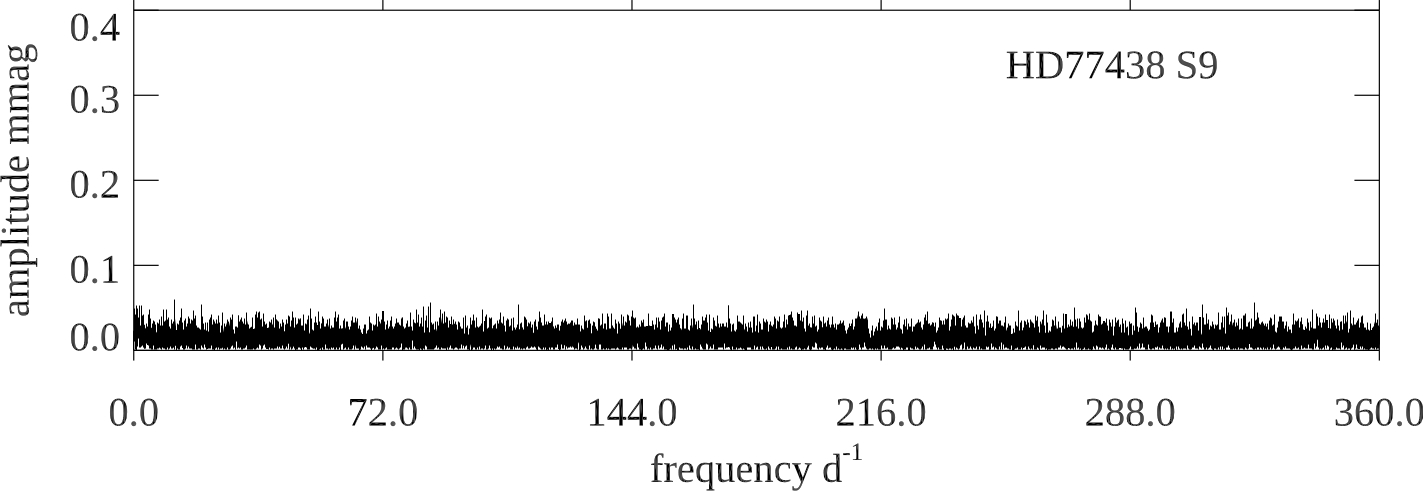}	
\includegraphics[width=0.48\linewidth,angle=0]{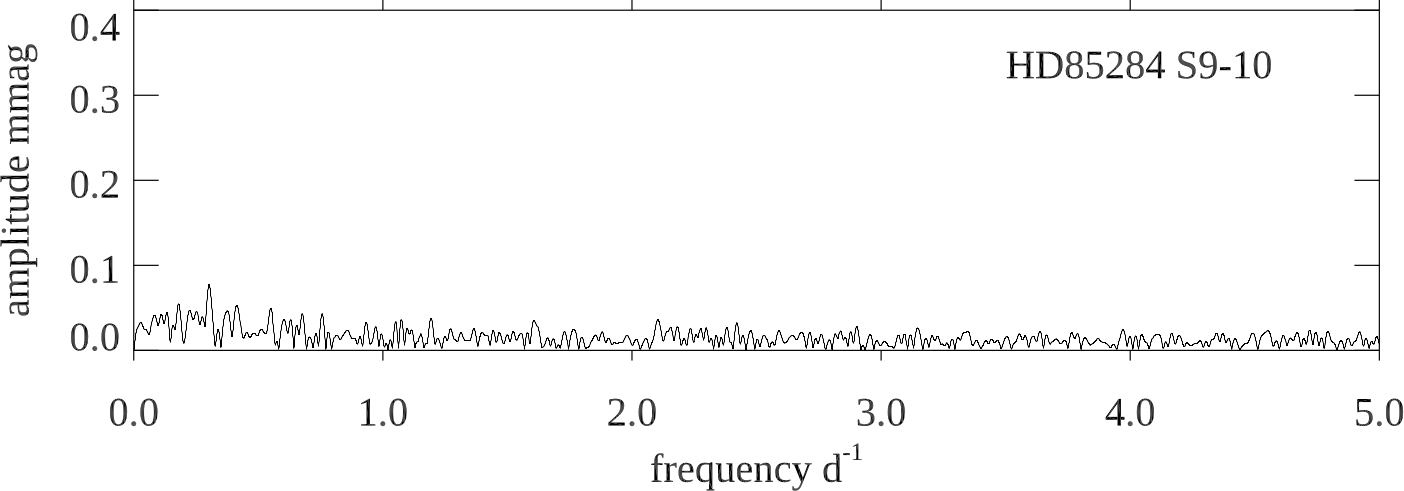}	
\includegraphics[width=0.48\linewidth,angle=0]{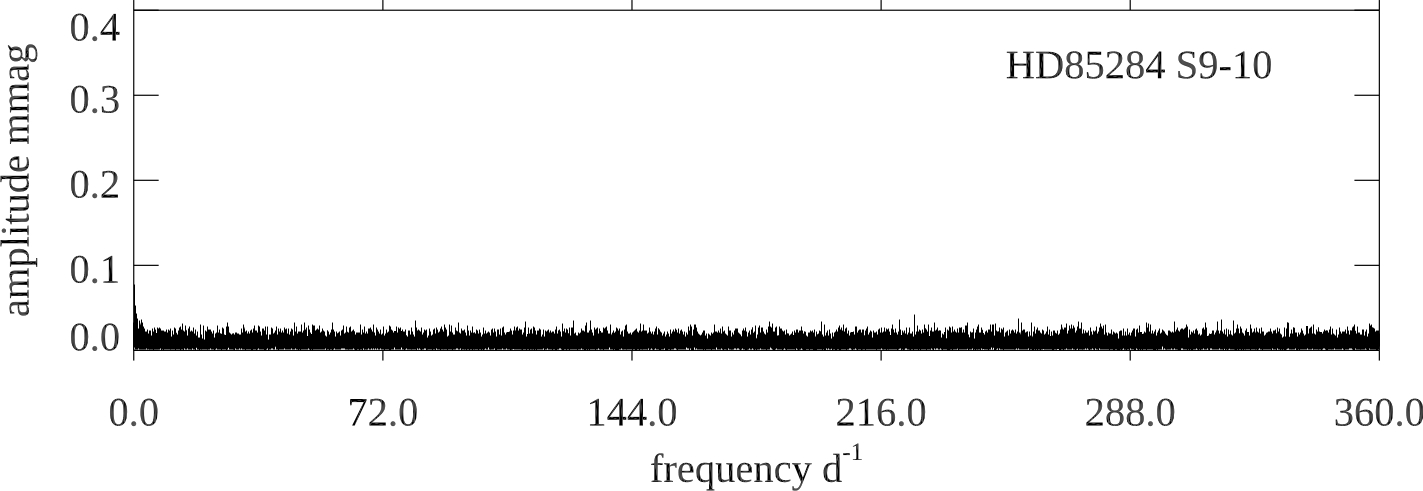}	
\includegraphics[width=0.48\linewidth,angle=0]{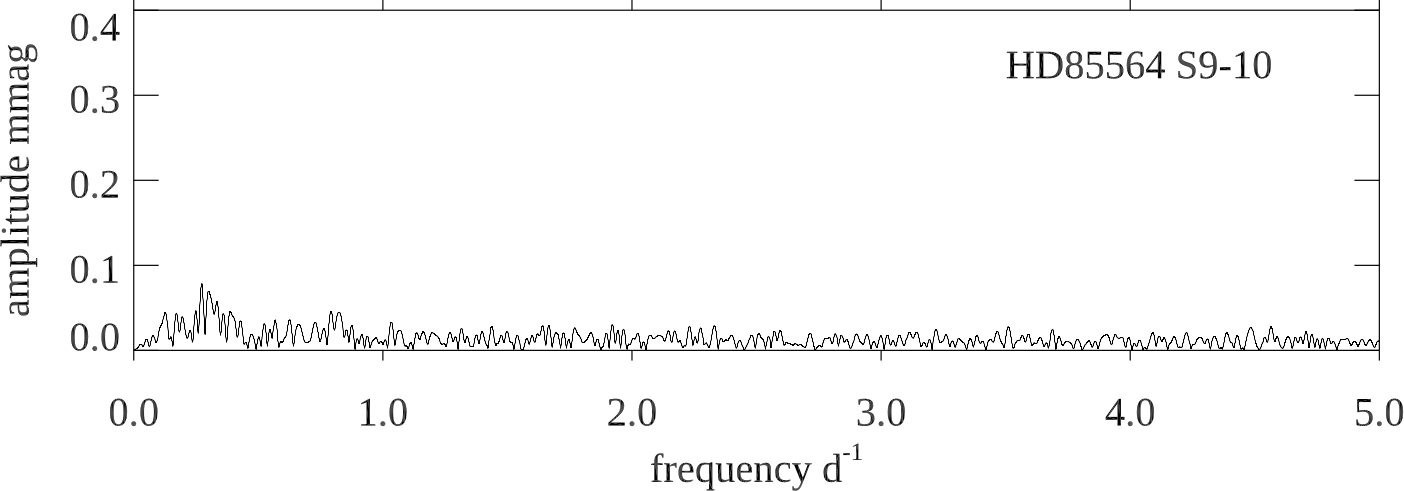}	
\includegraphics[width=0.48\linewidth,angle=0]{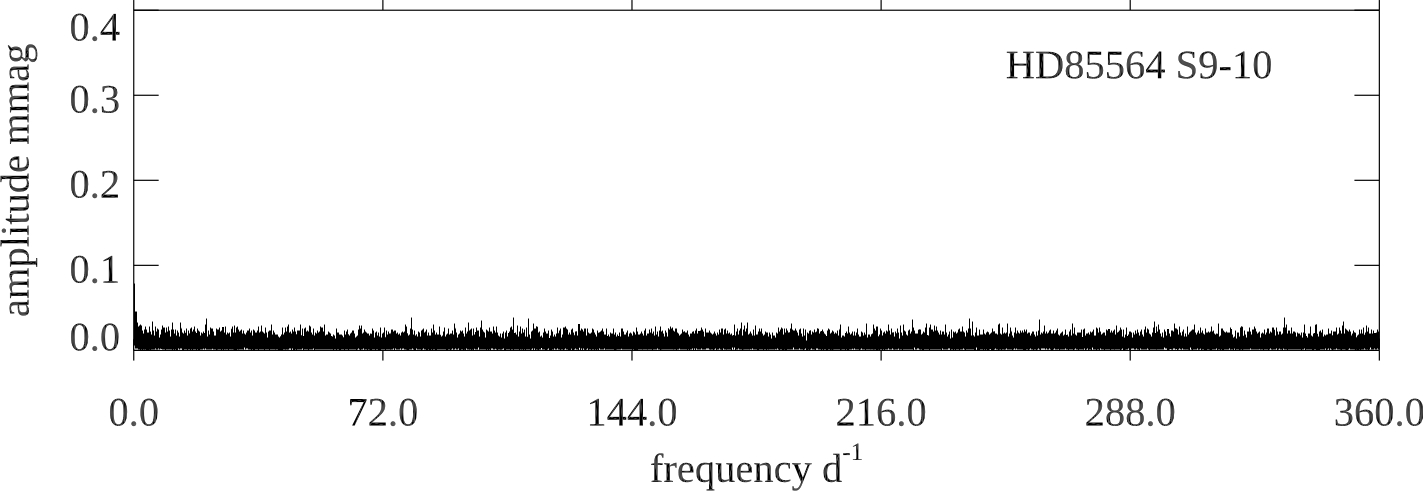}	
\includegraphics[width=0.48\linewidth,angle=0]{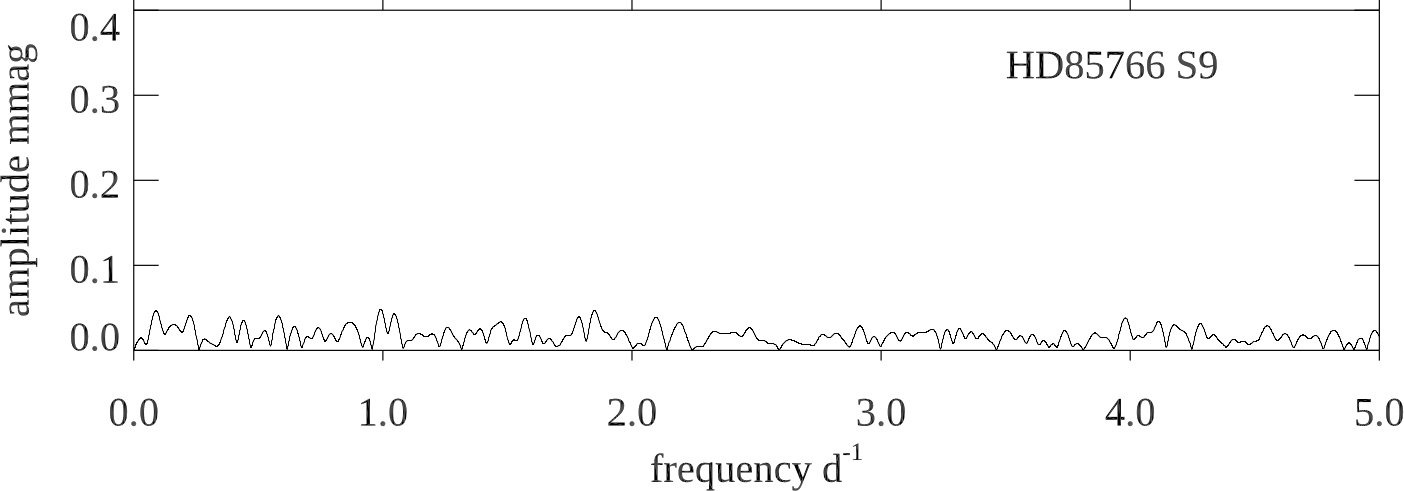}	
\includegraphics[width=0.48\linewidth,angle=0]{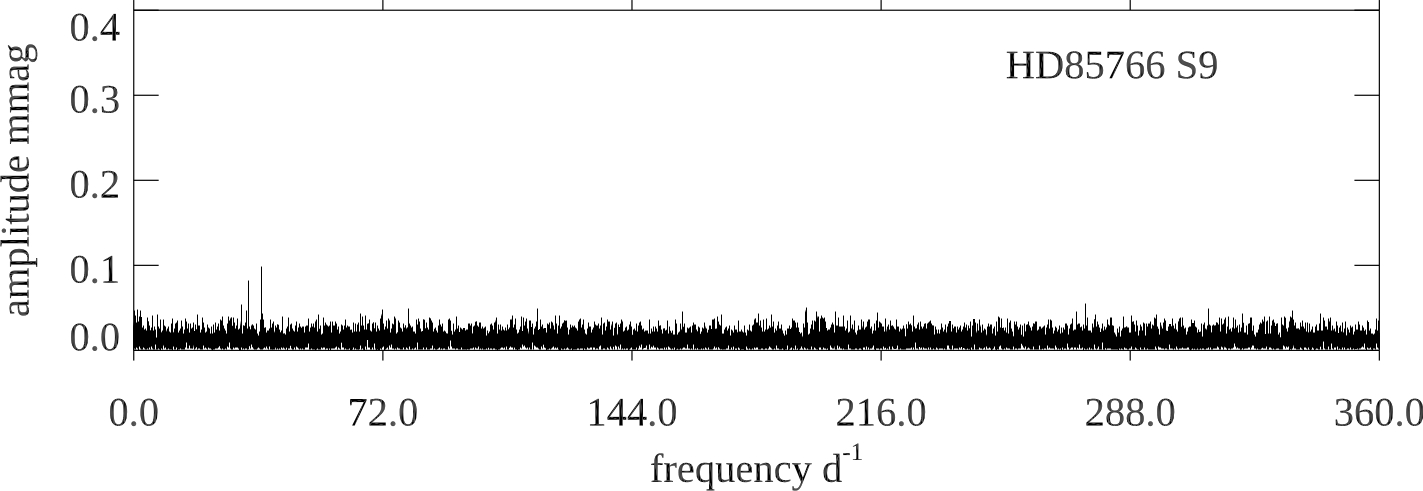}	
\includegraphics[width=0.48\linewidth,angle=0]{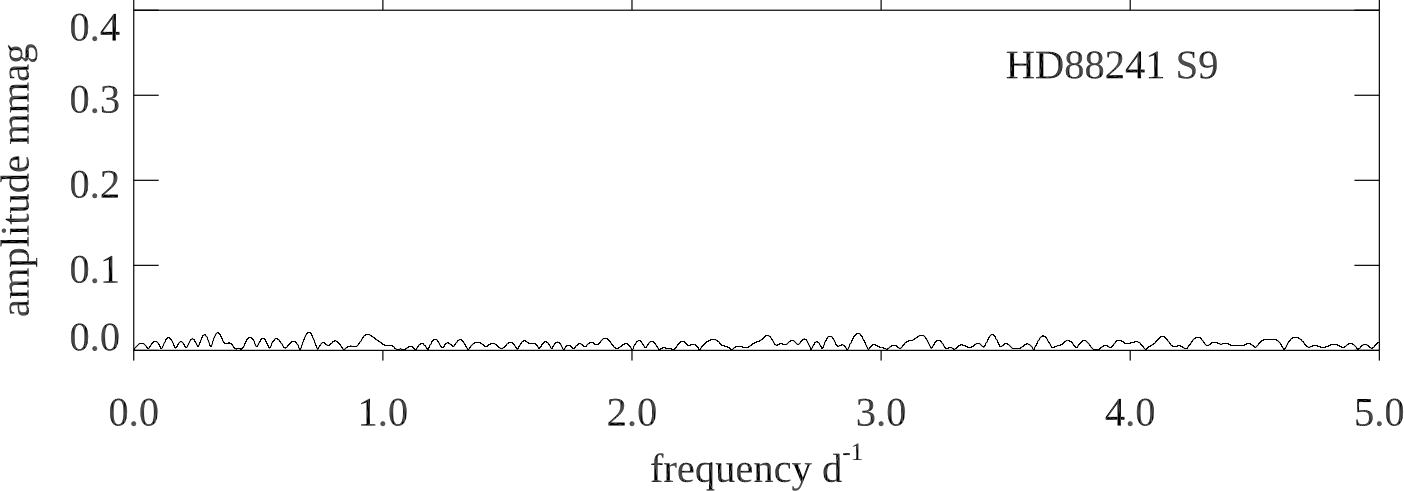}	
\includegraphics[width=0.48\linewidth,angle=0]{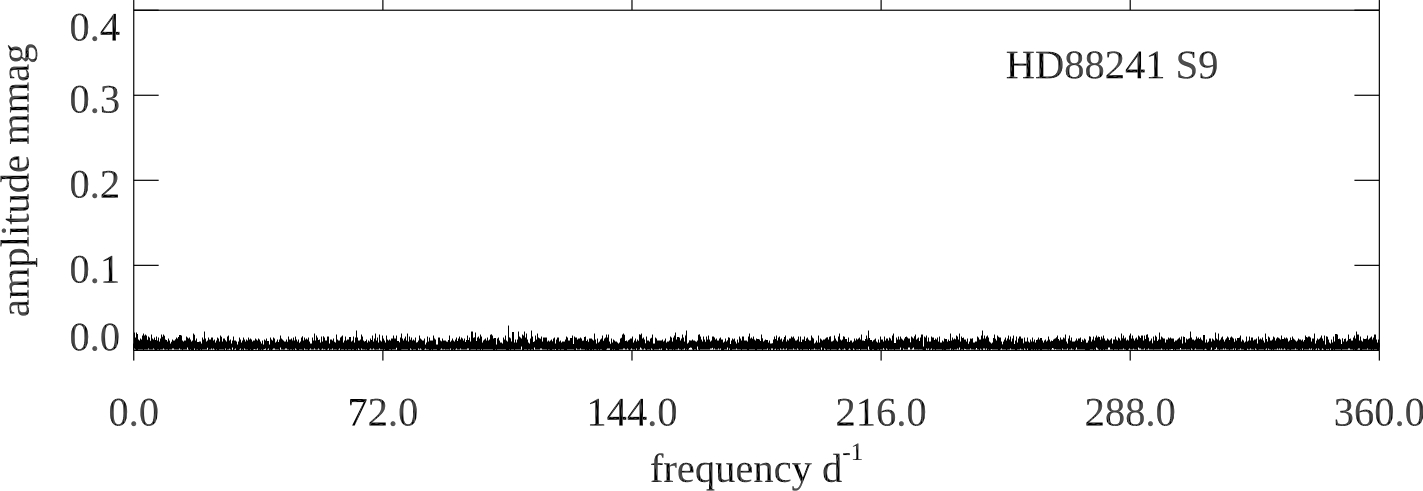}	
\includegraphics[width=0.48\linewidth,angle=0]{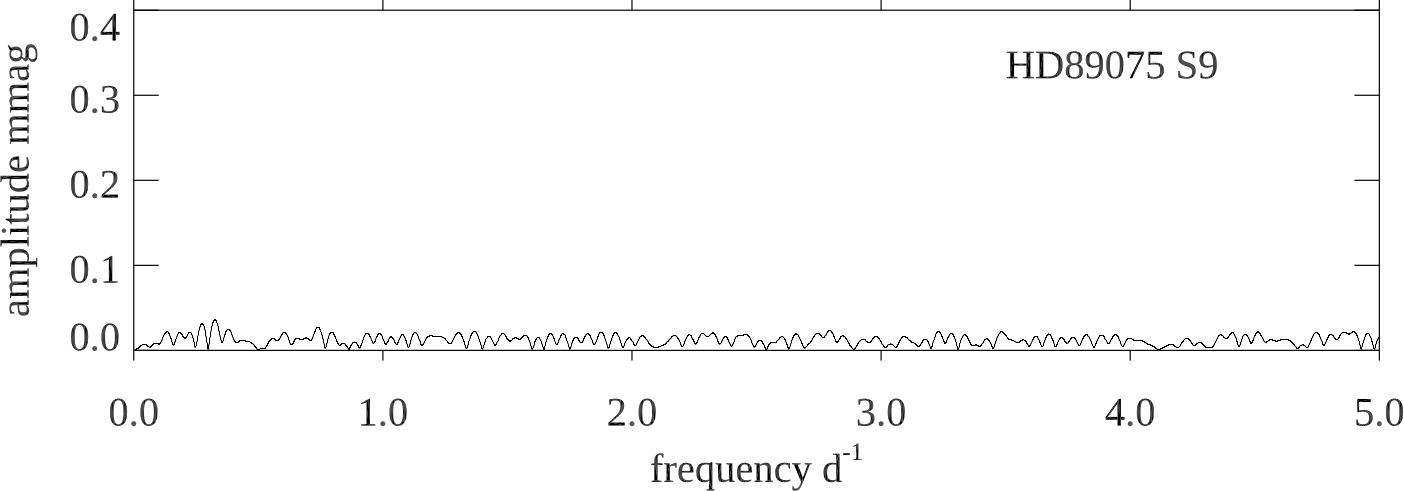}	
\includegraphics[width=0.48\linewidth,angle=0]{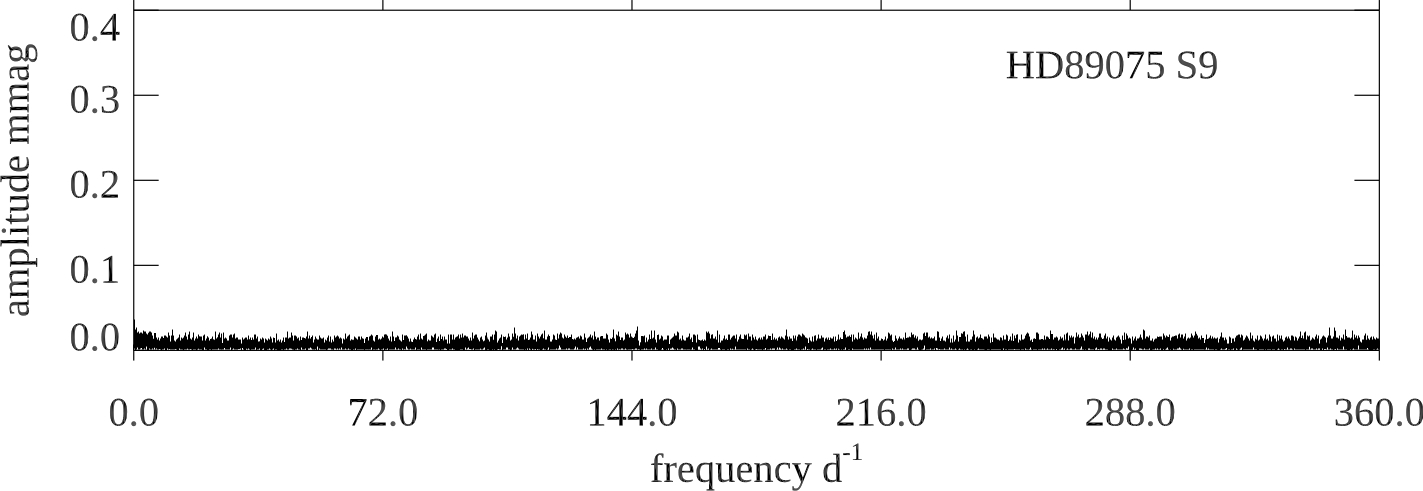}	
\caption{Same as Fig.~\ref{fig:fts_1}.   HD~76460 has a single peak at 55.866\,d$^{-1}$ ($P = 26$\,min) not previously noted in the literature with significant amplitude of $58 \pm 7$\,$\mu$mag. It could be a high-frequency $\delta$~Sct star, or a very low frequency roAp star.  HD~85766 has no rotational variation, but does appear to have two, or more, very low amplitude $\delta$~Sct peaks not previously noted in the literature; the highest amplitude two are at $36.8877$\,d$^{-1}$ ($A = 97$\,$\mu$mag) and $33.2056$\,d$^{-1}$ ($A = 82$\,$\mu$mag). Most $\delta$~Sct stars show low frequency peaks in the g-mode range, but this star does not, supporting the Ap~(SiCr) classification.}
\label{fig:fts_6}
\end{figure*}

\begin{figure*}[p]
\centering
\includegraphics[width=0.48\linewidth,angle=0]{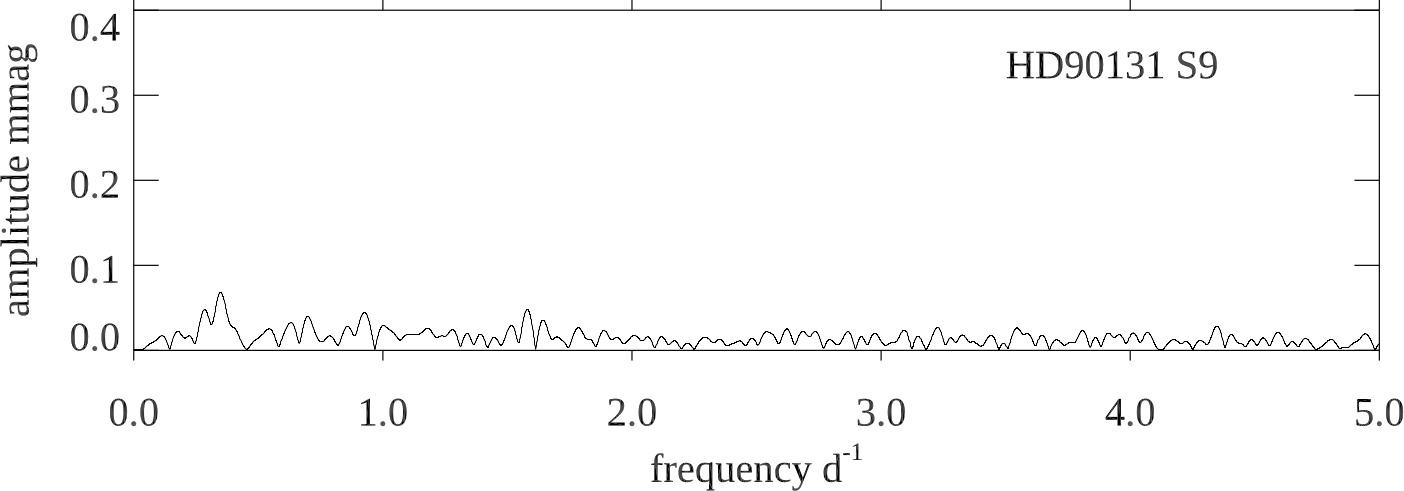}	
\includegraphics[width=0.48\linewidth,angle=0]{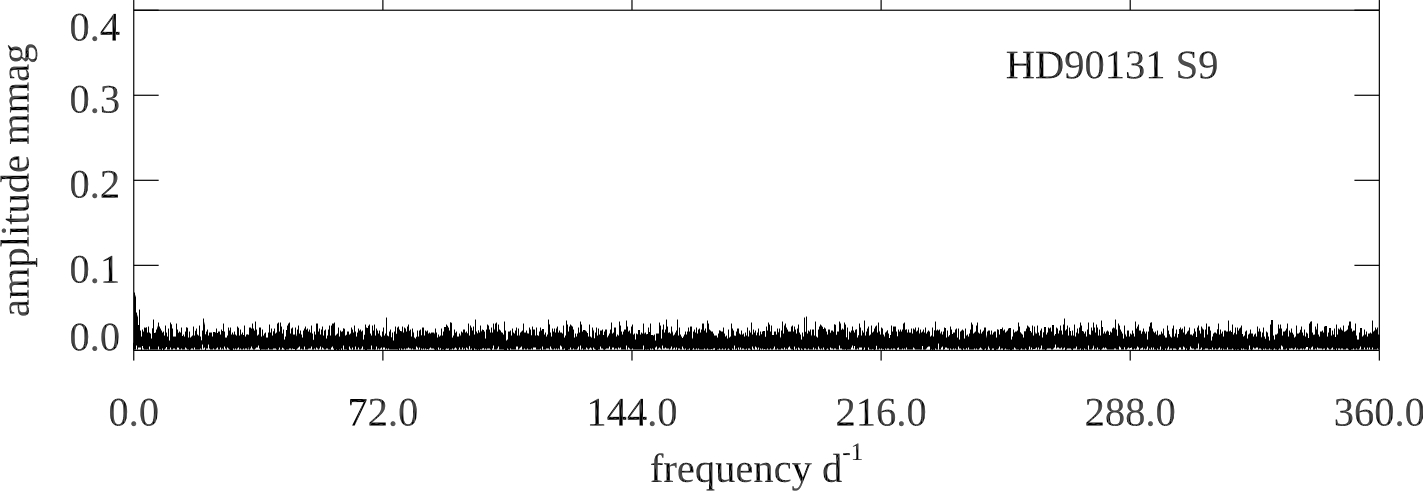}	
\includegraphics[width=0.48\linewidth,angle=0]{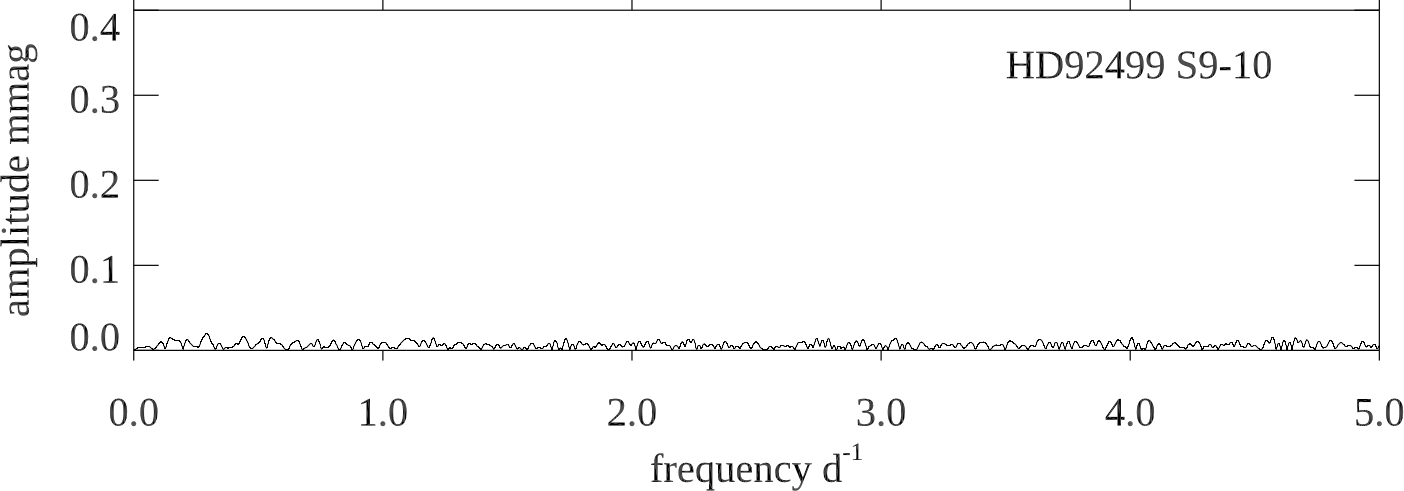}	
\includegraphics[width=0.48\linewidth,angle=0]{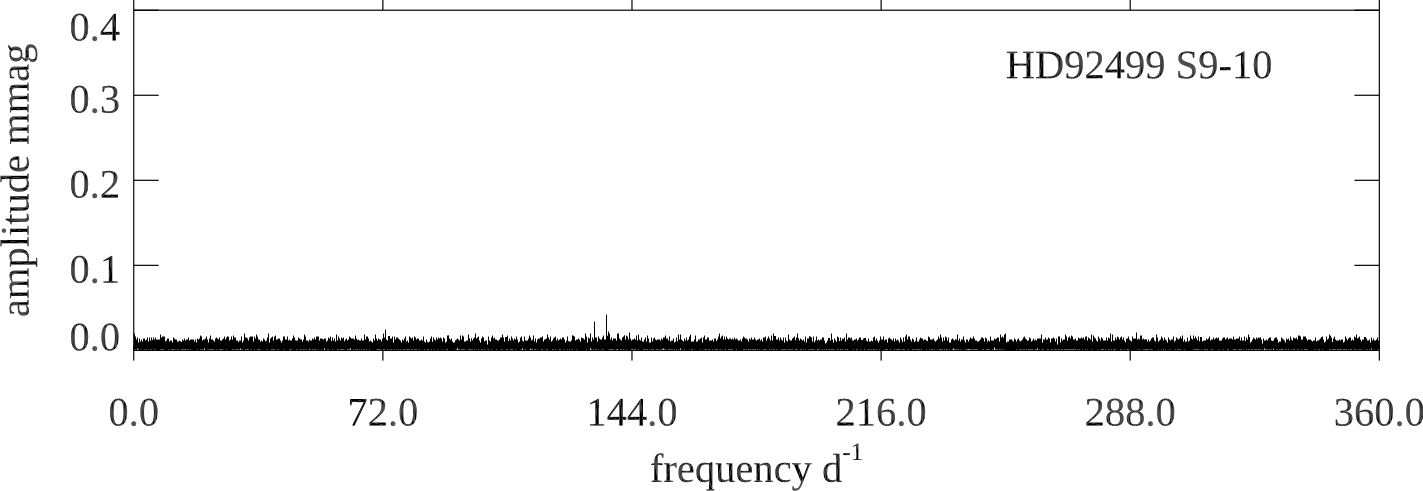}	
\includegraphics[width=0.48\linewidth,angle=0]{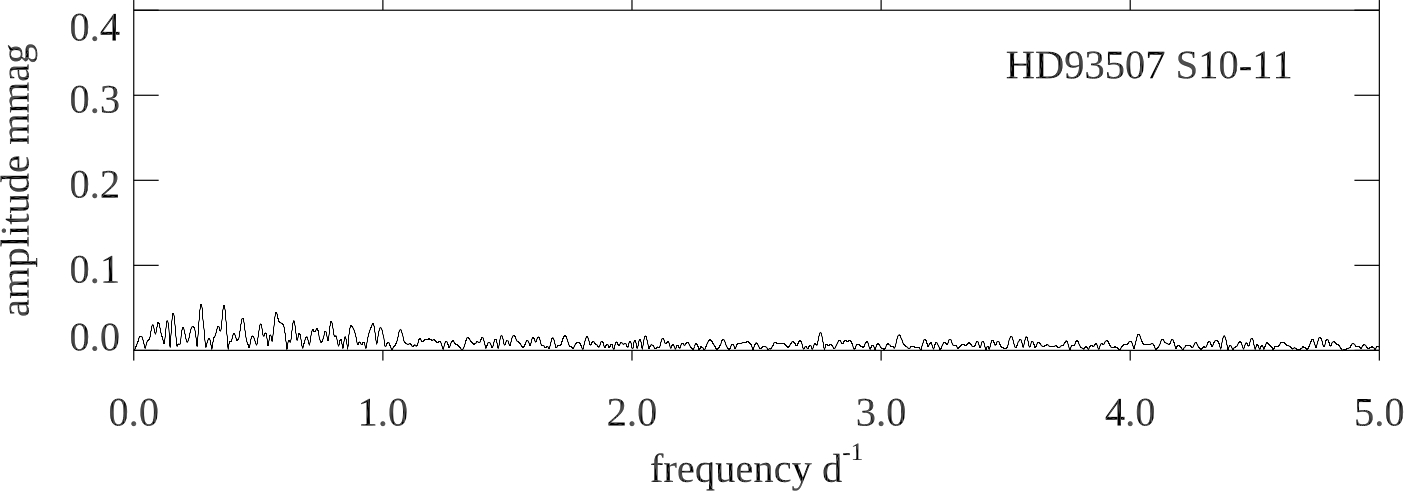}	
\includegraphics[width=0.48\linewidth,angle=0]{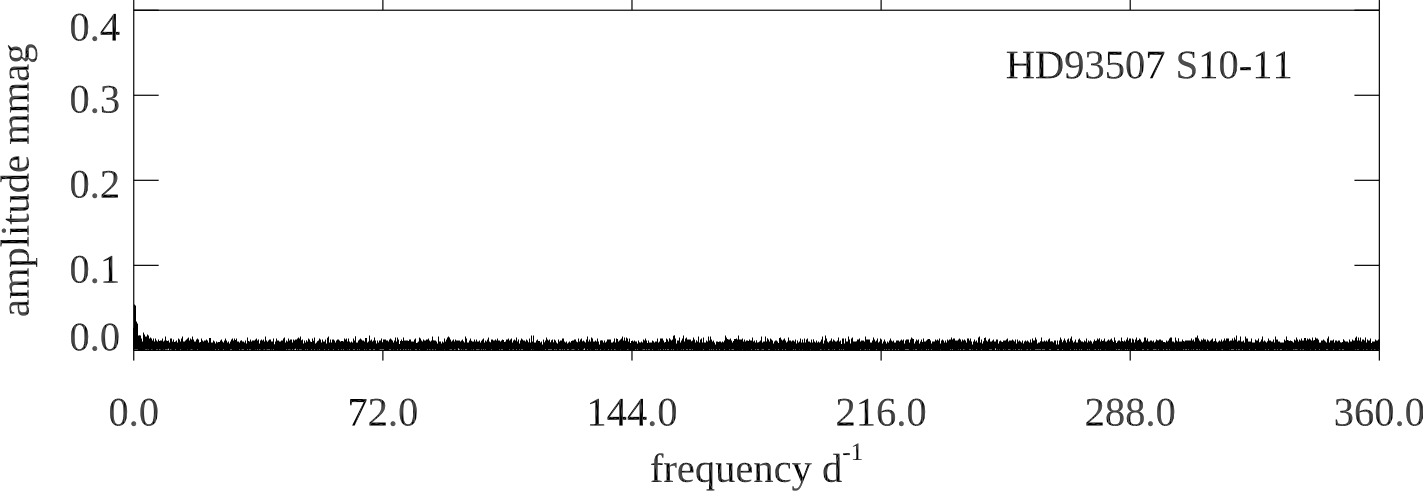}	
\includegraphics[width=0.48\linewidth,angle=0]{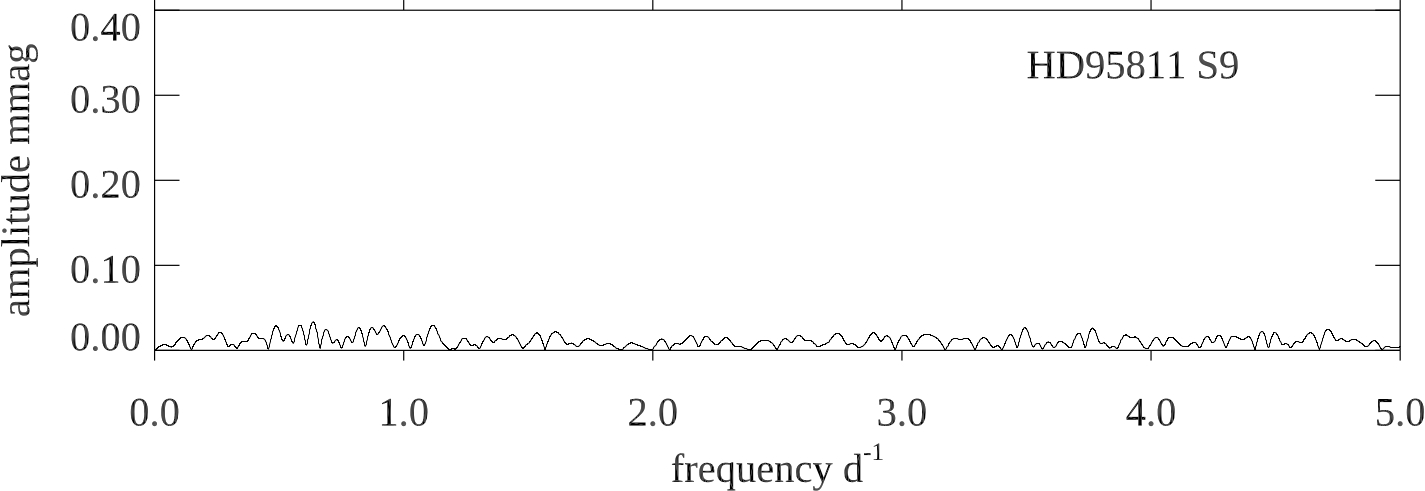}	
\includegraphics[width=0.48\linewidth,angle=0]{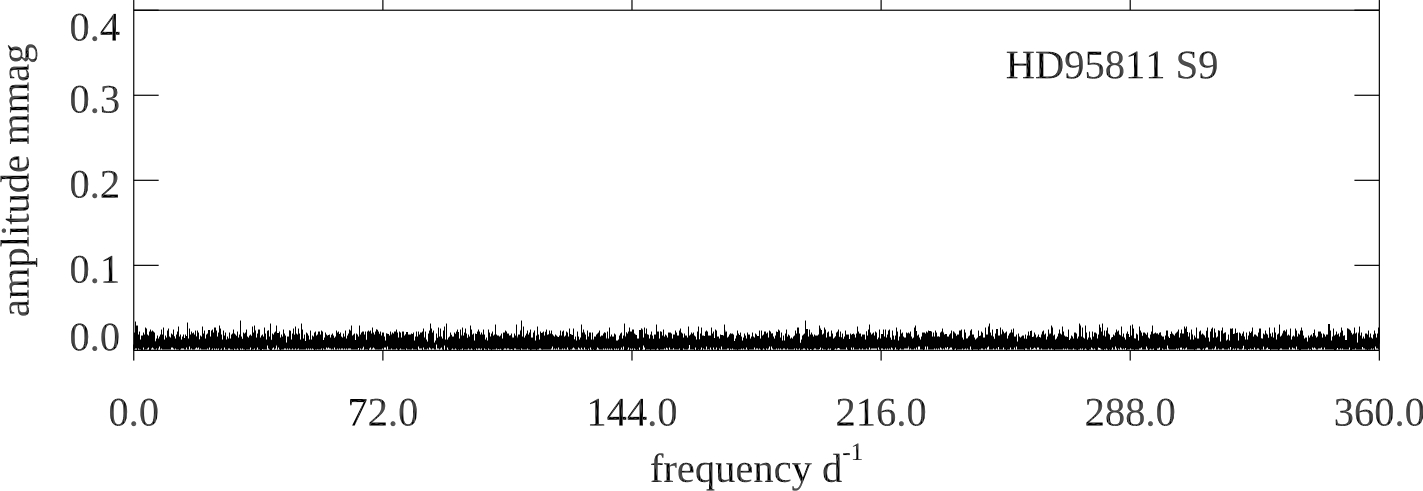}	
\includegraphics[width=0.48\linewidth,angle=0]{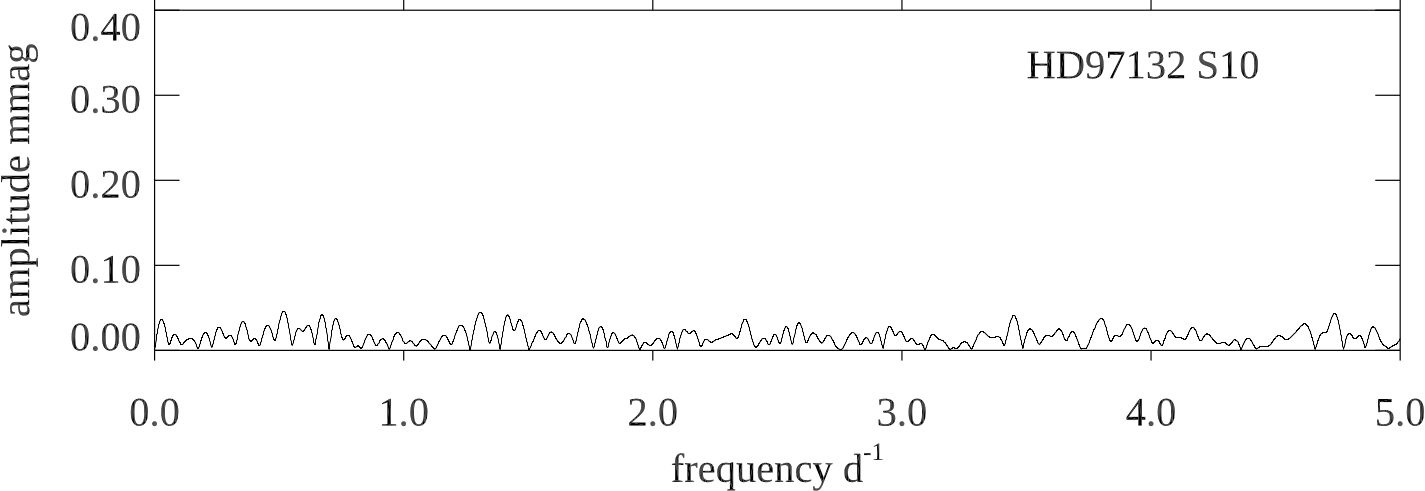}	
\includegraphics[width=0.48\linewidth,angle=0]{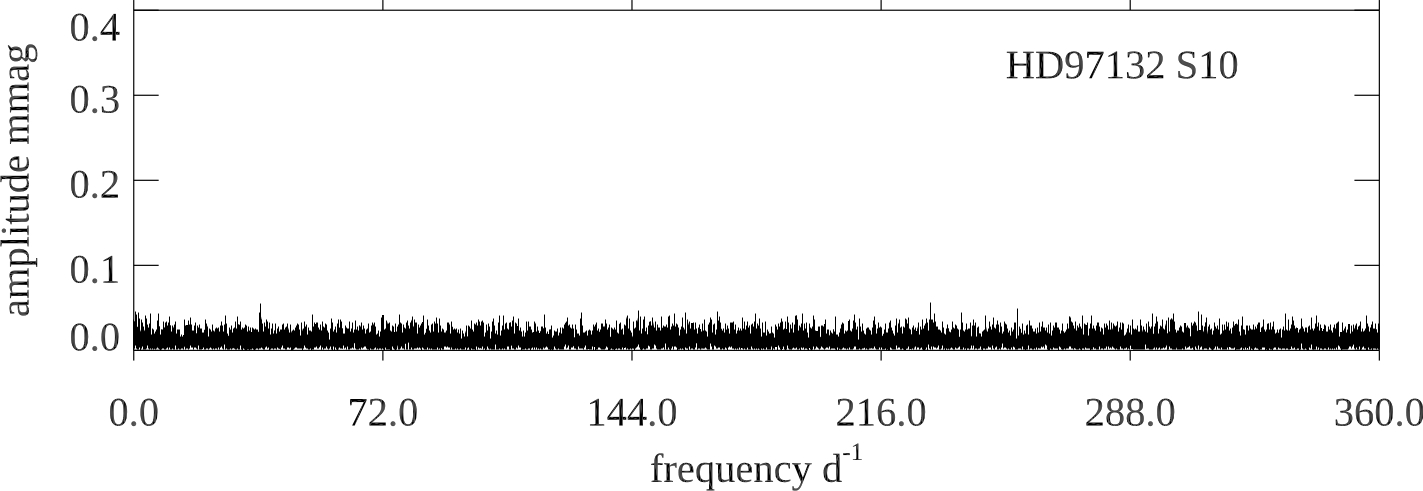}	
\includegraphics[width=0.48\linewidth,angle=0]{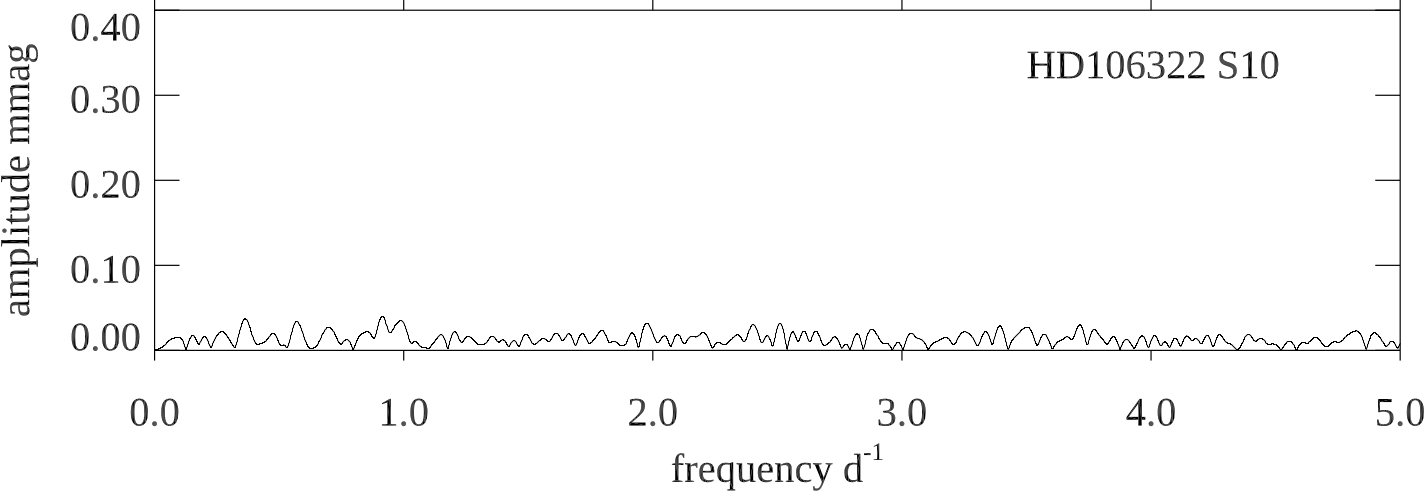}	
\includegraphics[width=0.48\linewidth,angle=0]{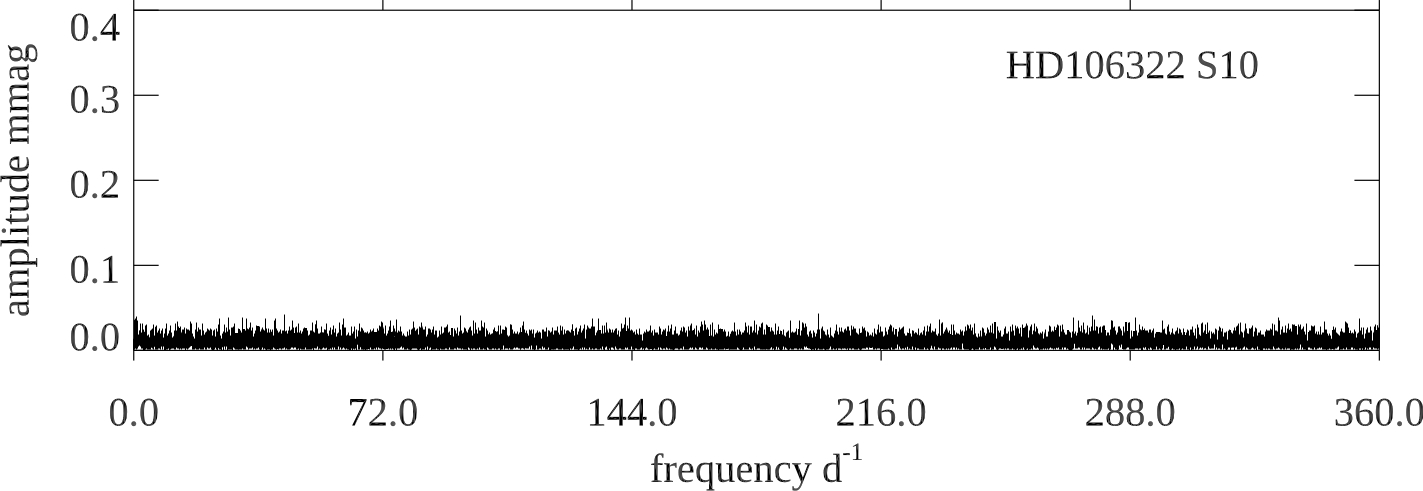}	
\includegraphics[width=0.48\linewidth,angle=0]{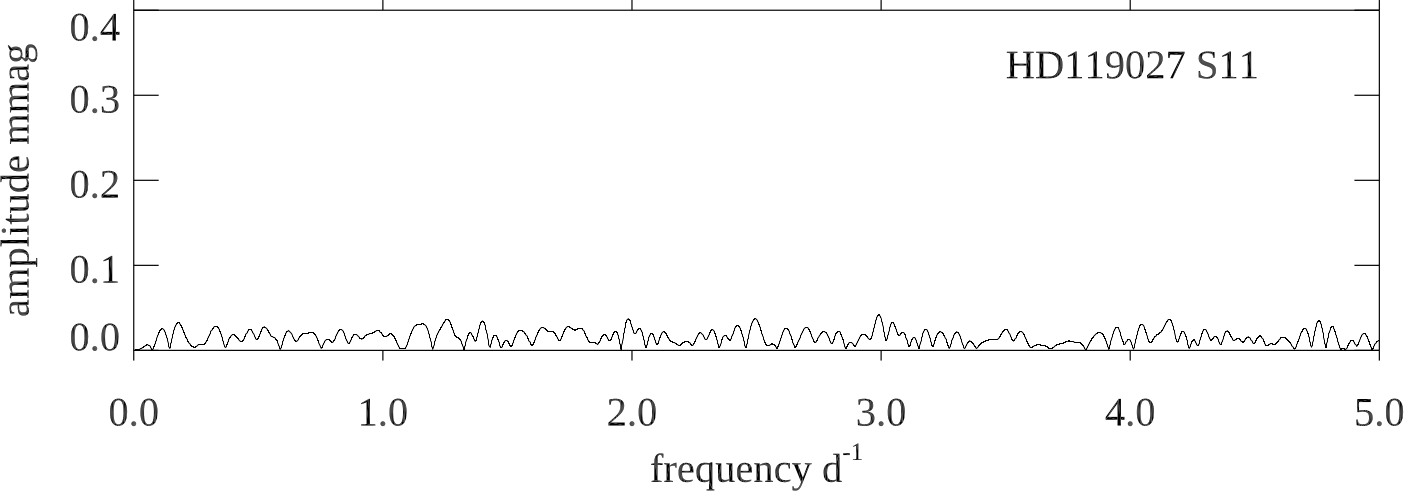}	
\includegraphics[width=0.48\linewidth,angle=0]{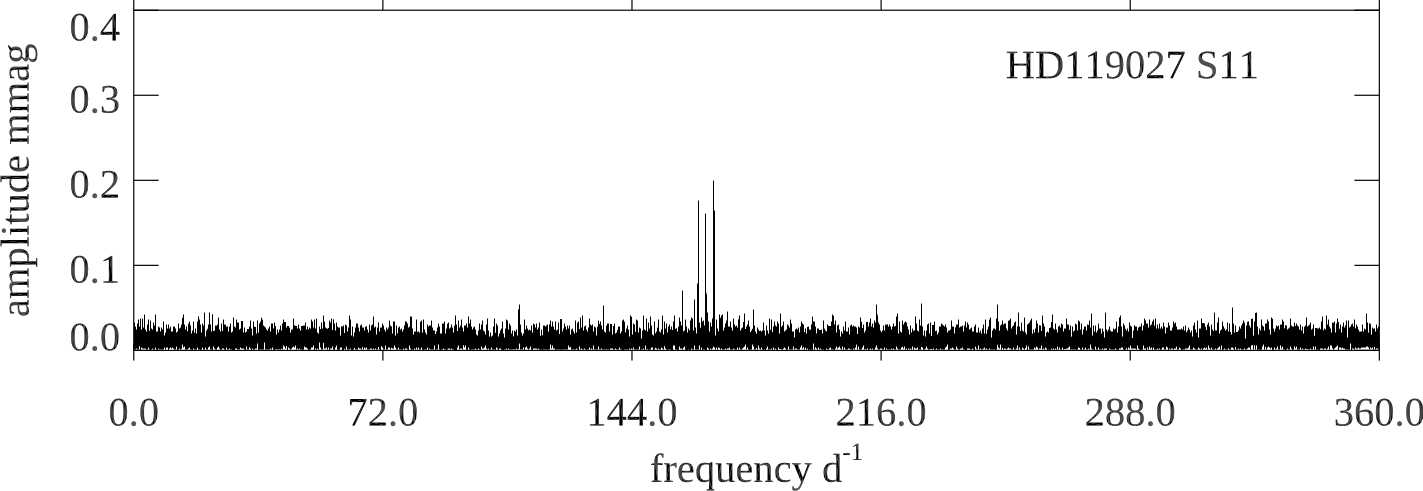}	
\caption{Same as Fig.~\ref{fig:fts_1}. HD~92499 has no rotational variation, but is a known roAp star discovered by \citet{2010MNRAS.404L.104E} with two roAp peaks at $133.2811$\,d$^{-1}$ ($A = 42$\,$\mu$mag) and $136.7034$\,d$^{-1}$ ($A = 34$\,$\mu$mag) in these TESS data; these two frequencies have a separation of 40\,$\mu$Hz, which is plausibly the large separation, or half the large separation. HD~119027 has no rotational variation, but is a known roAp star discovered by \citet{1991IBVS.3611....1M}. \citet{1998MNRAS.300..188M} found this star to multi-periodic, as can be seen in this figure, with a possible large separation 26\,$\mu$Hz. The principal peaks in this figure are separated by 26\,$\mu$Hz. The absolute magnitude and effective temperature given by \citet{2018MNRAS.476..601H} show this star to be near the ZAMS, hence the large separation is more probably 52\,$\mu$Hz, and the modes are of alternating even and odd degree.  }
\label{fig:fts_7}
\end{figure*}

\begin{figure*}[p]
\centering
\includegraphics[width=0.48\linewidth,angle=0]{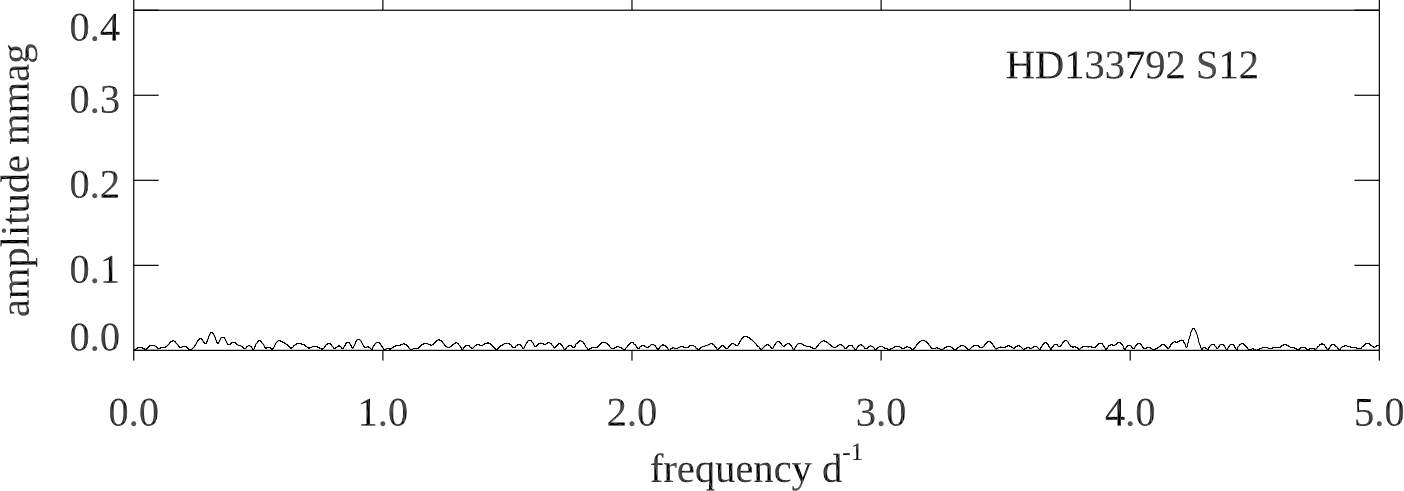}	
\includegraphics[width=0.48\linewidth,angle=0]{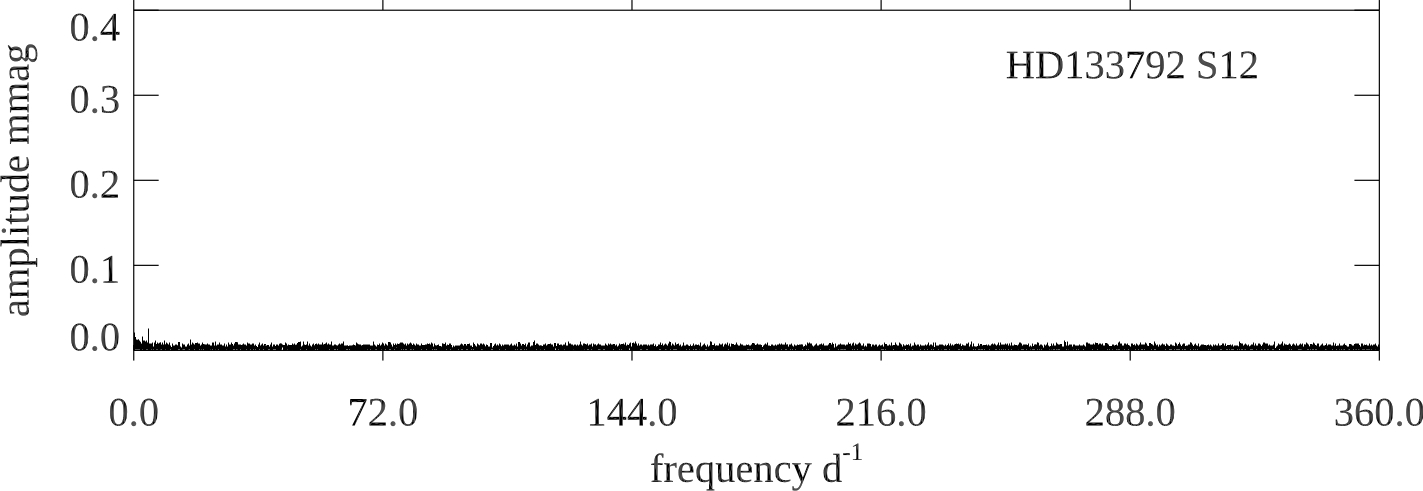}	
\includegraphics[width=0.48\linewidth,angle=0]{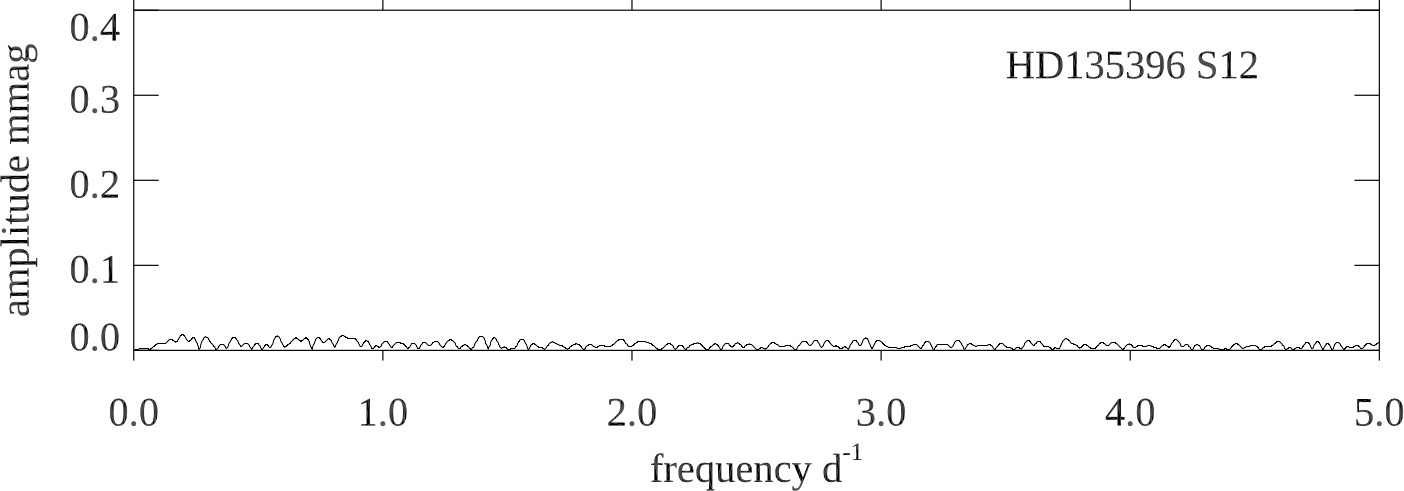}	
\includegraphics[width=0.48\linewidth,angle=0]{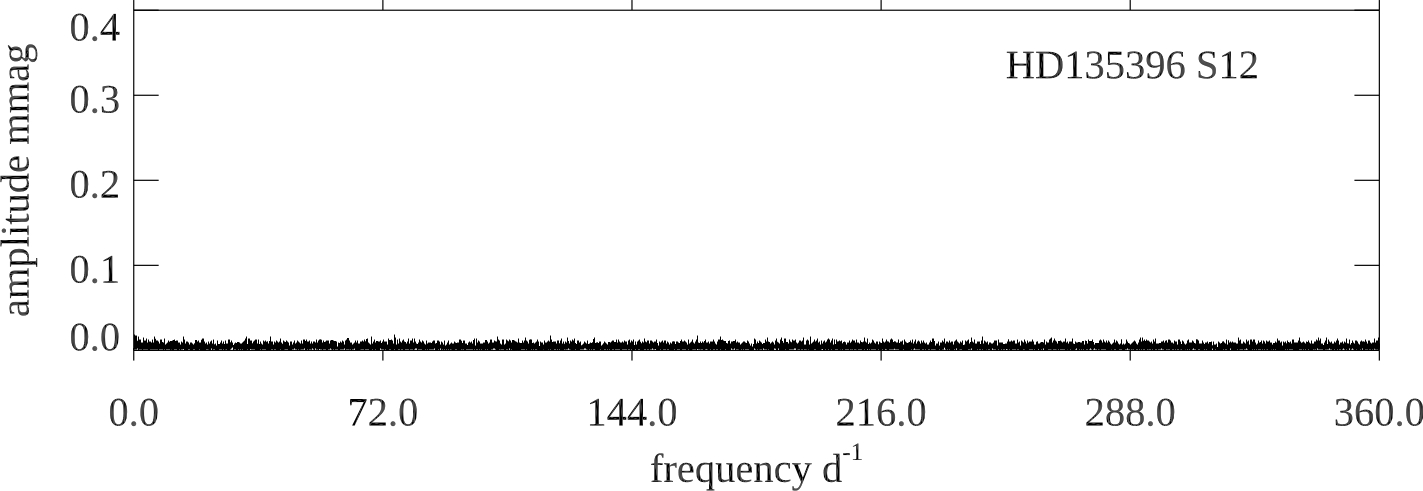}	
\includegraphics[width=0.48\linewidth,angle=0]{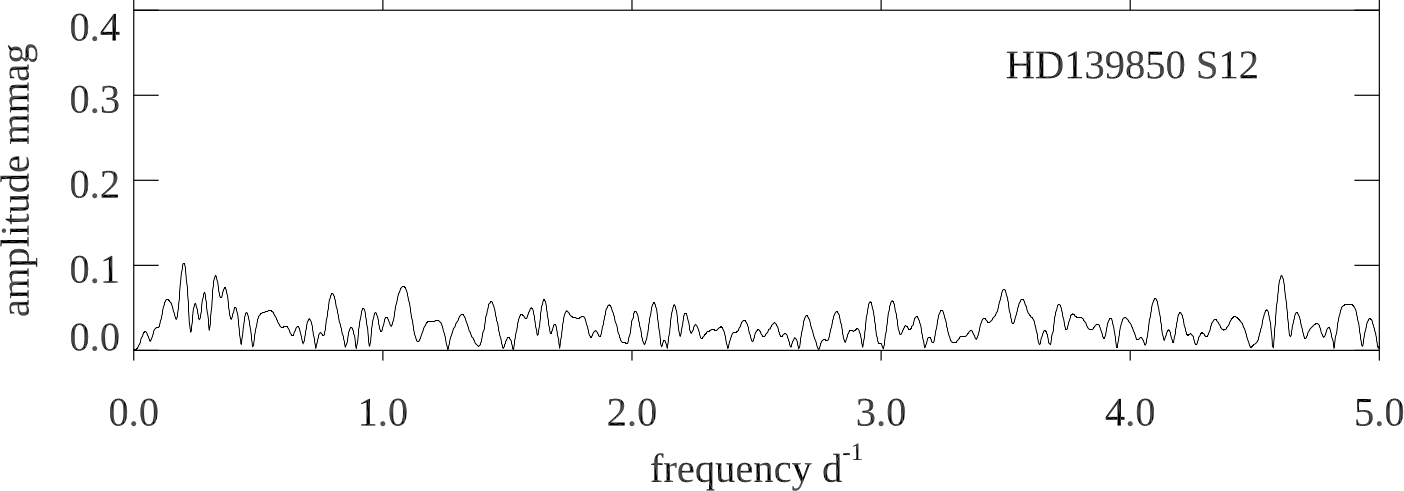}	
\includegraphics[width=0.48\linewidth,angle=0]{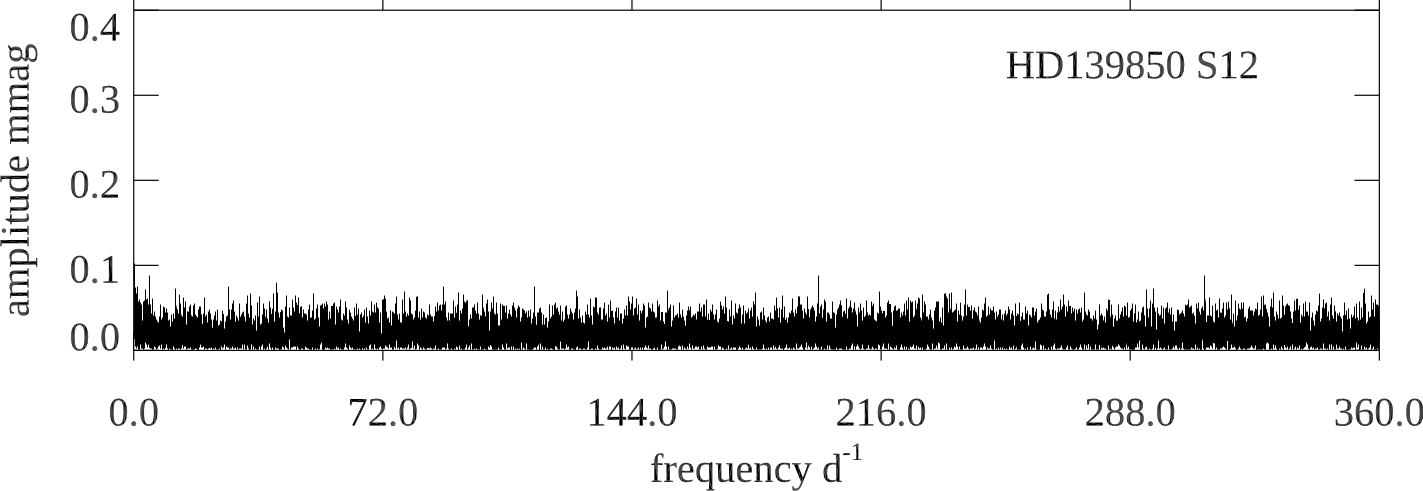}	
\includegraphics[width=0.48\linewidth,angle=0]{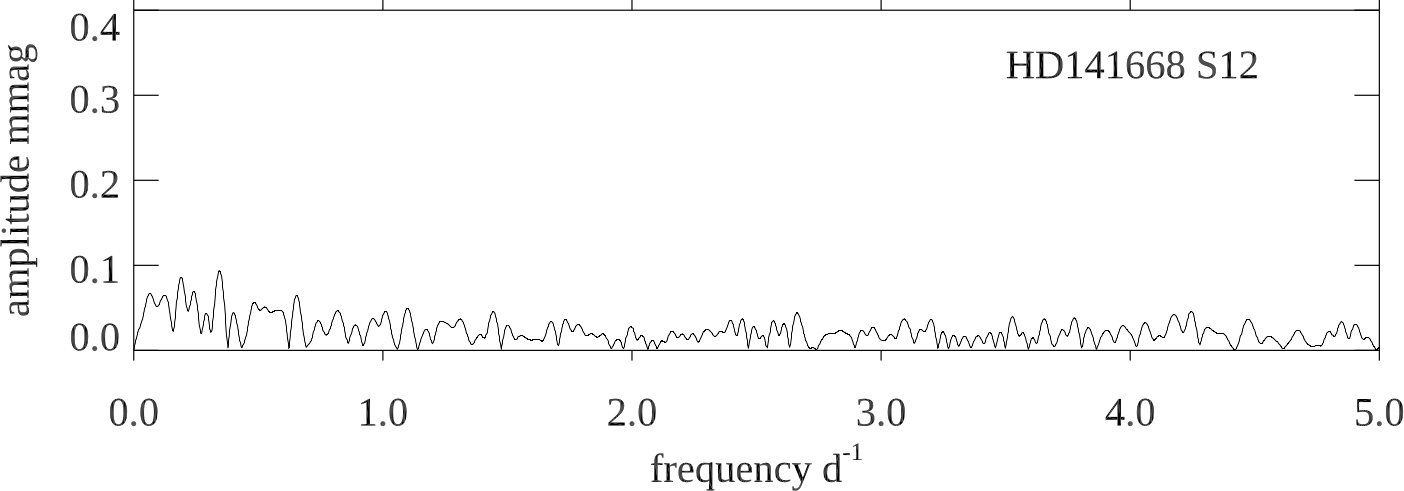}	 
\includegraphics[width=0.48\linewidth,angle=0]{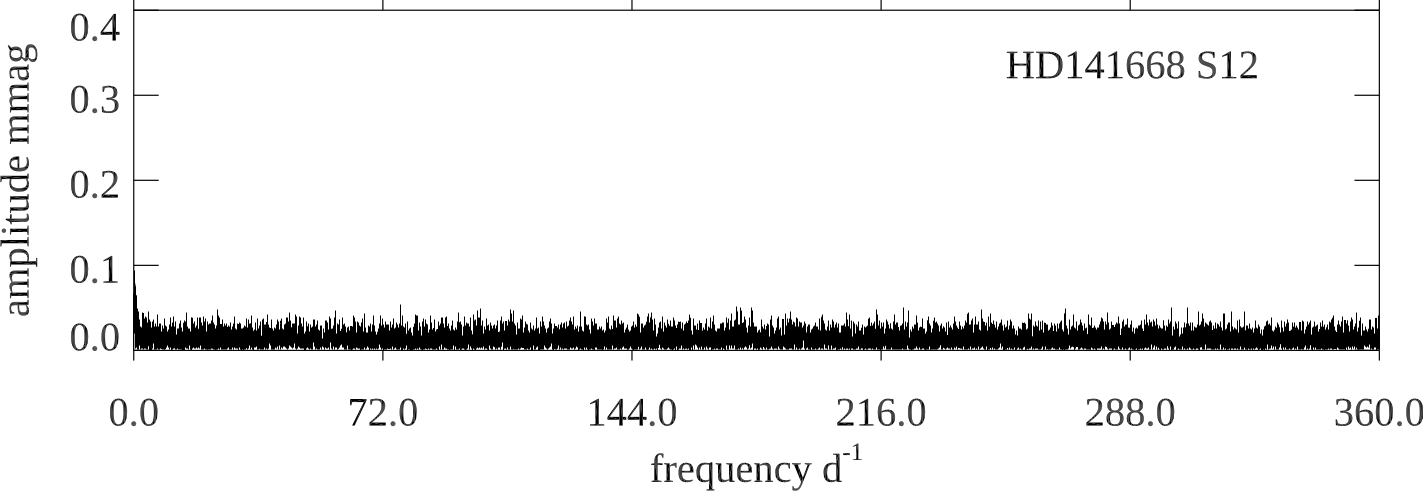}	
\includegraphics[width=0.48\linewidth,angle=0]{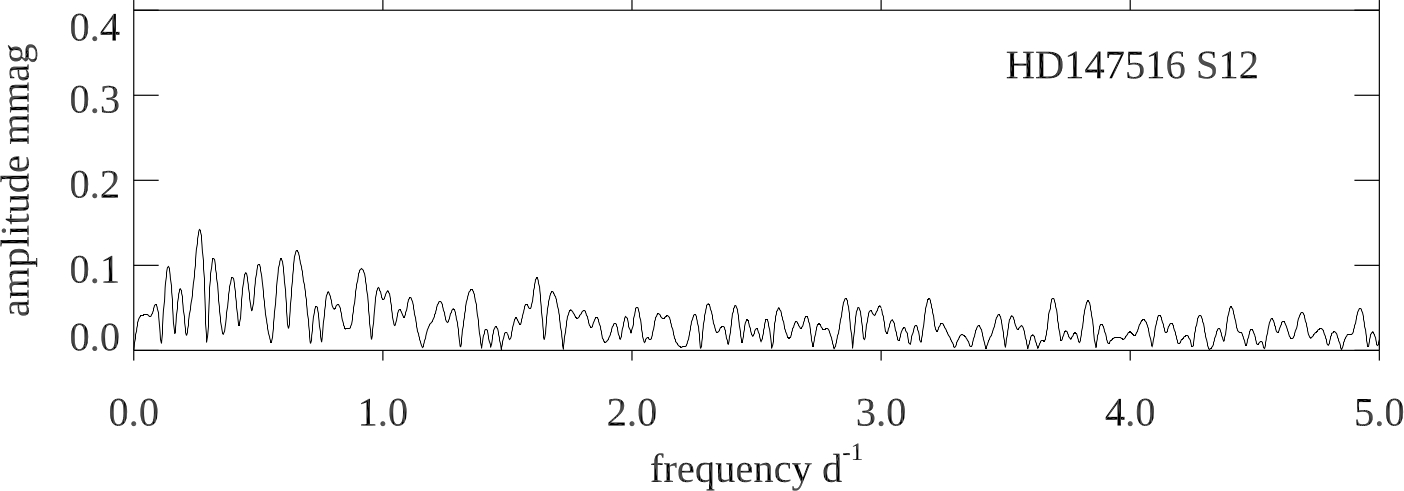}	
\includegraphics[width=0.48\linewidth,angle=0]{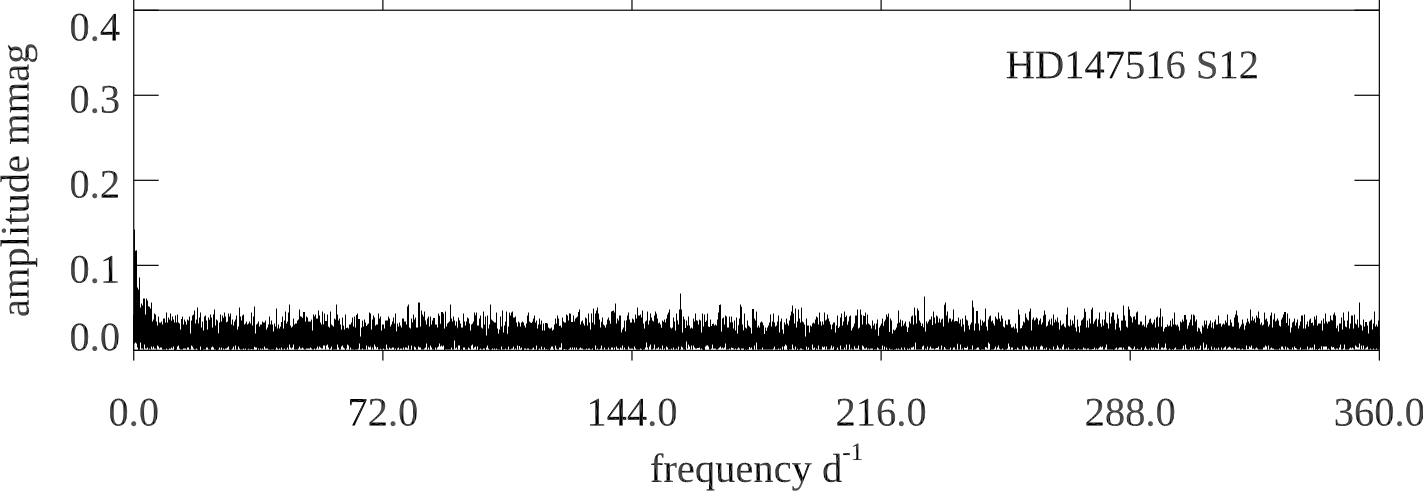}	
\includegraphics[width=0.48\linewidth,angle=0]{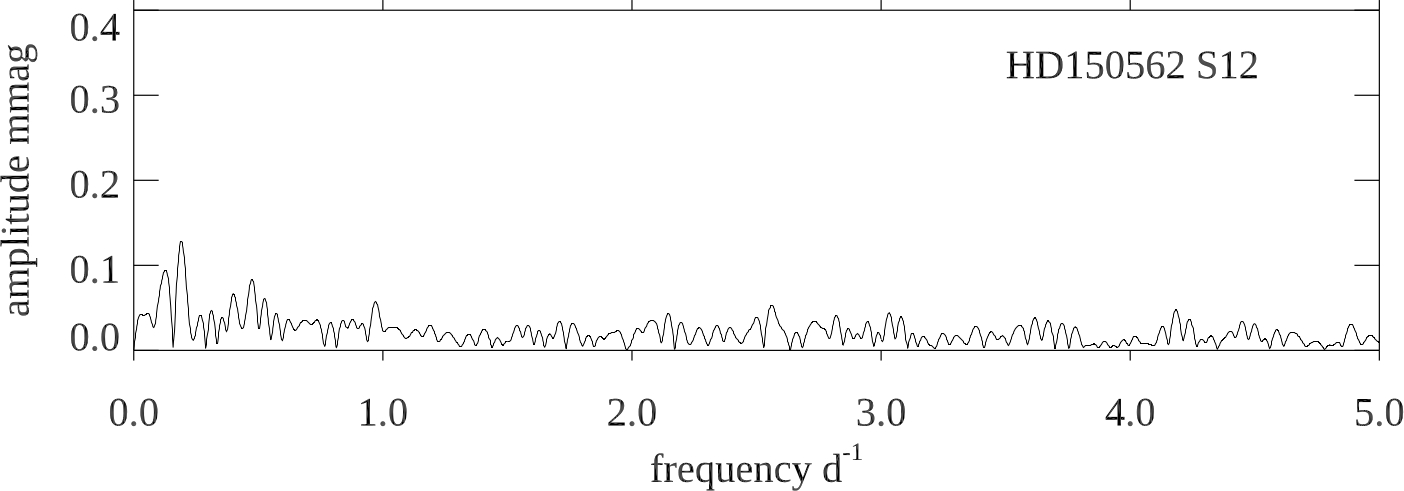}	
\includegraphics[width=0.48\linewidth,angle=0]{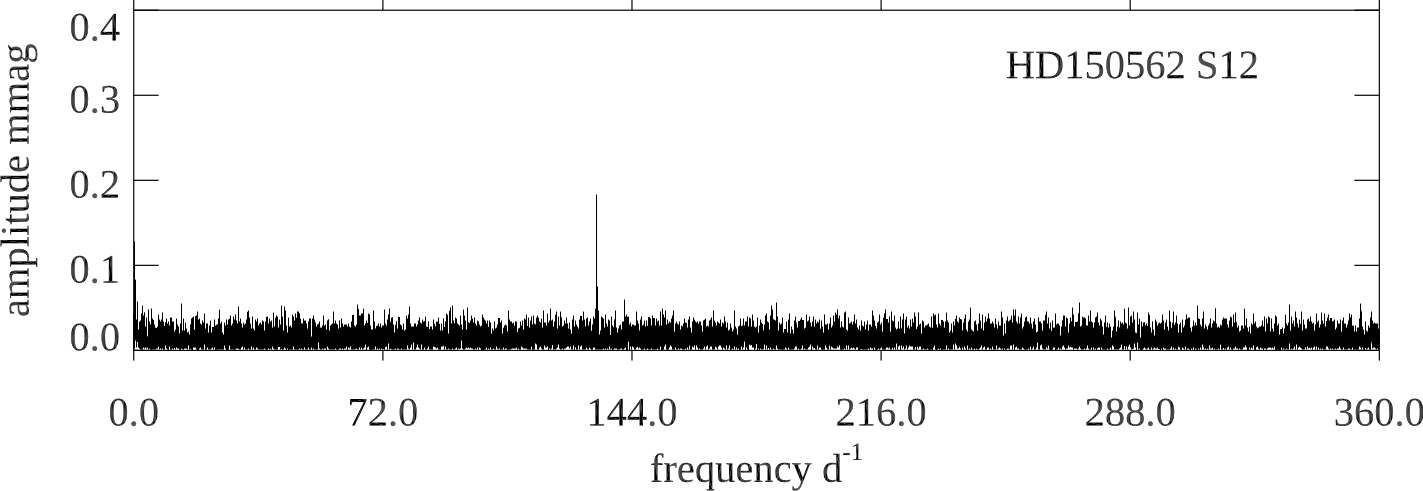}	
\includegraphics[width=0.48\linewidth,angle=0]{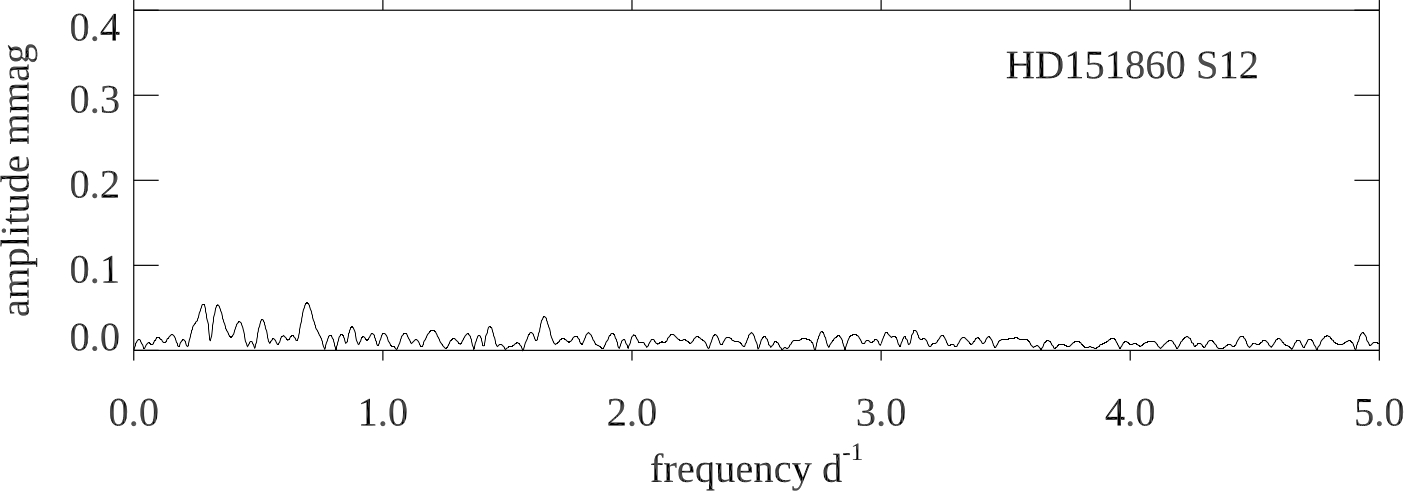}	
\includegraphics[width=0.48\linewidth,angle=0]{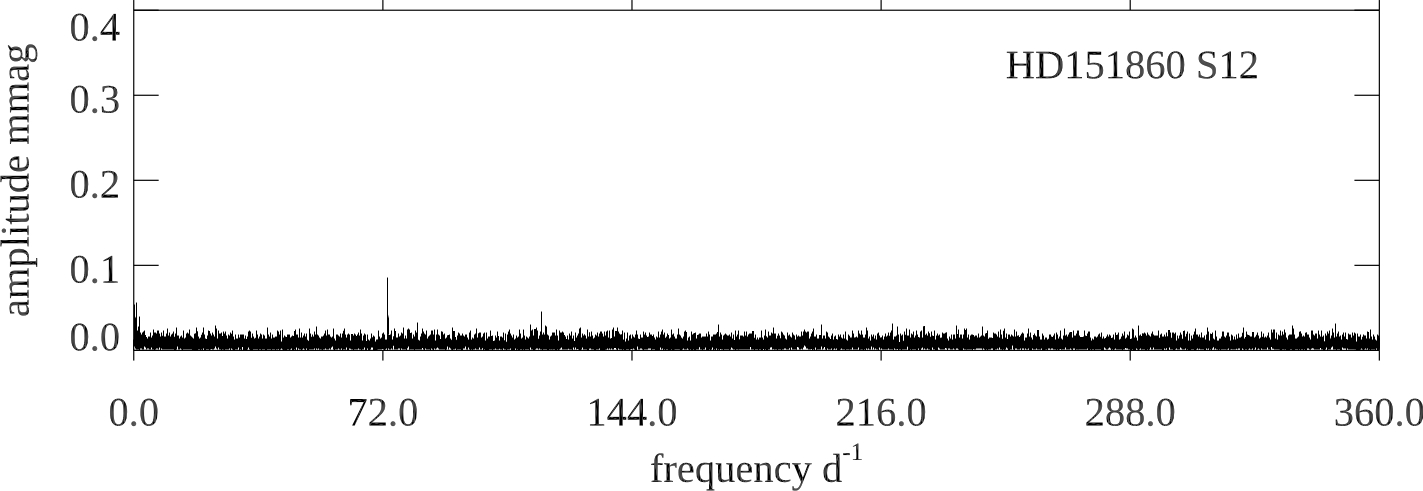}	
\caption{Same as Fig.~\ref{fig:fts_1}. HD~150562 has no clear rotation variation, but obvious roAp variations around 133.7\,d$^{-1}$, which were discovered by \citet{1992IBVS.3750....1M}. HD~151860 was discovered to be a multi-periodic roAp star by \citet{2013MNRAS.431.2808K} with radial velocity variations in Lanthanide lines with a highest amplitude peak near 117\,d$^{-1}$. The TESS photometry shows a highest amplitude pulsation peak at 73.4\,d$^{-1}$ and a second peak at 117.8\,d$^{-1}$, similar to the peak found in radial velocities. Photometry and radial velocities sample different depths in the atmospheres of roAp stars where different mode amplitudes are found, thus there is no inconsistency between the photometric results and the radial velocity study. Simultaneous radial velocity studies with TESS photometry would be synergistic in extracting more astrophysical understanding from these stars. }
\label{fig:fts_8}
\end{figure*}

\begin{figure*}[p]
\centering
\includegraphics[width=0.48\linewidth,angle=0]{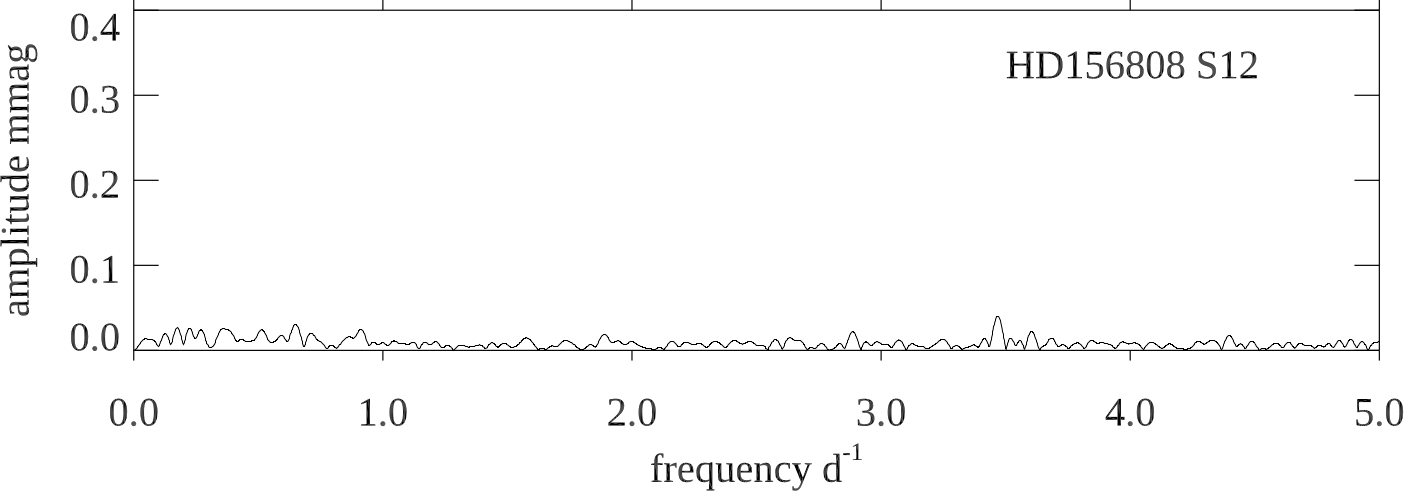}	
\includegraphics[width=0.48\linewidth,angle=0]{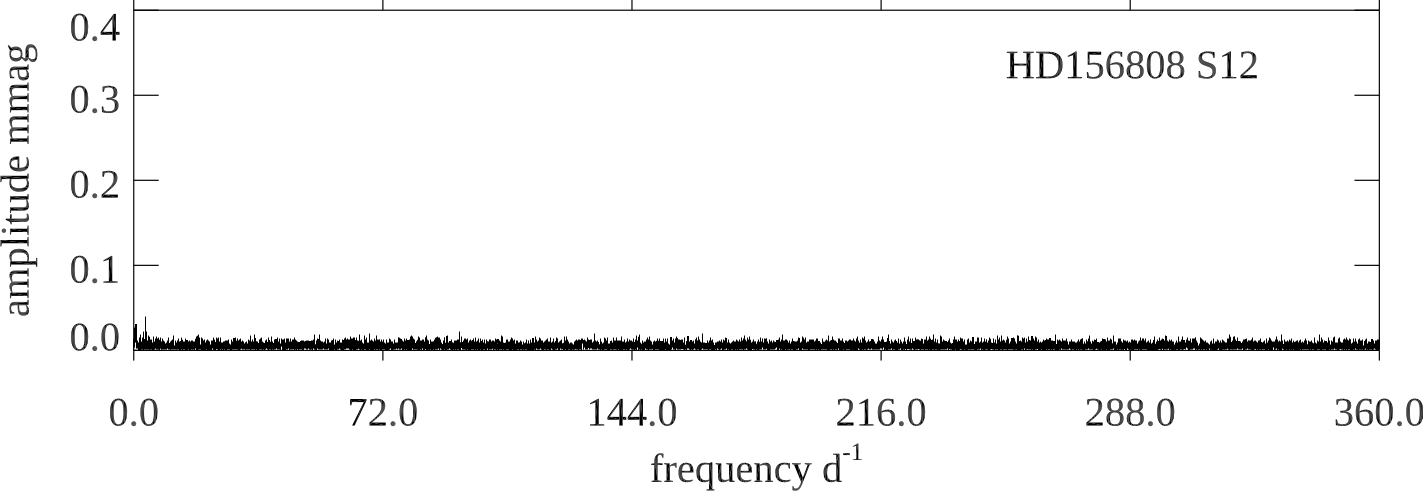}	
\includegraphics[width=0.48\linewidth,angle=0]{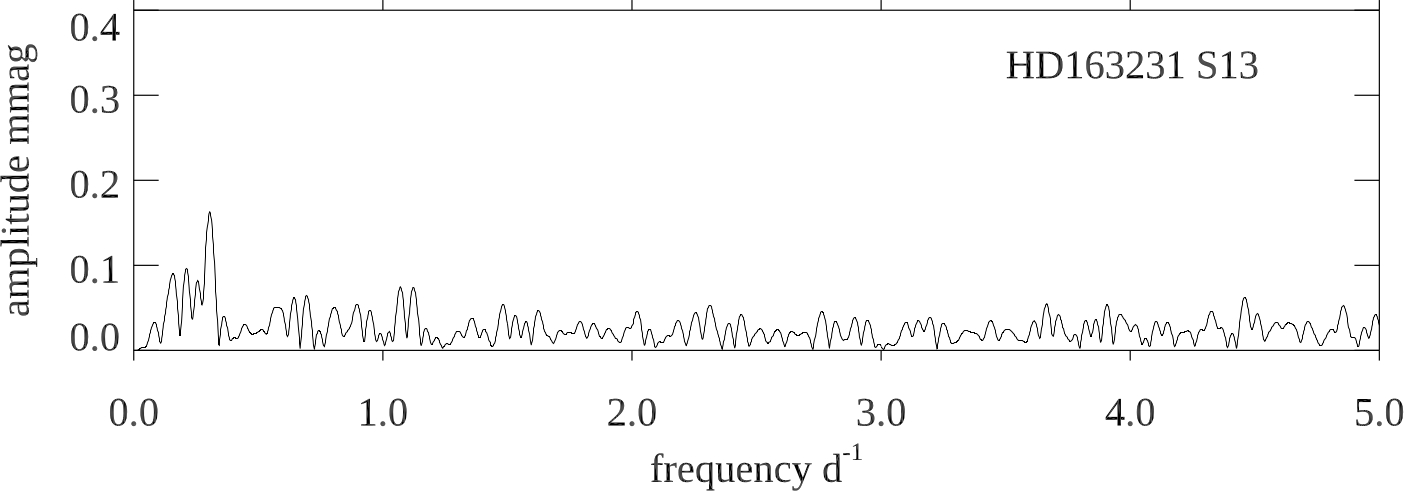}	
\includegraphics[width=0.48\linewidth,angle=0]{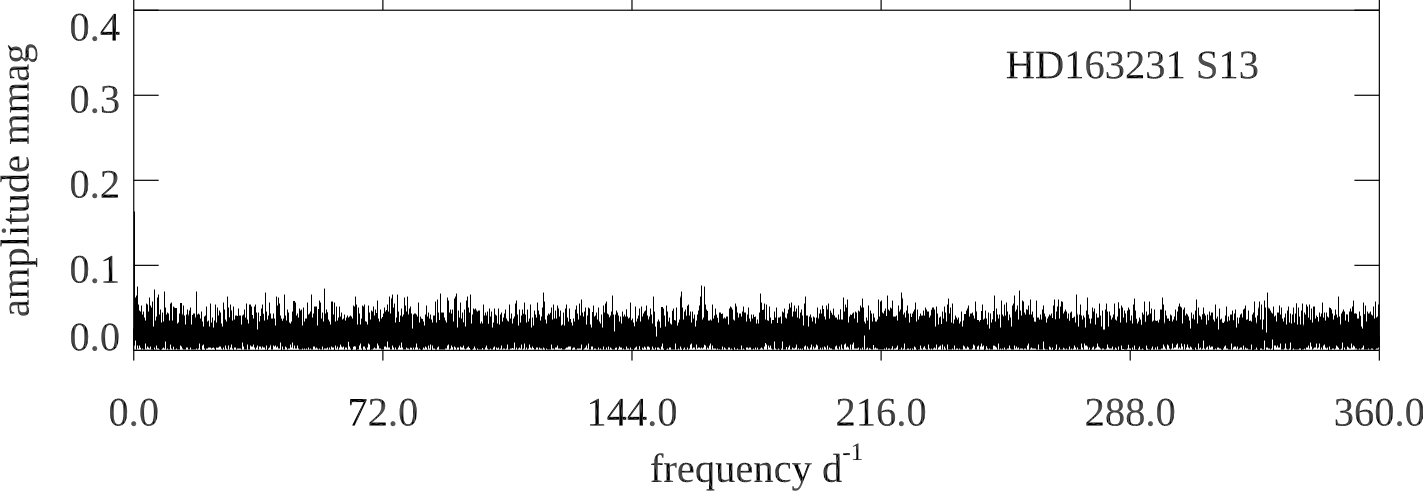}	
\includegraphics[width=0.48\linewidth,angle=0]{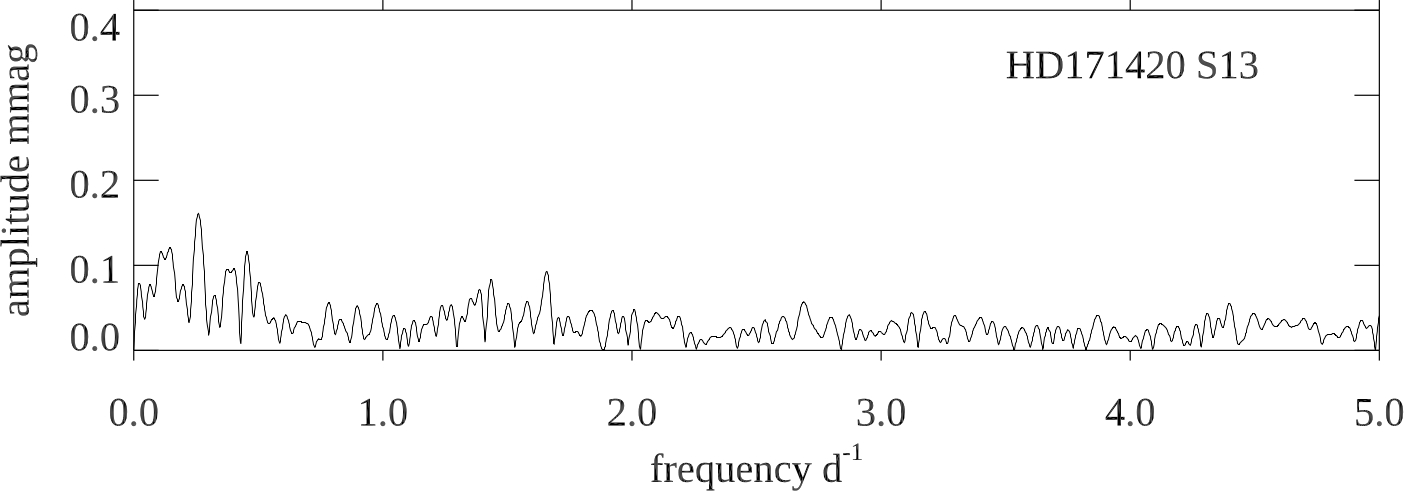}	
\includegraphics[width=0.48\linewidth,angle=0]{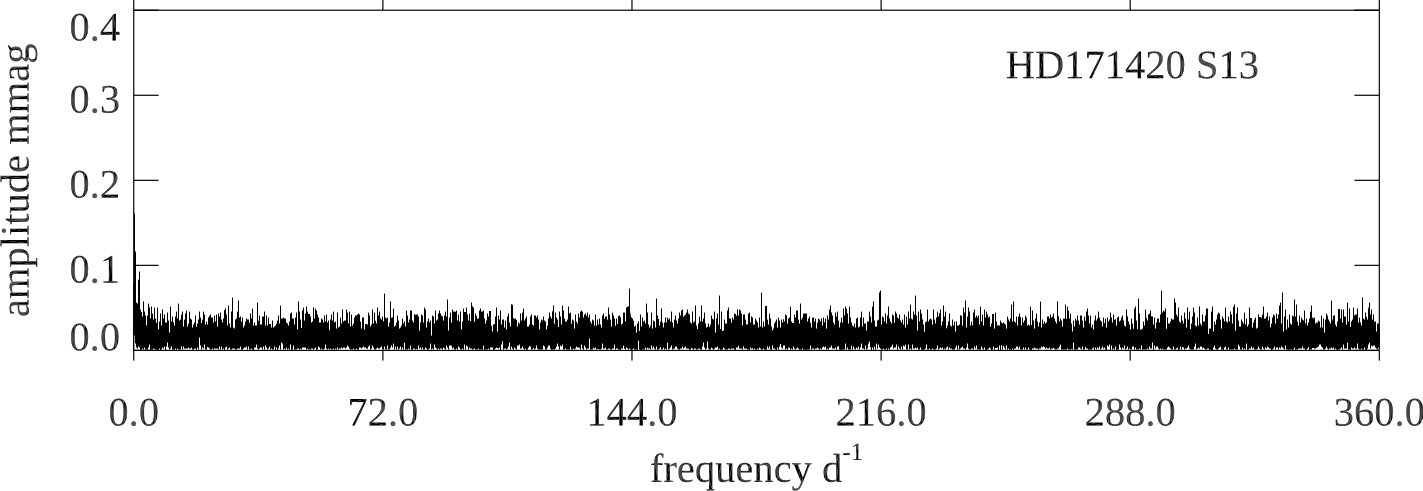}	
\includegraphics[width=0.48\linewidth,angle=0]{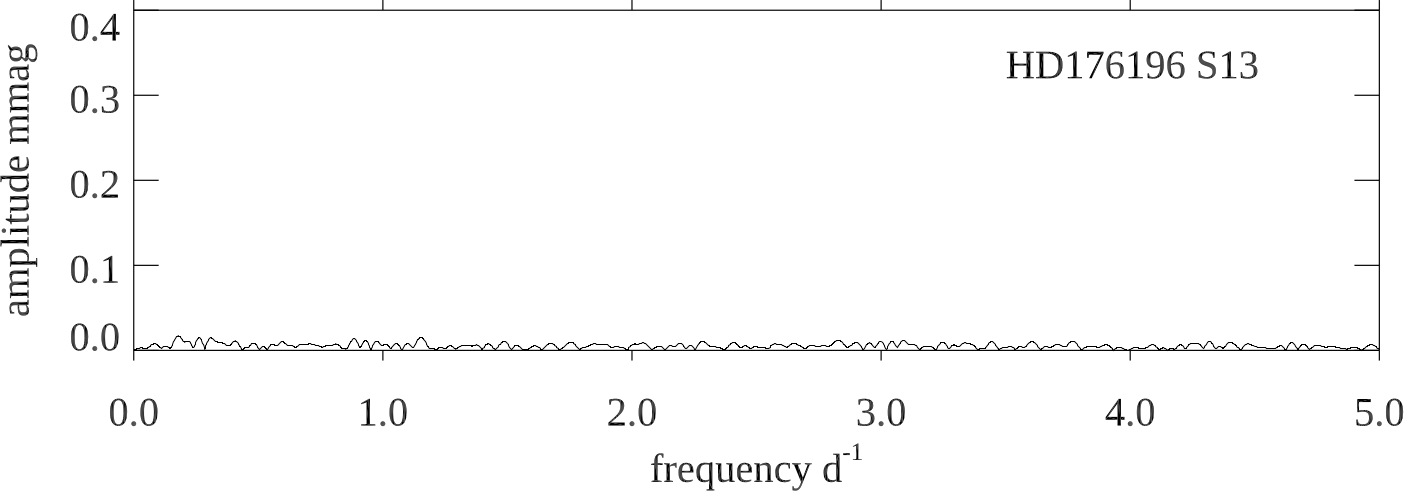}	
\includegraphics[width=0.48\linewidth,angle=0]{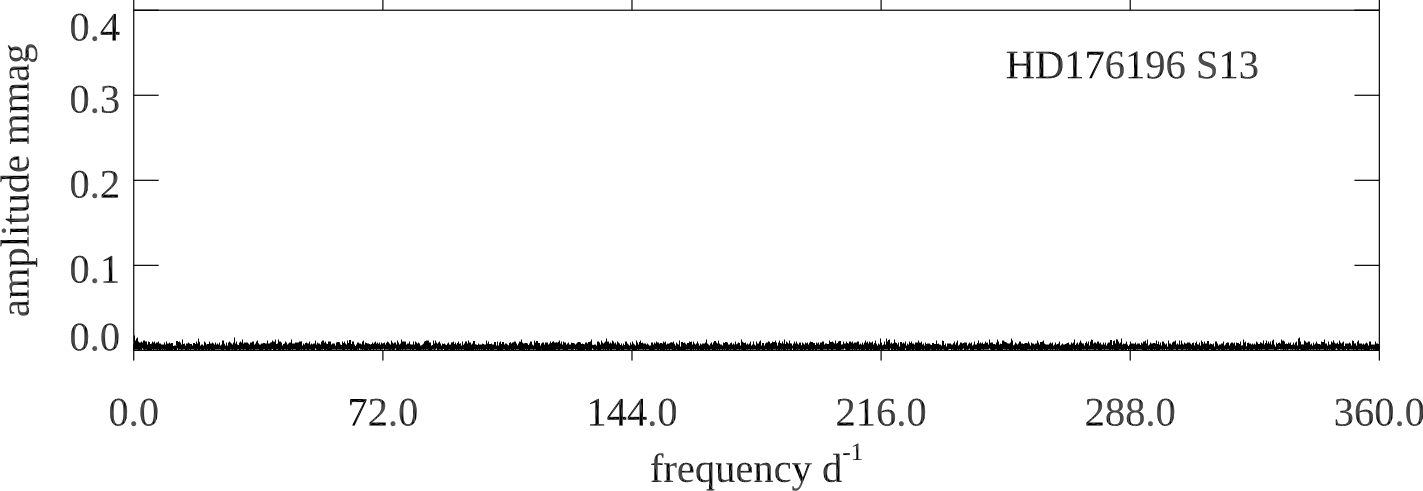}	
\includegraphics[width=0.48\linewidth,angle=0]{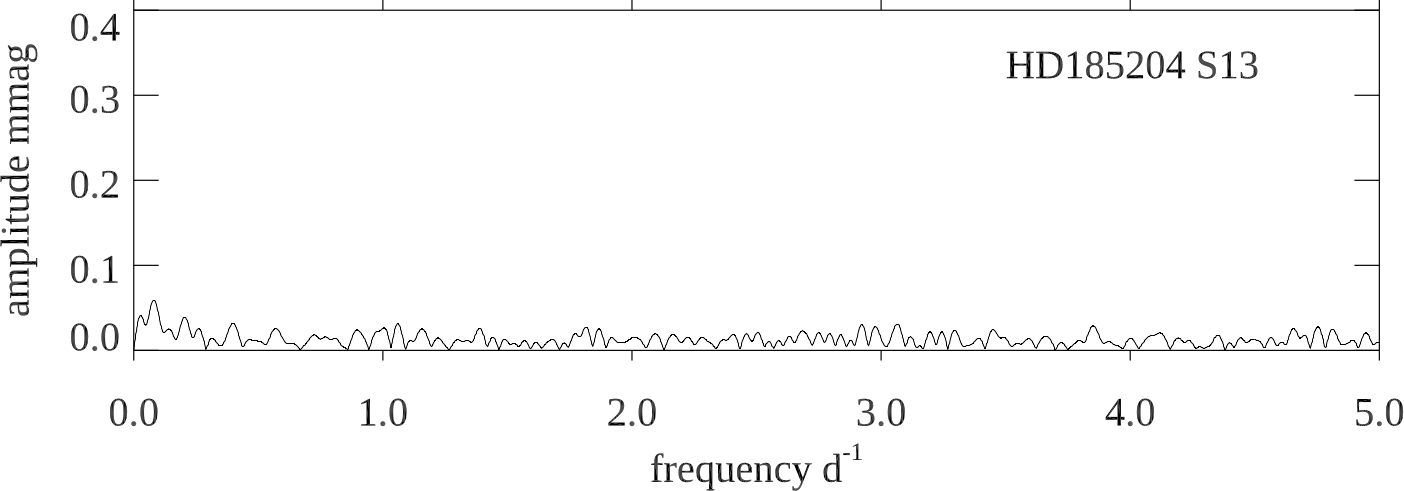}	
\includegraphics[width=0.48\linewidth,angle=0]{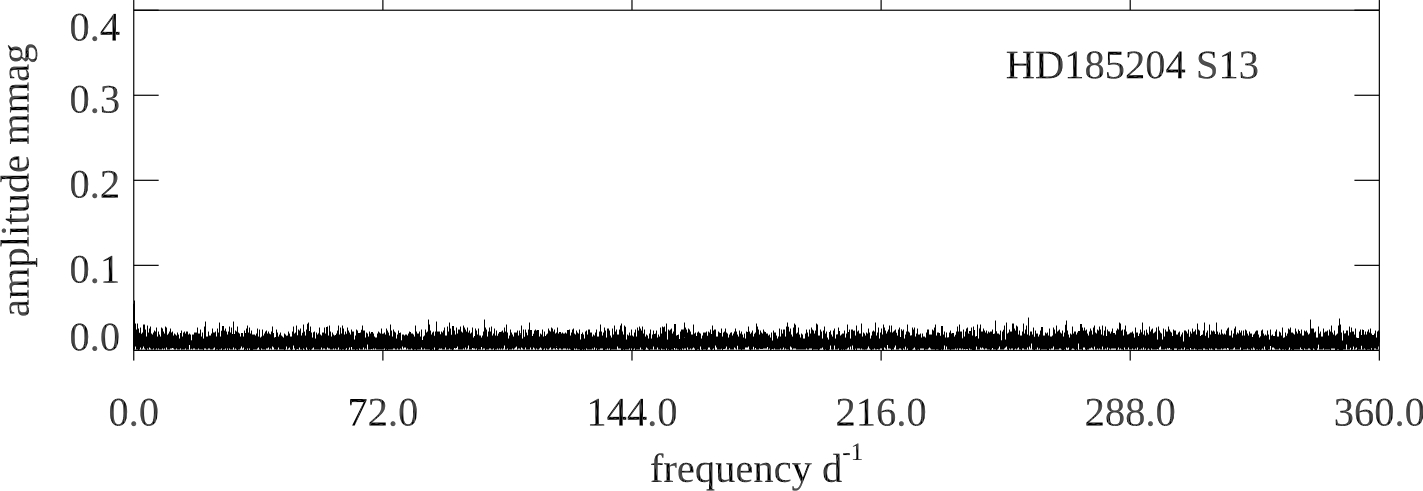}	
\includegraphics[width=0.48\linewidth,angle=0]{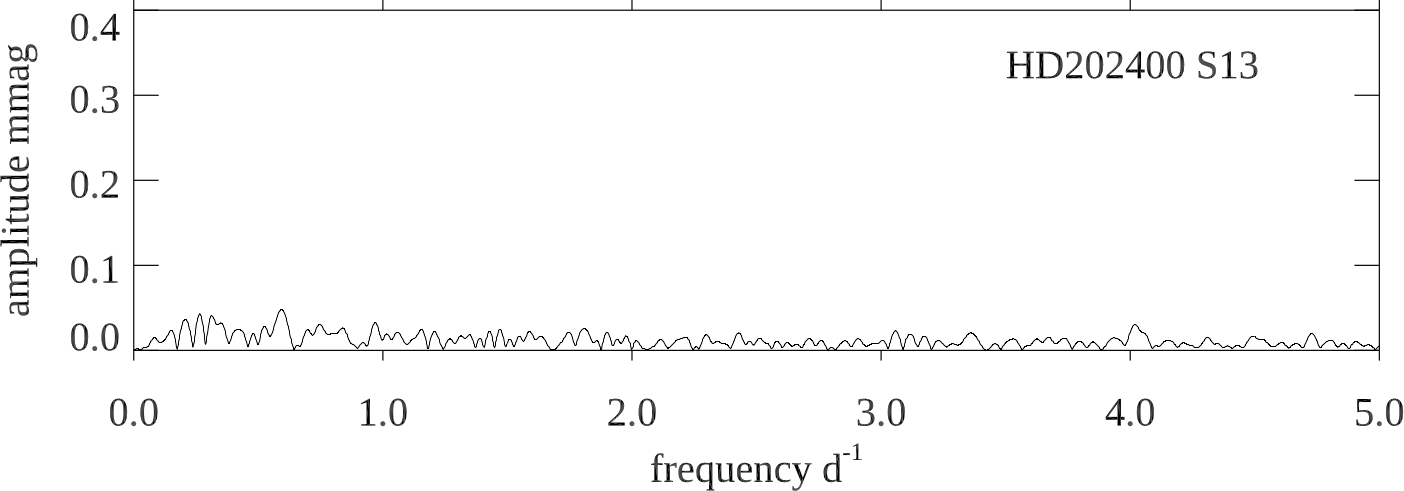}	
\includegraphics[width=0.48\linewidth,angle=0]{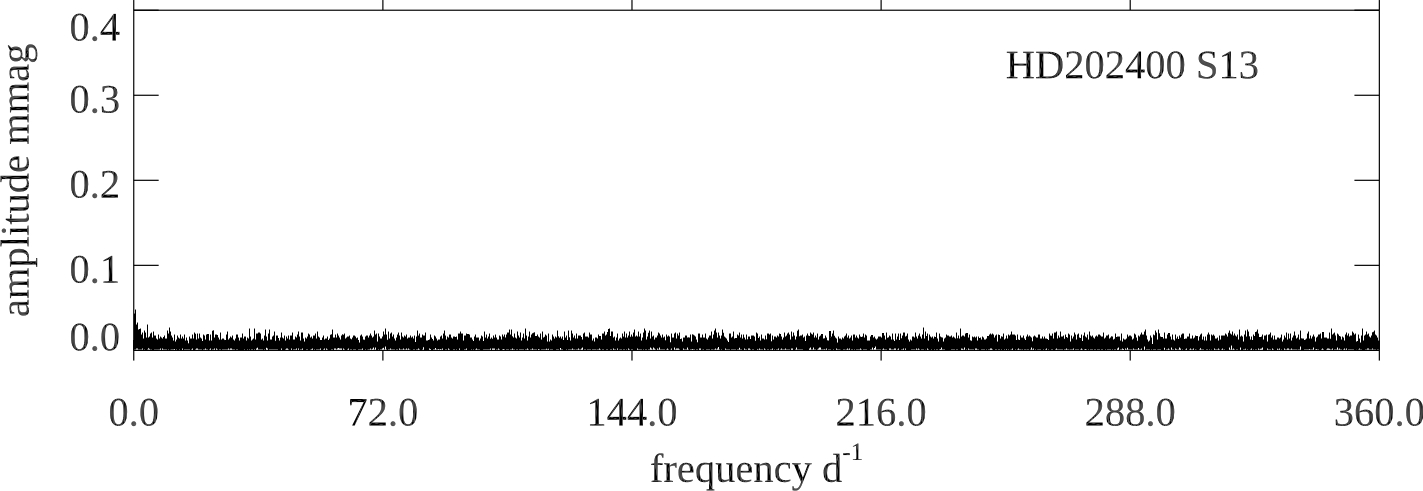}	
\includegraphics[width=0.48\linewidth,angle=0]{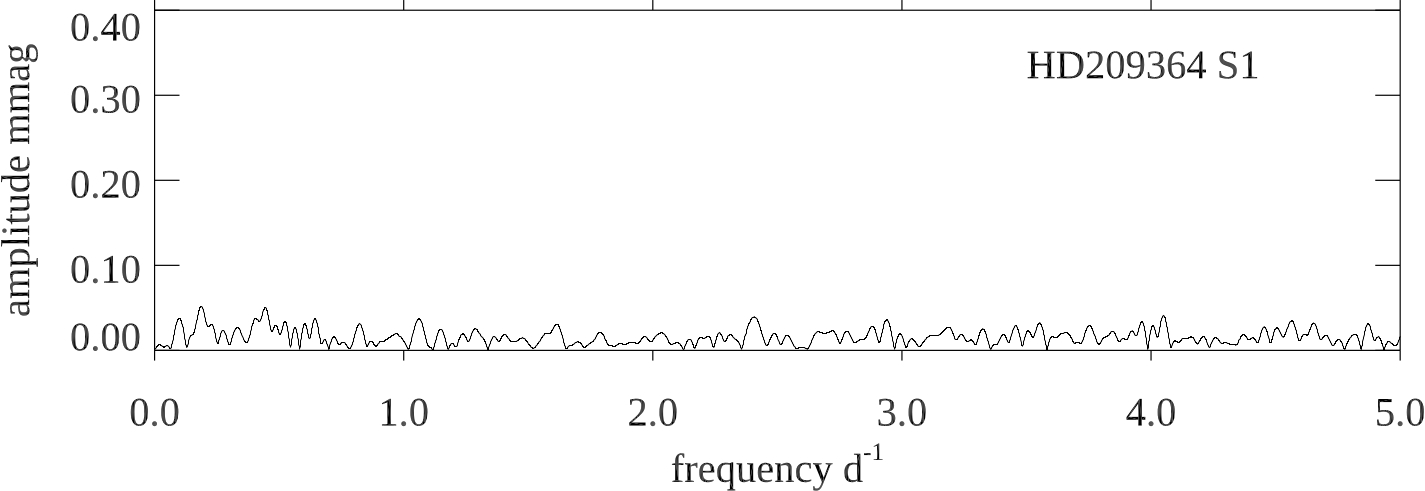}	
\includegraphics[width=0.48\linewidth,angle=0]{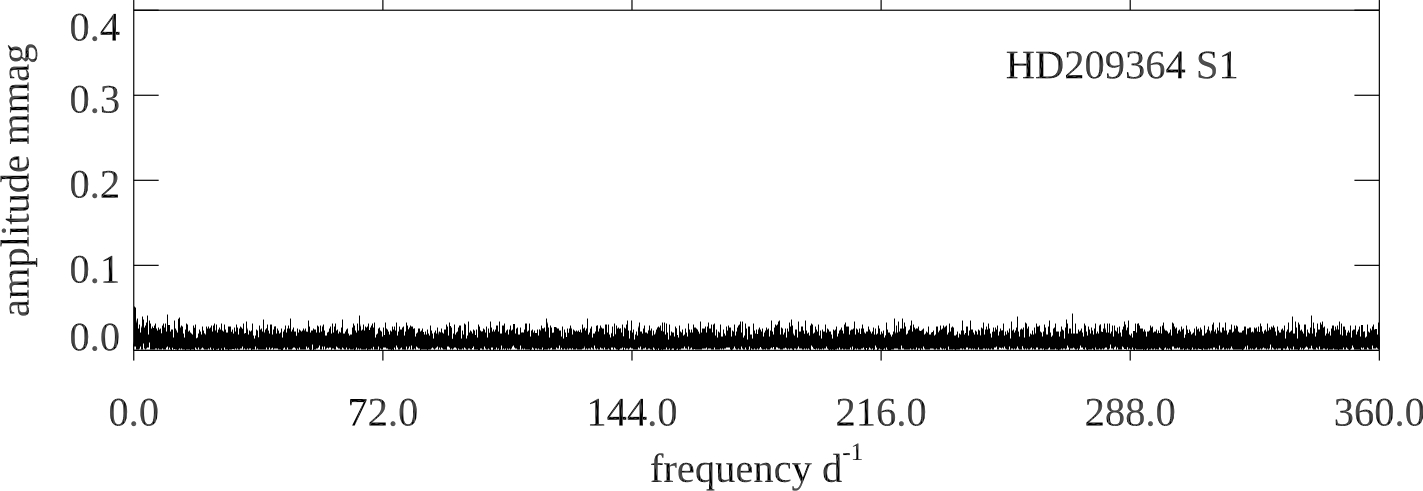}	
\caption{Same as Fig.~\ref{fig:fts_1}. }
\label{fig:fts_9}
\end{figure*}

Figures~\ref{fig:fts_2} to \ref{fig:fts_9} show the amplitude spectra of those Ap stars that we have identified as having long rotation periods and that are not shown in Fig.~\ref{fig:fts_1}.

\afterpage{\clearpage}

\afterpage{\clearpage} 

\afterpage{\clearpage} 

\afterpage{\clearpage} 

\afterpage{\clearpage} 

\afterpage{\clearpage} 

\afterpage{\clearpage} 

\afterpage{\clearpage} 

\end{document}